\documentclass[11pt]{article}
\pdfoutput=1
\usepackage{glas}
\let\ifpdf\relax
\usepackage{fullpage}
\usepackage[mathscr]{euscript}
\usepackage{cite}
\usepackage{graphicx}
\usepackage{amsmath}
\usepackage{amssymb}

\begin{document}
\bibliographystyle{utphys}
\newcommand{\msbar}{\ensuremath{\overline{\text{MS}}}}
\newcommand{\DIS}{\ensuremath{\text{DIS}}}
\newcommand{\abar}{\ensuremath{\bar{\alpha}_S}}
\newcommand{\bb}{\ensuremath{\bar{\beta}_0}}
\newcommand{\rc}{\ensuremath{r_{\text{cut}}}}
\newcommand{\Nd}{\ensuremath{N_{\text{d.o.f.}}}}
\setlength{\parindent}{0pt}

\begin{titlepage}{GLAS-PPE/2012-01}{\today}

\title{ NLO QCD corrections to $\mathbf{tW'}$ and $\mathbf{tZ'}$ production in forward-backward asymmetry models}

\author{J. Adelman$^1$, J. Ferrando$^2$, C. D. White$^2$
\\
$^1$ Department of Physics, Yale University, New Haven CT, USA\\
$^2$ SUPA, School of Physics and Astronomy, University of Glasgow, Glasgow, G12 8QQ, Scotland\\
 }

\begin{abstract}
We consider $Z'$ and $W'$ models recently proposed to explain the top forward-backward asymmetry at the Tevatron. We present the next-to-leading order QCD corrections to associated production of such vector bosons together with top quarks at the Large Hadron Collider, for centre-of-mass energies of 7 and 8 TeV. The corrections are significant, modifying the total production cross-section by 30-50\%. We consider the effects of the corrections on the top and vector-boson kinematics. The results are directly applicable to current experimental searches, for both the ATLAS and CMS collaborations.

\end{abstract}

\end{titlepage}

\section{Introduction}
\label{sec:intro}

Top-quark physics remains a highly active research area, both theoretically
and experimentally. The close proximity of the top-quark mass to the 
electroweak symmetry breaking scale, and the possible role of new physics in
explaining this symmetry breaking, mean that the top-quark sector may provide
a clear window through which to look for beyond the Standard Model effects.
Furthermore, the Large Hadron Collider offers top-quark production rates far
in excess of previous colliders, allowing detailed scrutiny of the top quark
and its interactions. A complementary view is provided by the Tevatron which,
despite having a lower centre of mass energy, has an antisymmetric ($p\bar{p}$)
initial state, which can probe different aspects of top-quark behaviour to the
LHC. One such feature is the forward-backward asymmetry of $t\bar{t}$ 
pairs, which has been measured to be in excess of the Standard Model prediction~\cite{kuhn-rodrigo:1998,almeida:2008,kidonakis:2011,ahrens:2011,hollik:20011,Manohar:2012",Halzen:1987xd} by both D0 and CDF~\cite{Aaltonen:2011kc,Abazov:2011rq,Abazov:2007ab}, and
which has generated much subsequent theoretical interest. \\

In this paper we focus on a particular new physics scenario motivated by the
forward-backward asymmetry, namely the existence of $W'$ and 
(flavour-changing) $Z'$ bosons. These have been considered in a number
of recent studies (see e.g.~\cite{Cheung:2009ch,Cheung:2011qa,
Bhattacherjee:2011nr,Barger:2011ih,Craig:2011an,Chen:2011mga,Yan:2011tf,
Knapen:2011hu,Jung:2009jz,Xiao:2010hm,Cao:2011ew,Berger:2011ua,
AguilarSaavedra:2011zy,Jung:2011ue,Duraisamy:2011pt,Ko:2012gj,Ko:2011vd,
Berger:2011sv,Gresham:2011dg,Duffty:2012zz,Ayazi:2012bb,Zhang,
AguilarSaavedra:2011ug,Jung:2011id,Berger:2011xk,Cao:2011hr,Jezo:2012rm,
Cao:2012ng}). In some of these works (e.g.~\cite{Barger:2011ih}), the new 
gauge bosons emerge from a complete beyond the Standard Model theory with 
extended gauge group. In other more phenomenological studies, the fundamental 
origin of the new gauge bosons is not spelled out explicitly. This is 
sufficient for the study of particular scattering processes, in that the 
interaction Lagrangian for a $W'$ or $Z'$ boson with Standard Model quarks 
can be written in a generic form which is independent of the underlying 
theory. Note also that coupling and parameter definitions (including overall 
normalisations) differ in the above literature. Here we will focus our 
discussion explicitly on the model of~\cite{Gresham:2011dg}. \\

Although motivated by the sizeable forward-backward asymmetry, $W'$ and $Z'$ 
models must face a number of stringent experimental constraints, including 
precision electroweak observables~\cite{EWWG:2010aa}; the neutron electric 
dipole moment~\cite{Altarev:1992cf,Altarev:1996xs}; same-sign top 
production~\cite{Chatrchyan:2011dk,Aad:2012bb}; the top-quark pair production cross-section
at the Tevatron~\cite{Abazov:2011cq,Aaltonen:2011zma} and 
LHC~\cite{Chatrchyan:2012vs,Aad}; and the top-quark charge asymmetry at the 
LHC~\cite{CMSchargeasym,ATLAS:2012an}. As recent analyses have pointed 
out~\cite{Zhang,Ayazi:2012bb,Chakrabortty:2012pp}, there is already tension between existing 
constraints and theoretical predictions, such that significant fractions of 
parameter space in such models are already being ruled out. In particular, 
there seems to be some incompatibility between the measured LHC charge 
asymmetry, and the Tevatron forward-backward asymmetry, if both are to have
a common origin in terms of $W'$- and $Z'$-boson exchange.\\

Given the above experimental constraints on $W'$ and $Z'$ models, it is also
important, experimentally, to explore every possible production 
mechanism involving $W'$ and $Z'$ bosons, so that cross-checks can be made of 
the conclusions reached from different observables. A crucial process in this
regard is the direct production of such a gauge boson in association with a
single top quark, and indeed this process is being actively sought by the
ATLAS, CDF~\cite{Aaltonen:2012qn} and CMS~\cite{CMSWtd} 
collaborations~\footnote{Analyses may differ in whether they choose to focus
explicitly on the kinematic region of resonant gauge boson production, as we 
discuss further in section~\ref{sec:imps}.}. These 
searches result in direct bounds on the  masses of the new gauge bosons, and 
on their couplings to the quarks. These are upper and lower bounds for the 
couplings and masses respectively, and in order for these to be as tight as 
possible, it is preferable to include higher order perturbative corrections to
the relevant cross-sections. \\

The aim of this paper is to present the complete NLO QCD corrections to the
production of a $W'$ or $Z'$ boson in association with a top (or anti-top) quark,
for both total cross sections and kinematic distributions of the top quark
and new gauge bosons. Our results can be used directly in relevant experimental
analyses, where they can be used to strengthen the bounds on the parameter 
space of models involving these bosons. Our intention is to be as 
model-independent as possible, so that our results can be combined to yield
total cross sections for various scenarios. We will see that NLO corrections 
are significant for total rates (potentially of the order of 50\% for central
values, and rising with the mass of the gauge boson), which by itself 
justifies the calculation of higher-order corrections, due to the potentially 
significant impact on extracted bounds on parameter space.\\

The structure of the paper is as follows. In section~\ref{sec:calc}, we 
describe technical details relating to our NLO calculation, noting similarities
to Standard Model $Wt$ production, which was first considered 
in~\cite{Tait:1999cf}, and calculated at NLO 
in~\cite{Campbell:2005bb,Zhu:2002uj}. In section~\ref{sec:total} we present
results for total cross sections, examining also kinematic distributions in
section~\ref{sec:dists}. In section~\ref{sec:imps} we consider the implications
of our results for current LHC searches. In section~\ref{sec:conclusion}, we
discuss our results before concluding. Various additional results are collected in the supplementary material accompanying this submission, which can be found in Appendix~{\ref{sec:supp}} of this preprint version.

\section{Calculation of  NLO corrections}
\label{sec:calc}
In this section, we describe the technical details of our NLO calculation. Note
that many of these details are very similar to the (Standard Model) $Wt$ 
calculation presented in~\cite{Frixione:2008yi}, thus we will be brief. 
Unless otherwise stated, we will explicitly refer to top-quark production as 
opposed to antitop-quark production. Similar remarks apply in the latter case.\\

\begin{figure}
\begin{center}
\scalebox{0.8}{\includegraphics{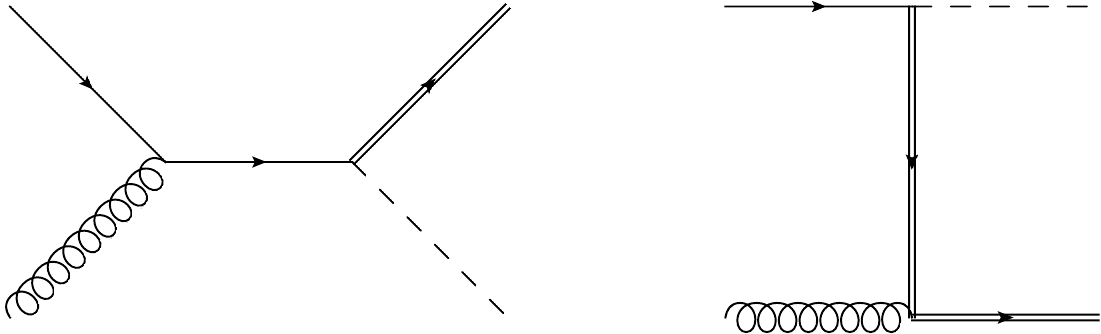}}
\caption{Leading order diagrams for associated production of a $W'$ or $Z'$
(dashed line) with a top quark (double line). The single fermion line 
represents an appropriate down or up-type quark.}
\label{LOdiags}
\end{center}
\end{figure}

We consider NLO QCD corrections to the process
\begin{equation}
q(p_1)+g(p_2)\rightarrow t(k_1)+X^-(k_2),
\label{proc}
\end{equation}
where $q$ is an appropriate quark, and $X$ is either a $W'$ or
$Z'$ boson. The LO diagrams are shown in figure~\ref{LOdiags}.
Following~\cite{Gresham:2011dg}, we define the coupling 
of these bosons to the top quark according to the Lagrangians
\begin{align}
{\cal L}_{W'}=\frac{P_{tD}}{\sqrt{2}}\bar{D}\gamma^\mu g_RP_Rt
W'+{\rm h.c.};\notag\\
{\cal L}_{Z'}=\frac{Q_{tU}}{\sqrt{2}}\bar{U}\gamma^\mu g_RP_Rt
Z'+{\rm h.c.},
\label{couplings}
\end{align}
where $g_R$ is a coupling constant, $P_R$ the right-handed projection 
operator, and $U$, $D$ denote generic up-type and down-type quarks. The model 
of~\cite{Gresham:2011dg} considers only couplings between the first and third 
quark generations. In order to increase the flexibility of our calculation, we
here generalise this by including also couplings between the top quark and arbitrary
up and down-type quarks. The additional coupling factors $P_{tD}$ and $Q_{tU}$
play a role analogous to CKM matrix elements. Note that we carry out our 
calculation in a five-flavour parton scheme, so that initial states involving
$b$ quarks are also included in the total $W't$ cross section.\\

Note that the Lagrangian of eq.~(\ref{couplings}) is specific in the sense
that the coupling involves the pure right-handed projection operator $P_R$.
In principle, there could be a mixture of left- and right-handed projectors,
with corresponding couplings $g_R$ and $g_L$. However, analogously to 
Standard Model $Wt$ production up to NLO~\cite{Campbell:2005bb,Zhu:2002uj}, the total cross section is 
proportional to the combination $|g_L|^2+|g_R|^2$. Hence, total cross sections
and kinematic distributions of the top and gauge bosons can be simply rescaled
in the presence of both left- and right-handed couplings. Note, however, that
the relative mixture of these would be important in considering distributions
relating to the decay products of the top quark and gauge boson.\\

As in~\cite{Frixione:2008yi,Weydert:2009vr}, we use the on-shell scheme for 
renormalisation of the top-quark mass and QCD coupling~\cite{Collins:1978wz} 
(this modifies the $\overline{\rm MS}$-scheme renormalisation of the latter, such
that the top-quark loop contribution is subtracted on-shell). 
The Feynman diagrams for the virtual corrections are exactly the same as those
for Standard Model $Wt$ production~\cite{Campbell:2005bb,Zhu:2002uj}, although
all finite parts of scalar integrals must be analytically continued to the 
kinematic region $m_X>m_t$. We use the results of~\cite{Ellis:2007qk}, as 
implemented in a previous $H^-t$ calculation involving one of the 
authors~\cite{Weydert:2009vr}. \\

For the real emission corrections, all Feynman diagrams have the same form
as Standard Model $Wt$ production, for which we use results from a previous
calculation involving one of the present authors~\cite{Frixione:2008yi}. 
In the latter case, however, an on-shell subtraction scheme is necessary in 
order to define the scattering process at NLO i.e. by removing the 
contribution from leading order top pair 
production~\cite{Belyaev:1998dn,Kauer:2001sp,Kersevan:2006fq,Frixione:2008yi,
White:2009yt,Re:2010bp}. Here no such difficulties arise, due to the fact that
we explicitly consider the regime $m_X>m_t$. Further differences with $Wt$
production arise in that the partonic channels are different in the present 
case. One must also replace the electroweak coupling and CKM matrix with the
relevant coupling factors from eq.~(\ref{couplings}). \\

In order to be able to numerically compute cross sections in a stable manner,
the real and virtual corrections must be regularised and combined using an
appropriate subtraction formalism for soft and collinear singularities, 
of which a number exist in the 
literature~\cite{Frixione:1995ms,Frixione:2011kh,Catani:1996jh,Catani:1996vz,
Chung:2010fx}. Here, as in~\cite{Frixione:2008yi}, we use the FKS 
formalism of~\cite{Frixione:1995ms}.

\section{Total cross section  for $\mathbf{tW'}$ and $\mathbf{tZ'}$ production}
\label{sec:total}
In this section, we present results for total NLO cross sections, for the 
process of eq.~(\ref{proc}). For ease of comparison with 
e.g.~\cite{Gresham:2011dg}, we begin by considering the Lagrangian of 
eq.~(\ref{couplings}) with a non-zero coupling between the first and third
generations only. That is, $P_{td}=Q_{tu}=1$ and $P_{ts}=P_{tb}=Q_{tc}=0$.
Furthermore, we fix $g_R=1$, and use default renormalisation and 
factorisation scale choices of $\mu_R=\mu_F=(m_t+m_X)/2$, where $m_X$ is the
mass of the $W'$ or $Z'$ boson, as appropriate, and $m_t=172.5$ GeV is the top
quark mass. We consider the LHC, with a centre of mass energy of 7 TeV.\\

\begin{figure}
\begin{center}
\scalebox{0.7}{\includegraphics{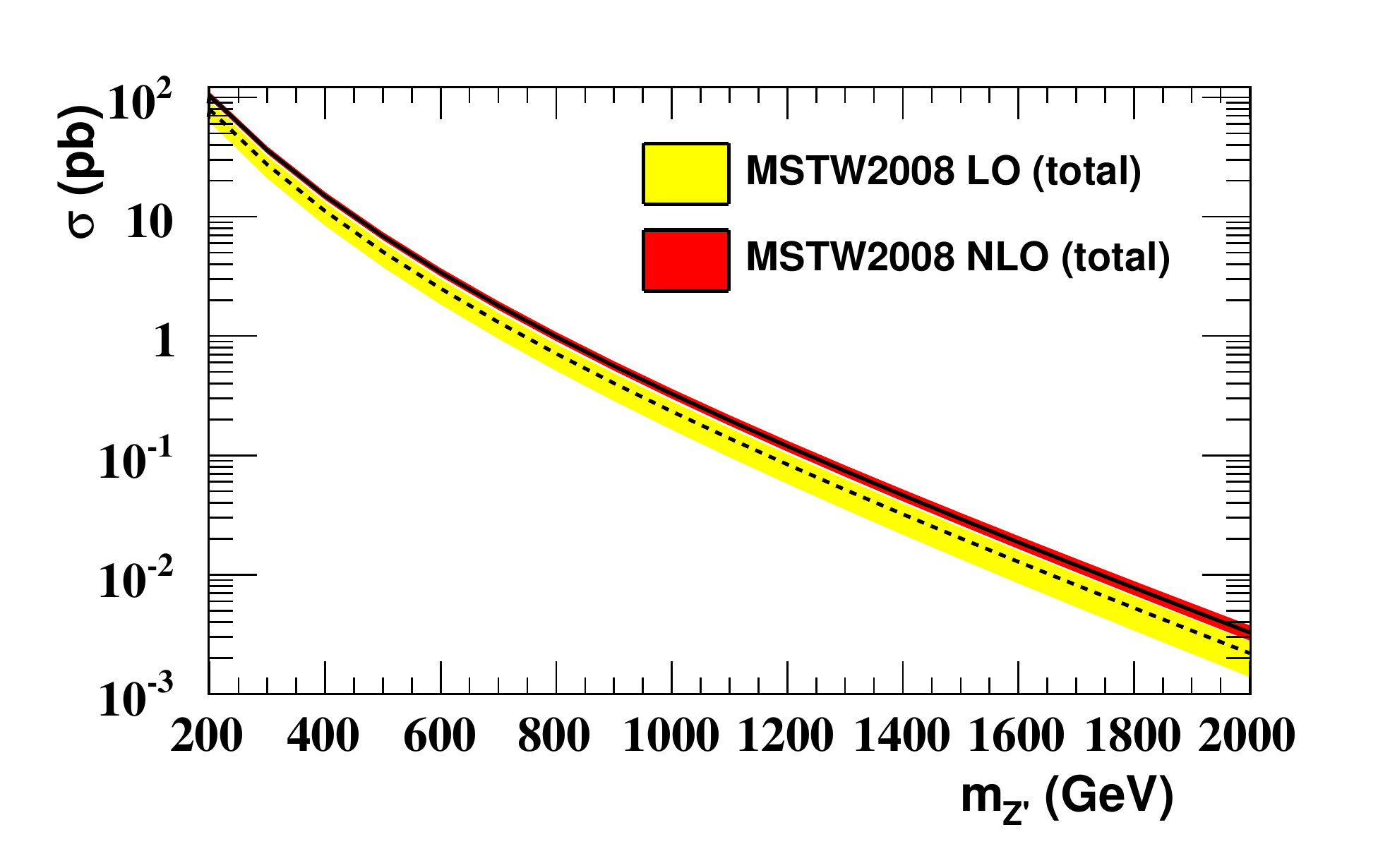}}
\caption{The total cross section for $Z'$ production at LO (dashed) and NLO
(solid), together with combined scale and PDF uncertainties.}
\label{fig:zp}
\end{center}
\end{figure}

\begin{figure}
\begin{center}
\scalebox{0.7}{\includegraphics{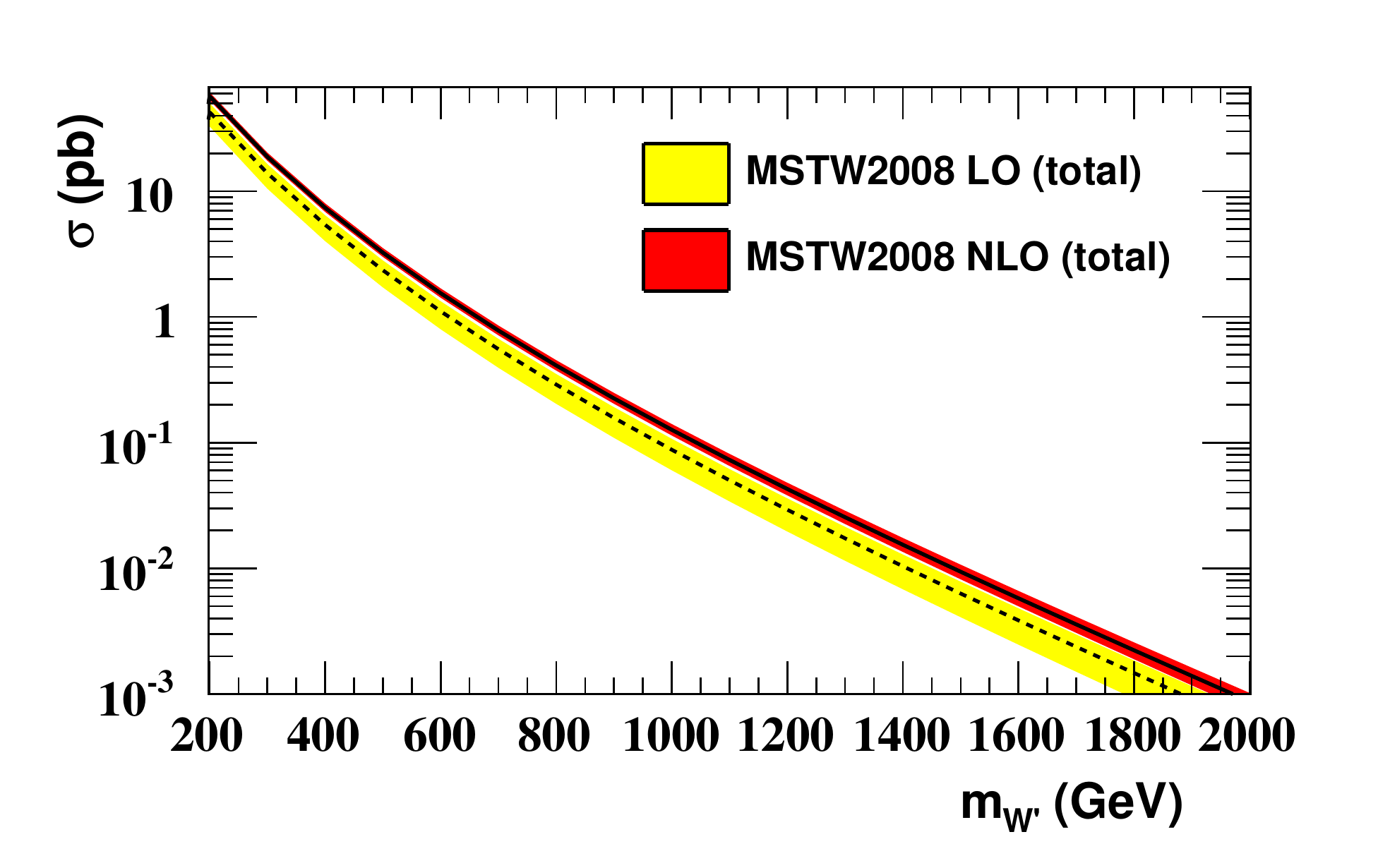}}
\caption{The total cross section for $W'$ production at LO (dashed) and NLO
(solid), together with combined  scale and PDF uncertainties.}
\label{fig:wp}
\end{center}
\end{figure}

Results for the total cross section (i.e. the sum of both top and antitop 
production) at LO and NLO are shown for both $Z'$ and $W'$ production in 
tables~\ref{tab:lo-zp}--\ref{tab:nlo-wp}. We show all results using both
CTEQ~\cite{Pumplin:2002vw,Lai:2010vv}\footnote{For ease of comparison with
existing literature, we provide results with CTEQ6L1 partons at LO.} 
and MSTW~\cite{Martin:2009iq} partons, 
where appropriate we calculate the parton distribution function (PDF) uncertainty using the relevant PDF 
error sets. In all subsequent plots, we show results using MSTW partons unless
otherwise stated. For each result, we also present the scale variation 
uncertainty obtained by varying both $\mu_F$ and $\mu_R$ independently in the 
range $\mu_0/2,2\mu_0$, where $\mu_0$ is the default scale. One sees that the
scale variation uncertainty is reduced at NLO, as expected. Furthermore, the
PDF uncertainty with MSTW partons is much smaller than the scale uncertainty. 
This is also to be
expected, given that the cross section is dominated by the production of top 
quarks involving a $u$ or $d$ quark in the initial state (for $Z'$ and $W'$
production respectively), namely valence quarks. Furthermore, the heavy nature
of the final state (involving both a top quark and a heavy vector boson) means
that the partons are evaluated at typically high momentum fractions $x$. 
Valence quark distributions at high $x$ are well constrained by global PDF 
fits, hence the small PDF uncertainty in the results of 
tables~\ref{tab:lo-zp}-\ref{tab:nlo-wp}. With CTEQ partons the PDF uncertainty is of a similar size to the scale uncertainty over most of the mass range, becoming larger than the scale uncertainty at very high mass. This larger PDF uncertainty is partly due to the fact that the CTEQ uncertainties represent 90\% confidence intervals whereas the MSTW uncertainties represent 68\% confidence intervals\footnote{This is typically accounted for by dividing the CTEQ uncertainties by a factor 1.645 \cite{Botje:2011sn}.}.  Plots of the total cross sections for
$Z'$ and $W'$ production are shown in figures~\ref{fig:zp} and~\ref{fig:wp},
where we have added the PDF and scale uncertainties in quadrature. \\

\begin{table}
\begin{center}
\begin{tabular}{|c||r@{.}ll|r@{.}lll|}
\hline
$M(Z')$\,(GeV) & \multicolumn{3}{|c|}{$\sigma^{\mathrm{CTEQ6L1}}_{\mathrm{born}} (tZ'+\bar{t}Z')$\,(pb)}& \multicolumn{4}{|c|}{$\sigma^{\mathrm{MSTW\ 2008\ LO}}_{\mathrm{born}} (tZ'+\bar{t}Z')$\,(pb)}\\ 
\hline 
200 & 77&9 & $^{+16.3}_{-12.5}$& 79&3 & $^{+17.4}_{-13.2}$ & $^{+ 0.8}_{- 1.2}$\\ 
300 & 26&65 & $^{+ 5.93}_{- 4.50}$& 27&28 & $^{+ 6.37}_{- 4.79}$ & $^{+ 0.31}_{- 0.44}$\\ 
400 & 10&81 & $^{+ 2.52}_{- 1.90}$& 11&15 & $^{+ 2.74}_{- 2.04}$ & $^{+ 0.15}_{- 0.19}$\\ 
500 & 4&89 & $^{+1.18}_{-0.88}$& 5&08 & $^{+1.30}_{-0.96}$ & $^{+0.08}_{-0.10}$\\ 
600 & 2&39 & $^{+0.60}_{-0.44}$& 2&50 & $^{+0.66}_{-0.49}$ & $^{+0.04}_{-0.05}$\\ 
700 & 1&234 & $^{+0.316}_{-0.233}$& 1&306 & $^{+0.358}_{-0.260}$ & $^{+0.025}_{-0.030}$\\ 
800 & 0&667 & $^{+0.175}_{-0.128}$& 0&713 & $^{+0.201}_{-0.145}$ & $^{+0.015}_{-0.018}$\\ 
900 & 0&373 & $^{+0.100}_{-0.073}$& 0&402 & $^{+0.117}_{-0.084}$ & $^{+0.010}_{-0.011}$\\ 
1000 & 0&215 & $^{+0.059}_{-0.043}$& 0&234 & $^{+0.069}_{-0.050}$ & $^{+0.006}_{-0.007}$\\ 
1100 & 0&126 & $^{+0.035}_{-0.025}$& 0&139 & $^{+0.042}_{-0.030}$ & $^{+0.004}_{-0.005}$\\ 
1200 & 0&0759 & $^{+0.0214}_{-0.0155}$& 0&0840 & $^{+0.0261}_{-0.0185}$ & $^{+0.0028}_{-0.0030}$\\ 
1300 & 0&0463 & $^{+0.0133}_{-0.0096}$& 0&0516 & $^{+0.0164}_{-0.0115}$ & $^{+0.0019}_{-0.0020}$\\ 
1400 & 0&0286 & $^{+0.0083}_{-0.0060}$& 0&0321 & $^{+0.0104}_{-0.0073}$ & $^{+0.0013}_{-0.0014}$\\ 
1500 & 0&0178 & $^{+0.0053}_{-0.0038}$& 0&0202 & $^{+0.0067}_{-0.0047}$ & $^{+0.0009}_{-0.0009}$\\ 
1600 & 0&0112 & $^{+0.0034}_{-0.0024}$& 0&0128 & $^{+0.0043}_{-0.0030}$ & $^{+0.0006}_{-0.0006}$\\ 
1700 & 0&0071 & $^{+0.0022}_{-0.0015}$& 0&0082 & $^{+0.0028}_{-0.0019}$ & $^{+0.0004}_{-0.0004}$\\ 
1800 & 0&00455 & $^{+0.00139}_{-0.00099}$& 0&00527 & $^{+0.00184}_{-0.00126}$ & $^{+0.00030}_{-0.00029}$\\ 
1900 & 0&00292 & $^{+0.00090}_{-0.00064}$& 0&00340 & $^{+0.00121}_{-0.00083}$ & $^{+0.00021}_{-0.00020}$\\ 
2000 & 0&00188 & $^{+0.00059}_{-0.00042}$& 0&00220 & $^{+0.00080}_{-0.00054}$ & $^{+0.00015}_{-0.00014}$\\ 
\hline 
\end{tabular} 

\end{center}
\caption{The sum of the leading-order cross sections for $pp \rightarrow tZ'$ and  $pp \rightarrow \bar{t}Z'$ at the 7 TeV LHC. Scale and PDF uncertainties are given first and second respectively. In the case of the CTEQ6L1 PDFs no PDF uncertainty exists. \label{tab:lo-zp}}
\end{table}

\begin{table}
\begin{center}
\begin{tabular}{|c||r@{.}ll|r@{.}lll|}
\hline
$M(W')$\,(GeV) & \multicolumn{3}{|c|}{$\sigma^{\mathrm{CTEQ6L1}}_{\mathrm{born}} (tW'^{-}+\bar{t}W'^{+})$\,(pb)}& \multicolumn{4}{|c|}{$\sigma^{\mathrm{MSTW\ 2008\ LO}}_{\mathrm{born}} (tW'^{-}+\bar{t}W'^{+})$\,(pb)}\\ 
\hline 
200 & 41&8 & $^{+ 9.1}_{-7.0}$& 42&7 & $^{+ 9.8}_{- 7.4}$ & $^{+ 0.6}_{- 0.7}$\\ 
300 & 13&60 & $^{+ 3.15}_{- 2.37}$& 13&92 & $^{+ 3.42}_{- 2.54}$ & $^{+ 0.22}_{- 0.26}$\\ 
400 & 5&26 & $^{+1.28}_{-0.95}$& 5&41 & $^{+1.40}_{-1.03}$ & $^{+0.10}_{-0.11}$\\ 
500 & 2&28 & $^{+0.57}_{-0.42}$& 2&35 & $^{+0.63}_{-0.46}$ & $^{+0.05}_{-0.05}$\\ 
600 & 1&068 & $^{+0.277}_{-0.203}$& 1&109 & $^{+0.309}_{-0.223}$ & $^{+0.026}_{-0.028}$\\ 
700 & 0&531 & $^{+0.141}_{-0.103}$& 0&554 & $^{+0.159}_{-0.114}$ & $^{+0.015}_{-0.016}$\\ 
800 & 0&276 & $^{+0.075}_{-0.055}$& 0&290 & $^{+0.086}_{-0.061}$ & $^{+0.009}_{-0.009}$\\ 
900 & 0&149 & $^{+0.041}_{-0.030}$& 0&157 & $^{+0.048}_{-0.034}$ & $^{+0.005}_{-0.005}$\\ 
1000 & 0&0829 & $^{+0.0234}_{-0.0169}$& 0&0879 & $^{+0.0273}_{-0.0193}$ & $^{+0.0033}_{-0.0033}$\\ 
1100 & 0&0472 & $^{+0.0135}_{-0.0098}$& 0&0502 & $^{+0.0160}_{-0.0112}$ & $^{+0.0021}_{-0.0021}$\\ 
1200 & 0&0274 & $^{+0.0080}_{-0.0057}$& 0&0293 & $^{+0.0095}_{-0.0067}$ & $^{+0.0014}_{-0.0013}$\\ 
1300 & 0&0161 & $^{+0.0048}_{-0.0034}$& 0&0173 & $^{+0.0058}_{-0.0040}$ & $^{+0.0009}_{-0.0009}$\\ 
1400 & 0&0097 & $^{+0.0029}_{-0.0021}$& 0&0104 & $^{+0.0035}_{-0.0024}$ & $^{+0.0006}_{-0.0006}$\\ 
1500 & 0&0058 & $^{+0.0018}_{-0.0013}$& 0&0063 & $^{+0.0022}_{-0.0015}$ & $^{+0.0004}_{-0.0004}$\\ 
1600 & 0&00356 & $^{+0.00109}_{-0.00078}$& 0&00386 & $^{+0.00136}_{-0.00093}$ & $^{+0.00027}_{-0.00024}$\\ 
1700 & 0&00218 & $^{+0.00068}_{-0.00048}$& 0&00238 & $^{+0.00085}_{-0.00058}$ & $^{+0.00018}_{-0.00016}$\\ 
1800 & 0&00135 & $^{+0.00042}_{-0.00030}$& 0&00147 & $^{+0.00054}_{-0.00036}$ & $^{+0.00012}_{-0.00010}$\\ 
1900 & 0&00084 & $^{+0.00027}_{-0.00019}$& 0&00092 & $^{+0.00034}_{-0.00023}$ & $^{+0.00008}_{-0.00007}$\\ 
2000 & 0&00052 & $^{+0.00017}_{-0.00012}$& 0&00057 & $^{+0.00022}_{-0.00015}$ & $^{+0.00006}_{-0.00005}$\\ 
\hline 
\end{tabular} 

\end{center}
\caption{The sum of the leading-order cross sections for $pp \rightarrow tW'^-$ and $pp \rightarrow \bar{t}W^+$ at the 7 TeV 
LHC. Other details as in 
table~\ref{tab:lo-zp}. \label{tab:lo-wp}}
\end{table}

\begin{table}
\begin{center}
\begin{tabular}{|c||r@{.}lll|r@{.}lll|}
\hline
$M(Z')$\,(GeV) & \multicolumn{4}{|c|}{$\sigma^{\mathrm{CT10}}_{\mathrm{nlo}} (tZ'+\bar{t}Z')$\,(pb)}& \multicolumn{4}{|c|}{$\sigma^{\mathrm{MSTW\ 2008\ NLO}}_{\mathrm{nlo}} (tZ'+\bar{t}Z')$\,(pb)}\\ 
\hline 
200 & 101&2 & $^{+  6.7}_{-  7.1}$ & $^{+  2.1}_{-  3.1}$& 103&6 & $^{+  7.0}_{-  7.5}$ & $^{+  0.7}_{-  1.1}$\\ 
300 & 35&32 & $^{+ 2.29}_{- 2.59}$ & $^{+ 0.93}_{- 1.26}$& 36&10 & $^{+ 2.42}_{- 2.71}$ & $^{+ 0.26}_{- 0.41}$\\ 
400 & 14&61 & $^{+ 0.96}_{- 1.12}$ & $^{+ 0.50}_{- 0.60}$& 14&92 & $^{+ 1.01}_{- 1.17}$ & $^{+ 0.13}_{- 0.20}$\\ 
500 & 6&74 & $^{+0.46}_{-0.54}$ & $^{+0.30}_{-0.32}$& 6&87 & $^{+0.48}_{-0.56}$ & $^{+0.08}_{-0.10}$\\ 
600 & 3&35 & $^{+0.23}_{-0.28}$ & $^{+0.18}_{-0.19}$& 3&41 & $^{+0.25}_{-0.29}$ & $^{+0.05}_{-0.06}$\\ 
700 & 1&769 & $^{+0.128}_{-0.153}$ & $^{+0.115}_{-0.114}$& 1&795 & $^{+0.134}_{-0.158}$ & $^{+0.033}_{-0.038}$\\ 
800 & 0&975 & $^{+0.073}_{-0.087}$ & $^{+0.075}_{-0.072}$& 0&987 & $^{+0.077}_{-0.090}$ & $^{+0.022}_{-0.024}$\\ 
900 & 0&557 & $^{+0.044}_{-0.051}$ & $^{+0.050}_{-0.046}$& 0&561 & $^{+0.045}_{-0.053}$ & $^{+0.015}_{-0.016}$\\ 
1000 & 0&327 & $^{+0.027}_{-0.031}$ & $^{+0.034}_{-0.030}$& 0&328 & $^{+0.028}_{-0.032}$ & $^{+0.010}_{-0.011}$\\ 
1100 & 0&197 & $^{+0.017}_{-0.019}$ & $^{+0.023}_{-0.020}$& 0&196 & $^{+0.017}_{-0.020}$ & $^{+0.007}_{-0.007}$\\ 
1200 & 0&121 & $^{+0.011}_{-0.012}$ & $^{+0.017}_{-0.014}$& 0&120 & $^{+0.011}_{-0.012}$ & $^{+0.005}_{-0.005}$\\ 
1300 & 0&0752 & $^{+0.0068}_{-0.0078}$ & $^{+0.0117}_{-0.0093}$& 0&0740 & $^{+0.0069}_{-0.0079}$ & $^{+0.0032}_{-0.0033}$\\ 
1400 & 0&0475 & $^{+0.0045}_{-0.0051}$ & $^{+0.0084}_{-0.0063}$& 0&0463 & $^{+0.0045}_{-0.0051}$ & $^{+0.0023}_{-0.0022}$\\ 
1500 & 0&0303 & $^{+0.0030}_{-0.0034}$ & $^{+0.0060}_{-0.0044}$& 0&0293 & $^{+0.0030}_{-0.0033}$ & $^{+0.0016}_{-0.0015}$\\ 
1600 & 0&0195 & $^{+0.0020}_{-0.0022}$ & $^{+0.0044}_{-0.0031}$& 0&0187 & $^{+0.0020}_{-0.0022}$ & $^{+0.0011}_{-0.0011}$\\ 
1700 & 0&0127 & $^{+0.0013}_{-0.0015}$ & $^{+0.0032}_{-0.0022}$& 0&0120 & $^{+0.0013}_{-0.0014}$ & $^{+0.0008}_{-0.0008}$\\ 
1800 & 0&0083 & $^{+0.0009}_{-0.0010}$ & $^{+0.0023}_{-0.0015}$& 0&0078 & $^{+0.0009}_{-0.0010}$ & $^{+0.0005}_{-0.0005}$\\ 
1900 & 0&0055 & $^{+0.0006}_{-0.0007}$ & $^{+0.0017}_{-0.0011}$& 0&0050 & $^{+0.0006}_{-0.0006}$ & $^{+0.0004}_{-0.0004}$\\ 
2000 & 0&00360 & $^{+0.00042}_{-0.00046}$ & $^{+0.00127}_{-0.00078}$& 0&00328 & $^{+0.00039}_{-0.00042}$ & $^{+0.00025}_{-0.00025}$\\ 
\hline 
\end{tabular} 

\end{center}
\caption{The sum of the next-to-leading-order cross sections for $pp \rightarrow tZ'$ and  $pp \rightarrow \bar{t}Z'$ at the 7 TeV LHC.  Scale and PDF uncertainties are given first and second respectively. \label{tab:nlo-zp}}

\end{table}

\begin{table}
\begin{center}
\begin{tabular}{|c||r@{.}lll|r@{.}lll|}
\hline
$M(W')$\,(GeV) & \multicolumn{4}{|c|}{$\sigma^{\mathrm{CT10}}_{\mathrm{nlo}} (tW'^{-}+\bar{t}W'^{+})$\,(pb)}& \multicolumn{4}{|c|}{$\sigma^{\mathrm{MSTW\ 2008\ NLO}}_{\mathrm{nlo}} (tW'^{-}+\bar{t}W'^{+})$\,(pb)}\\ 
\hline 
200 & 55&6 & $^{+ 3.7}_{- 4.0}$ & $^{+ 2.1}_{- 2.7}$& 57&1 & $^{+ 3.9}_{- 4.3}$ & $^{+ 0.5}_{- 0.9}$\\ 
300 & 18&48 & $^{+ 1.22}_{- 1.40}$ & $^{+ 0.83}_{- 1.00}$& 18&83 & $^{+ 1.29}_{- 1.47}$ & $^{+ 0.21}_{- 0.34}$\\ 
400 & 7&32 & $^{+0.49}_{-0.58}$ & $^{+0.40}_{-0.45}$& 7&40 & $^{+0.52}_{-0.61}$ & $^{+0.10}_{-0.16}$\\ 
500 & 3&24 & $^{+0.23}_{-0.27}$ & $^{+0.22}_{-0.23}$& 3&25 & $^{+0.24}_{-0.28}$ & $^{+0.06}_{-0.08}$\\ 
600 & 1&555 & $^{+0.112}_{-0.134}$ & $^{+0.122}_{-0.123}$& 1&546 & $^{+0.117}_{-0.137}$ & $^{+0.033}_{-0.045}$\\ 
700 & 0&792 & $^{+0.060}_{-0.071}$ & $^{+0.073}_{-0.070}$& 0&780 & $^{+0.061}_{-0.072}$ & $^{+0.020}_{-0.026}$\\ 
800 & 0&423 & $^{+0.033}_{-0.039}$ & $^{+0.046}_{-0.042}$& 0&411 & $^{+0.034}_{-0.039}$ & $^{+0.013}_{-0.015}$\\ 
900 & 0&234 & $^{+0.019}_{-0.022}$ & $^{+0.029}_{-0.026}$& 0&225 & $^{+0.019}_{-0.022}$ & $^{+0.008}_{-0.009}$\\ 
1000 & 0&134 & $^{+0.011}_{-0.013}$ & $^{+0.019}_{-0.016}$& 0&127 & $^{+0.011}_{-0.013}$ & $^{+0.005}_{-0.006}$\\ 
1100 & 0&0781 & $^{+0.0069}_{-0.0080}$ & $^{+0.0127}_{-0.0103}$& 0&0729 & $^{+0.0068}_{-0.0077}$ & $^{+0.0034}_{-0.0037}$\\ 
1200 & 0&0466 & $^{+0.0043}_{-0.0049}$ & $^{+0.0086}_{-0.0067}$& 0&0428 & $^{+0.0041}_{-0.0047}$ & $^{+0.0022}_{-0.0024}$\\ 
1300 & 0&0283 & $^{+0.0027}_{-0.0031}$ & $^{+0.0059}_{-0.0044}$& 0&0255 & $^{+0.0026}_{-0.0029}$ & $^{+0.0015}_{-0.0016}$\\ 
1400 & 0&0174 & $^{+0.0017}_{-0.0019}$ & $^{+0.0041}_{-0.0030}$& 0&0154 & $^{+0.0016}_{-0.0018}$ & $^{+0.0010}_{-0.0010}$\\ 
1500 & 0&0108 & $^{+0.0011}_{-0.0012}$ & $^{+0.0029}_{-0.0020}$& 0&0094 & $^{+0.0010}_{-0.0011}$ & $^{+0.0007}_{-0.0007}$\\ 
1600 & 0&0068 & $^{+0.0007}_{-0.0008}$ & $^{+0.0020}_{-0.0014}$& 0&0058 & $^{+0.0007}_{-0.0007}$ & $^{+0.0004}_{-0.0005}$\\ 
1700 & 0&00434 & $^{+0.00048}_{-0.00053}$ & $^{+0.00143}_{-0.00094}$& 0&00358 & $^{+0.00042}_{-0.00045}$ & $^{+0.00030}_{-0.00030}$\\ 
1800 & 0&00277 & $^{+0.00032}_{-0.00035}$ & $^{+0.00102}_{-0.00065}$& 0&00223 & $^{+0.00027}_{-0.00029}$ & $^{+0.00020}_{-0.00020}$\\ 
1900 & 0&00178 & $^{+0.00021}_{-0.00023}$ & $^{+0.00073}_{-0.00045}$& 0&00139 & $^{+0.00017}_{-0.00018}$ & $^{+0.00013}_{-0.00013}$\\ 
2000 & 0&00115 & $^{+0.00014}_{-0.00015}$ & $^{+0.00052}_{-0.00031}$& 0&00087 & $^{+0.00011}_{-0.00012}$ & $^{+0.00009}_{-0.00009}$\\ 
\hline 
\end{tabular} 

\end{center}
\caption{The sum of the next-to-leading-order cross sections for $pp \rightarrow tW'^-$ and  $pp \rightarrow \bar{t}W'^+$  at the 7 TeV LHC. Other details as in table \ref{tab:nlo-zp}. \label{tab:nlo-wp}}
\end{table}

\begin{figure}
\begin{center}
\scalebox{0.7}{\includegraphics{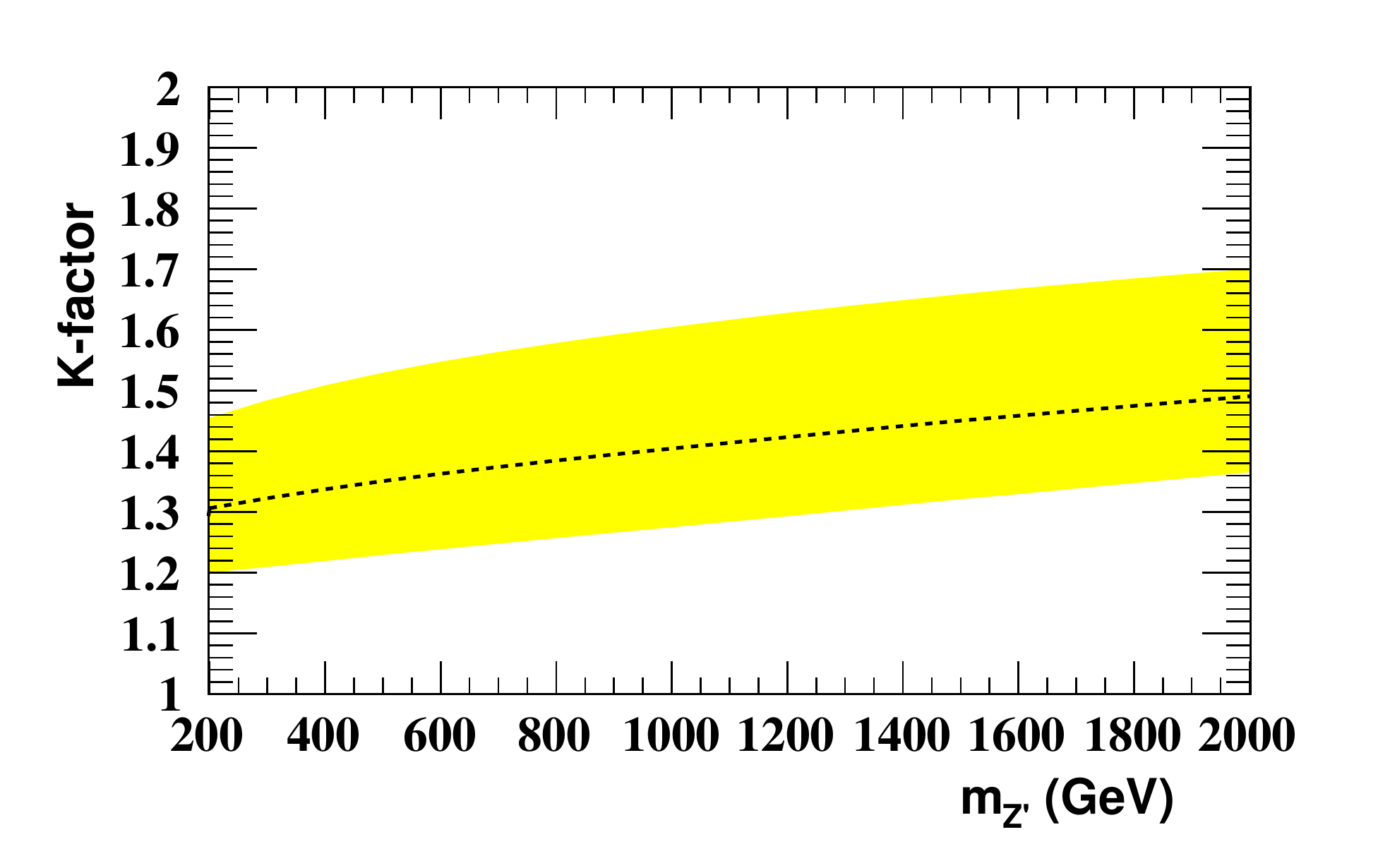}}
\caption{The $K$-factor for $Z'$ production as defined in eq.~(\ref{Kfacdef}),
with combined scale and PDF uncertainty.}
\label{fig:zpK}
\end{center}
\end{figure}

\begin{figure}
\begin{center}
\scalebox{0.7}{\includegraphics{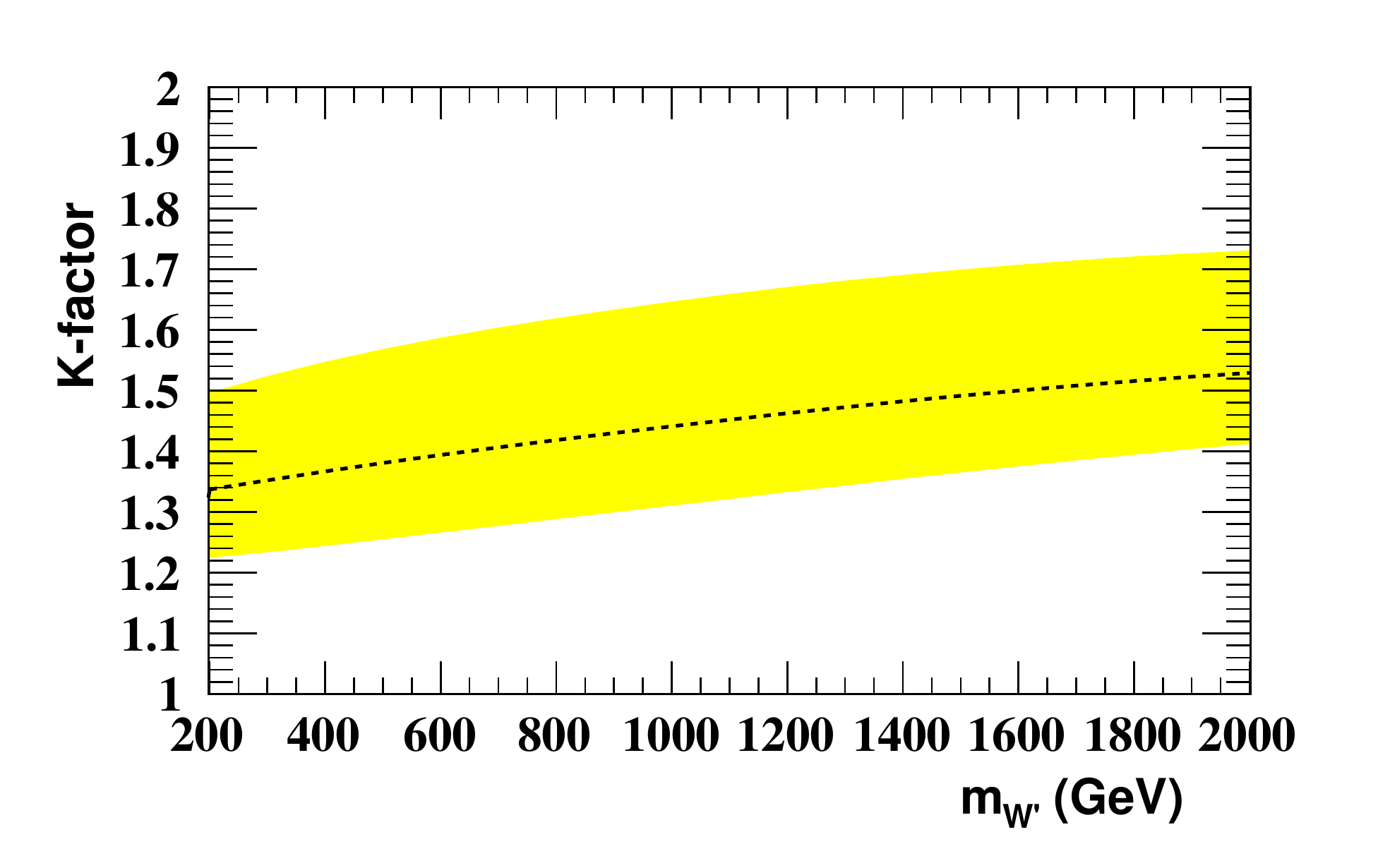}}
\caption{The $K$-factor for $W'$ production as defined in eq.~(\ref{Kfacdef}),
with combined scale and PDF uncertainty.}
\label{fig:wpK}
\end{center}
\end{figure}

In figures~\ref{fig:zpK} and~\ref{fig:wpK}, we show the $K$-factor for both
$Z'$ and $W'$ production as a function of the mass of the gauge boson $m_X$, 
evaluated according to
\begin{equation}
K=\frac{\sigma_{\rm NLO}}{\sigma_{\rm LO}},
\label{Kfacdef}
\end{equation}
where we have used the results obtained with MSTW partons from 
tables~\ref{tab:lo-zp}-\ref{tab:nlo-wp}. We see that the $K$-factor is sizeable
and increasing with $m_X$. For the values of $m_X$ we consider, the NLO 
corrections range from $\simeq 30\%-50\%$  in both cases. This can be compared
with Standard Model $Wt$ production, for which the $K$-factor is 
$\simeq 25\%$~\cite{Campbell:2005bb,Zhu:2002uj}, subject to an appropriate 
on-shell subtraction formalism being employed to separate the $Wt$ mode from
top pair production. The higher $K$ factor in the present case, and the fact
that the $K$-factor rises with increasing gauge boson mass, have partonic 
origins. As stated above, both $W'$ and $Z'$ production have valence quarks
in the initial state. Furthermore, as $m_X$ increases, these distributions
are probed at typically higher $x$ values, and relatively more so at NLO. 
Given that valence quark distributions remain significant at higher $x$ values,
the $K$-factor is therefore sizable. One may find the opposite trend if sea or 
$b$ quarks are involved in the initial state i.e. a 
$K$-factor which decreases with increasing $m_X$, due to the sharp fall-off in
the partons as $x\rightarrow 1$ (we consider other partonic couplings in
the supplementary document). Note that there are
also extra partonic subchannels which open up at NLO (i.e. the $q\bar{q}$ and
$gg$ initial states), which also act to increase the $K$-factor. \\

From tables~\ref{tab:lo-zp}--\ref{tab:nlo-wp} and 
figures~\ref{fig:zpK}--\ref{fig:wpK}, one sees that the $K$-factor for 
$W'$ production is slightly larger than that for $Z'$ production. 
This can be explained by the fact that the former case has $d$ quarks in the 
initial state rather than $u$ quarks. The heavy final state means that these
partons are typically probed at high $x$ values, where the $u$ quark 
distribution (at LO or NLO) has a pronounced shoulder (followed by a steep fall off)
relative to the $d$ quark distribution~\cite{Martin:2009iq}. At NLO, the 
partons are probed at higher $x$ values than at LO, which leads to a decrease 
in the parton luminosity. This is less marked for the $d$ distribution than for
the $u$ distribution, due to the shoulder in the latter. Hence, the $K$-factor 
for $W'$ production is slightly larger than that for $Z'$ 
production~\footnote{Another reason for the $K$-factor difference is due to 
additional partonic subchannels opening up at NLO, with $gg$ and $q\bar{q}$
initial states, which contribute different fractions of the total 
cross-sections in $W't$ and $Z't$ production. However, this has a small effect
on the $K$-factor difference, due to the fact that the NLO cross-section 
remains dominated by $qg$ initial states.}. \\


\begin{figure}
\begin{center}
\scalebox{0.7}{\includegraphics{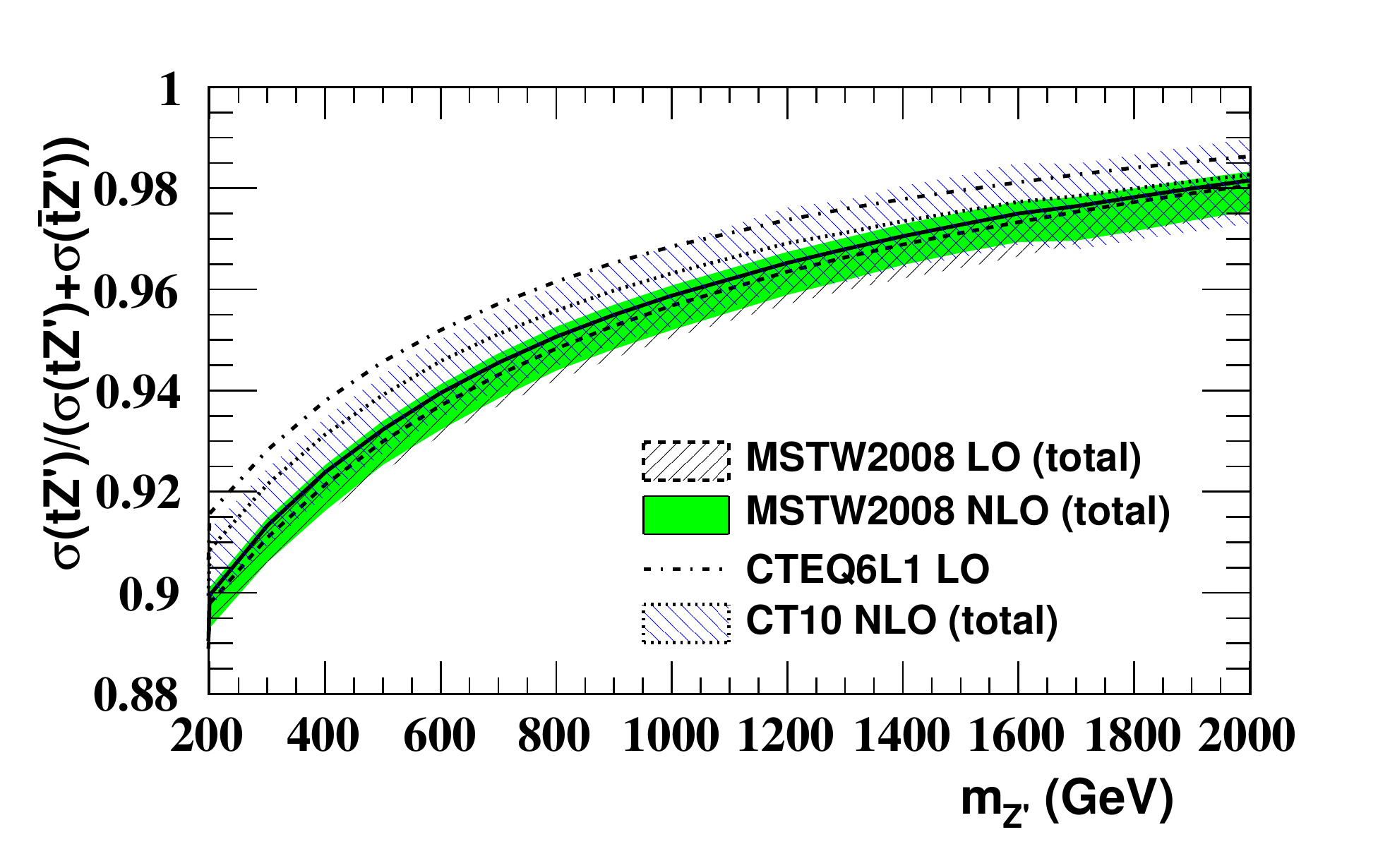}}
\caption{Fraction of $Z'$ events containing a top (rather than anti-top) quark,
together with combined scale and PDF uncertainties.}
\label{fig:tfracZ}
\end{center}
\end{figure}

\begin{figure}
\begin{center}
\scalebox{0.7}{\includegraphics{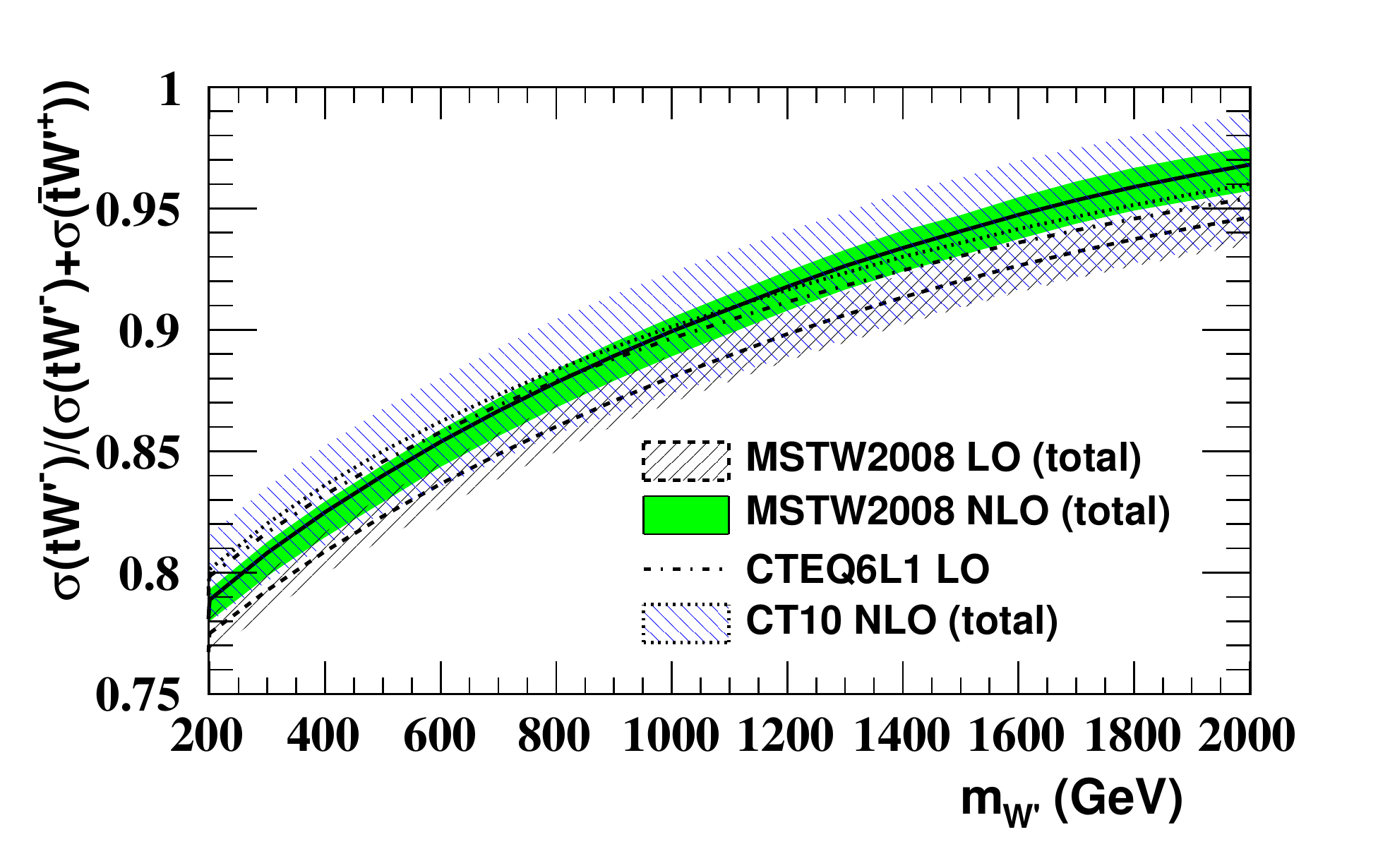}}
\caption{Fraction of $W'$ events containing a top (rather than anti-top) quark,
together with combined scale and PDF uncertainties.}
\label{fig:tfracW}
\end{center}
\end{figure}

As well as the total cross section, it is also interesting to consider the 
fraction of events containing a top quark, rather than an anti-top quark (in
the case of $W'$ production, this changes the sign of the charge of the 
accompanying boson). 
This fraction is shown for $Z'$ and $W'$ production in 
figures~\ref{fig:tfracZ} and~\ref{fig:tfracW} respectively, at both LO and NLO.
We see in both cases that the total cross section is dominated by $t$ quark
production, due to the presence of valence rather than sea quarks in the 
initial state at LO. Furthermore, we note that the theoretical uncertainty is dominated by PDF uncertainties rather than scale variation, as the latter effect is similar for top and anti-top production, and cancels in the ratio.\\

The top quark fraction is higher for $Z'$ rather than $W'$ 
production, due to the dominance of up quarks over down quarks in the proton.
Note that the fractions increase with the gauge boson mass, reflecting the
fact that the partons are evaluated at higher $x$ values on average as the
gauge boson mass increases, which causes sea-quark dominated processes to
fall off at the expense of those which are dominated by valence quarks.
We may also note that the fraction changes slightly at NLO, and can either 
decrease or increase, depending on the gauge boson mass. This is due to two
competing effects. Firstly, extra partonic subchannels open up at NLO which are
insensitive to the exchange of a top and anti-top quark. This acts to decrease
the fraction of top quark events. Secondly, the partons are evaluated at 
slightly higher $x$ values at NLO compared to LO. This increases the dominance
of valence quark-initiated processes, and acts to increase the proportion of
top quark events. The size of the latter effect is somewhat uncertain due to
the influence of high-$x$ sea quark distributions. This can clearly be seen
in both figures~\ref{fig:tfracZ} and~\ref{fig:tfracW}, where we show the 
top quark fraction for CTEQ as well as MSTW partons. The CTEQ results lead to
higher top-quark fractions, with a decrease at NLO (apart from $W'$ production
at low gauge boson masses). In the MSTW results, there is an increase in the
top-quark fraction at NLO, suggesting that valence quark distribution 
effects are more dominant. \\

In this section, we have considered a subset of possible couplings in 
eq.~(\ref{couplings}), in particular considering only the $u-t$ and $d-t$
interactions for $W'$ and $Z'$ respectively. In 
the supplementary material, we present tables of results for other
possible choices. Furthermore, we present results
for the LHC at 8 TeV. Having discussed the total cross-section results in
detail, we discuss NLO corrections to kinematic distributions of the top quark
and gauge bosons in the following section.

\section{Top and heavy gauge boson kinematics}
\label{sec:dists}

In this section, we examine the impact of NLO QCD corrections on various 
kinematic distributions relating to the top quark and gauge boson. Throughout,
we use the parameter choices described in the previous section, with MSTW LO
and NLO partons~\cite{Martin:2009iq} as appropriate. \\

\begin{figure}
\begin{center}
\scalebox{0.4}{\includegraphics{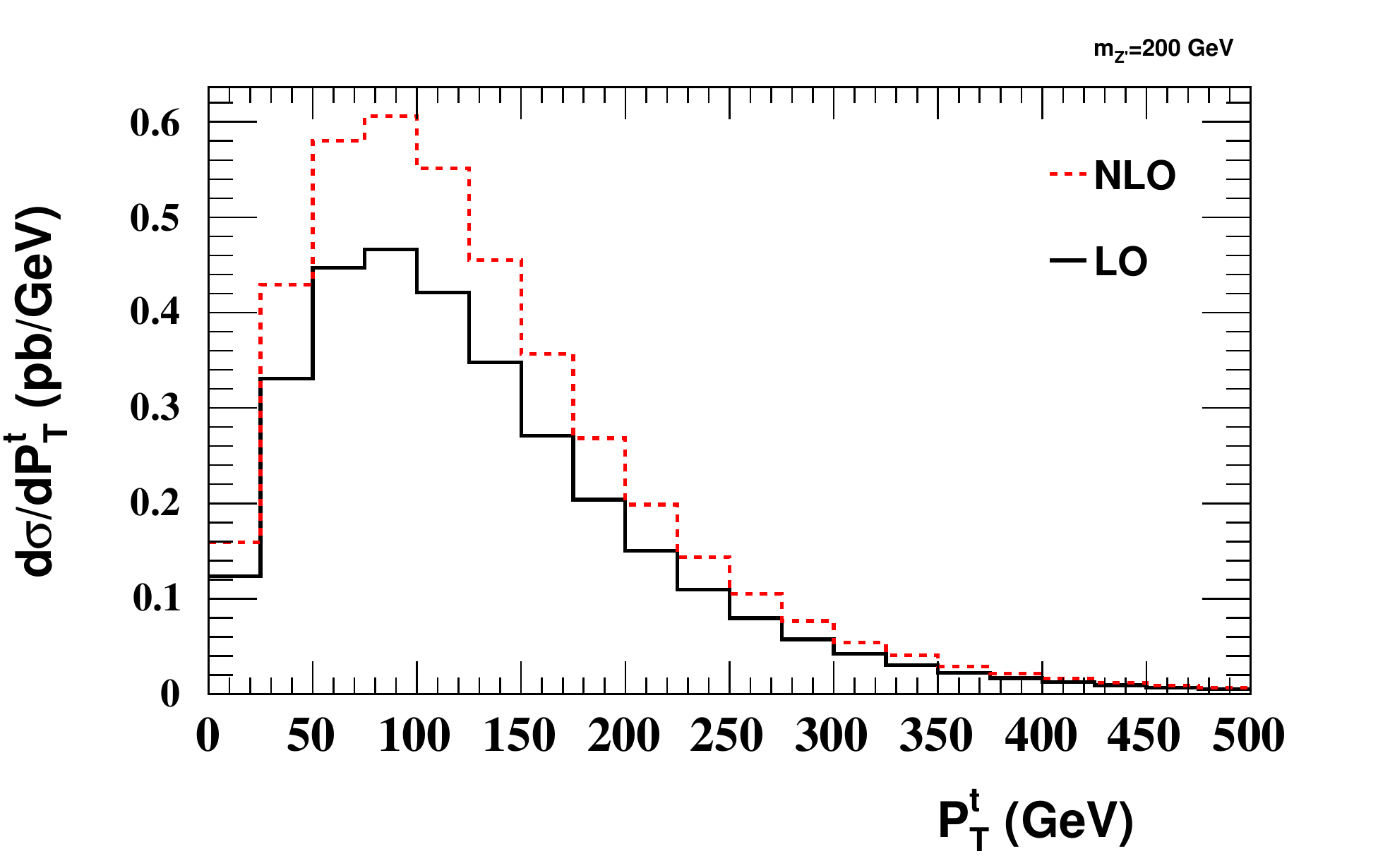}} \scalebox{0.4}{\includegraphics{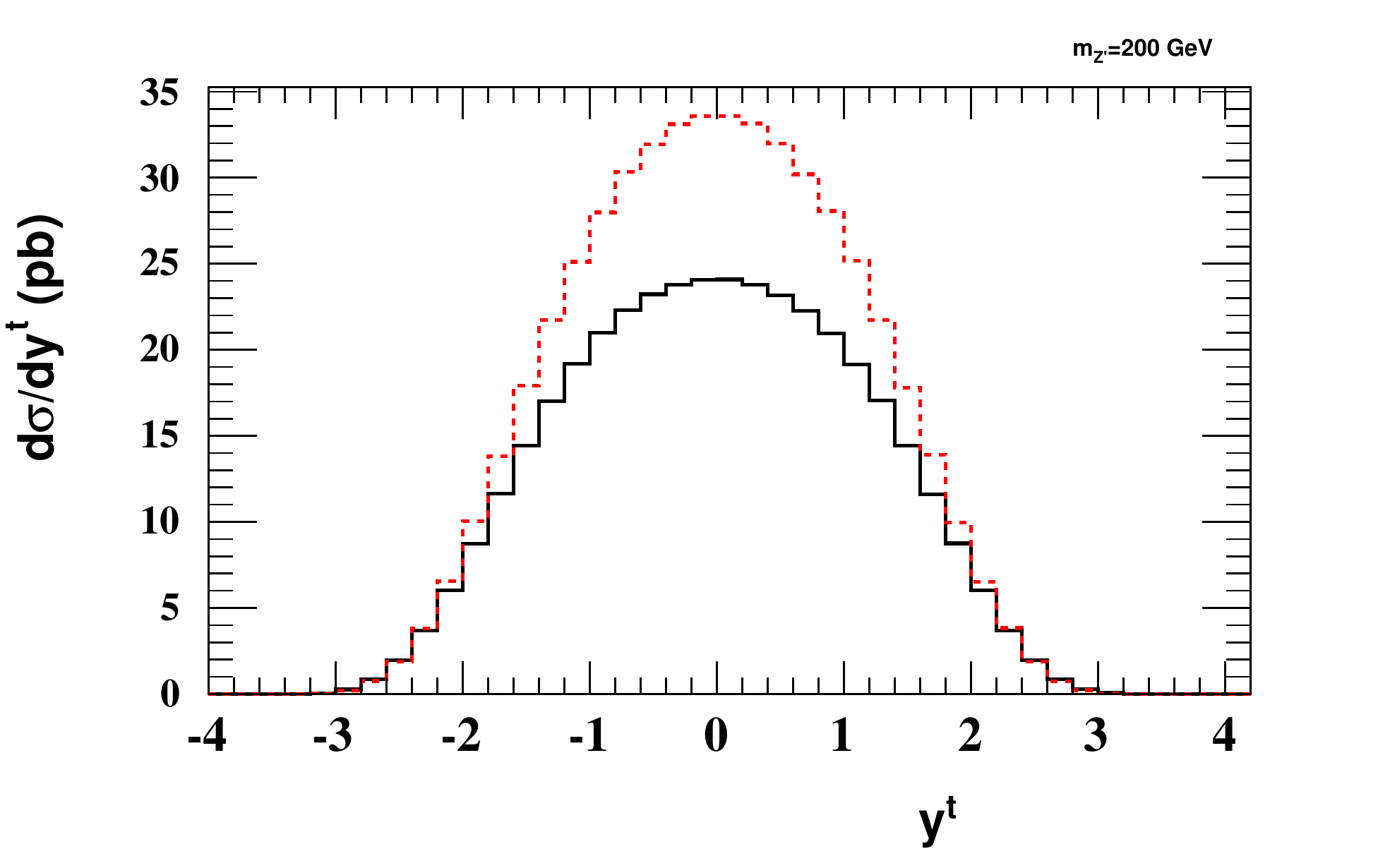}}\\
\scalebox{0.4}{\includegraphics{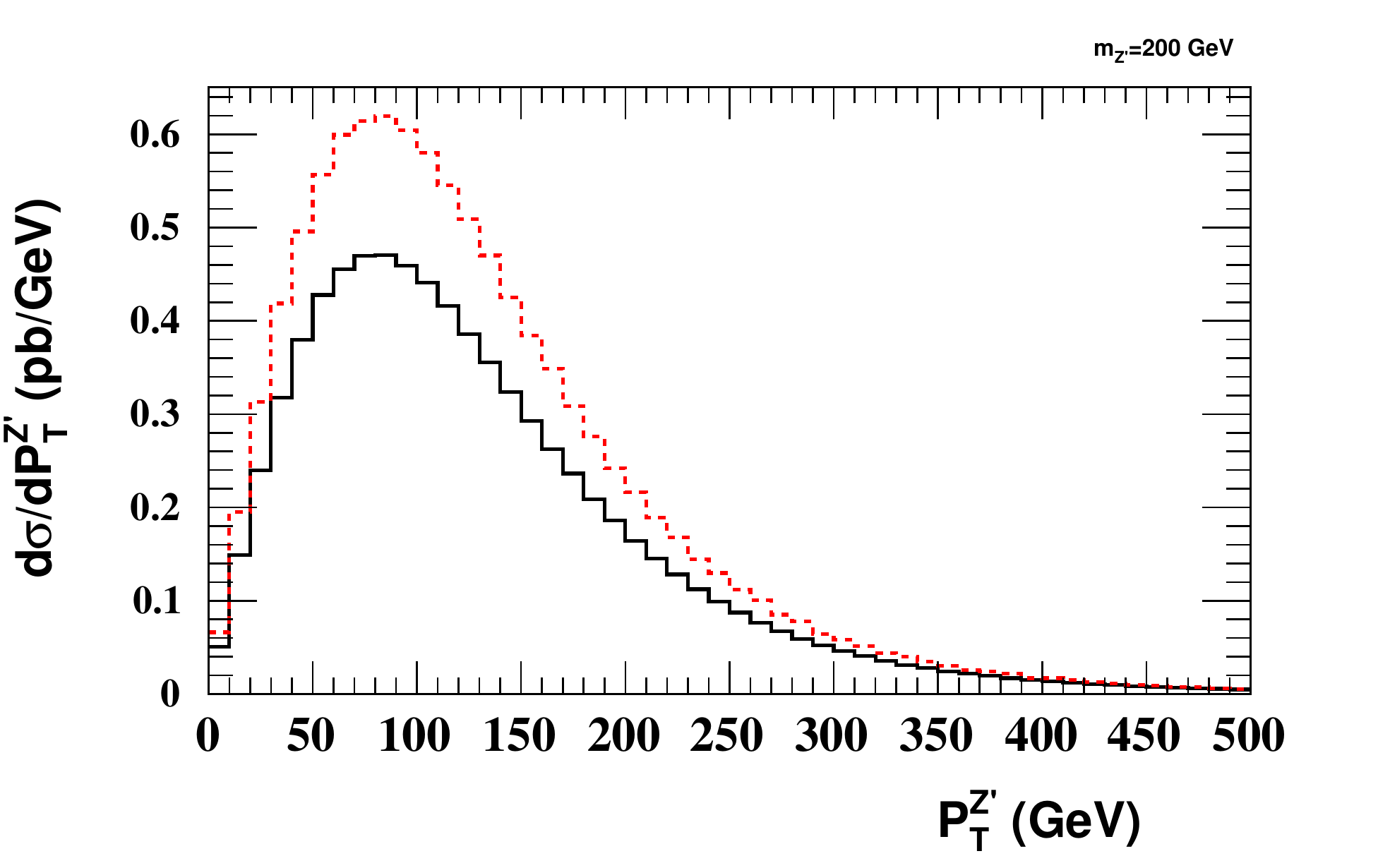}} \scalebox{0.4}{\includegraphics{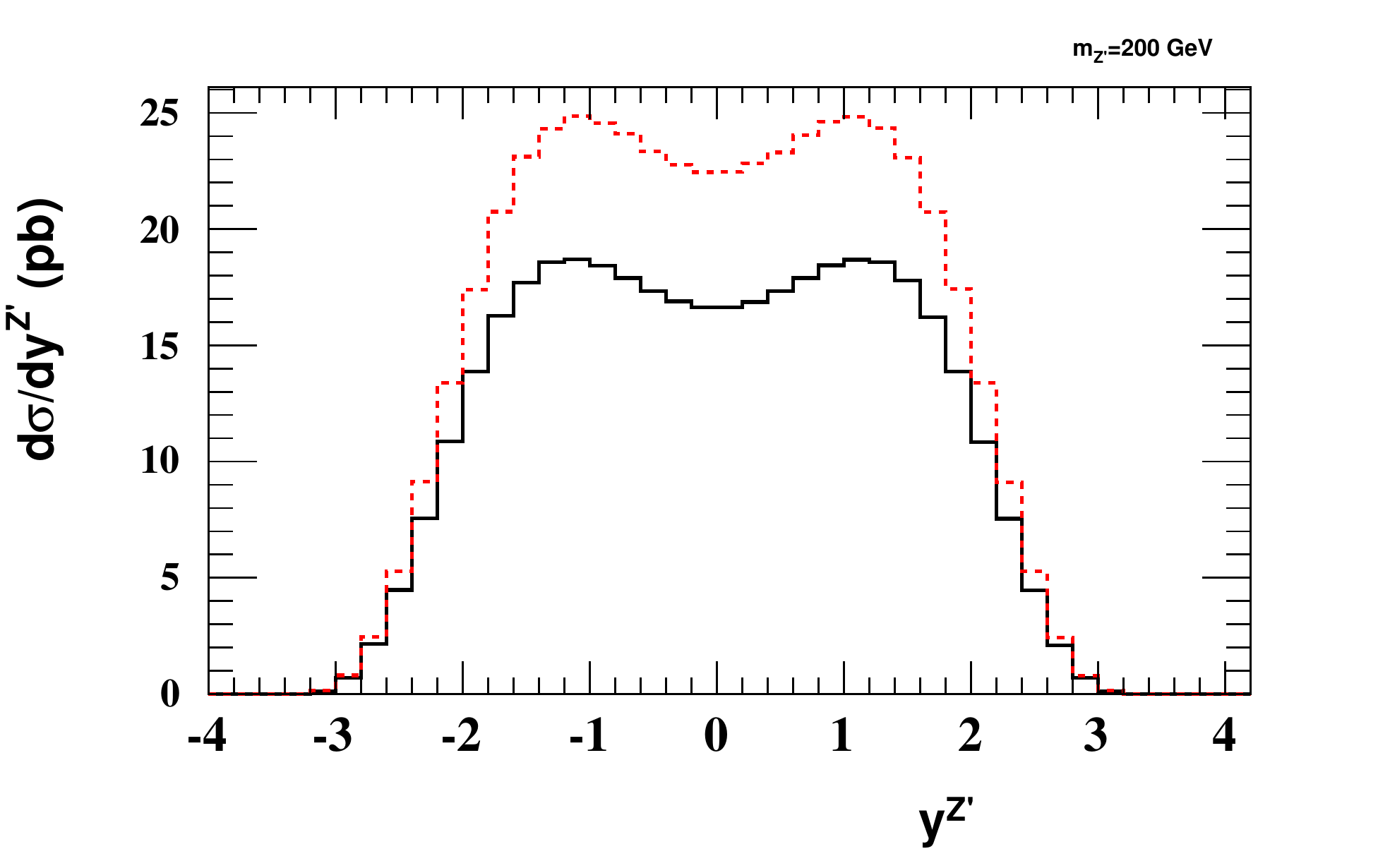}}\\
\scalebox{0.4}{\includegraphics{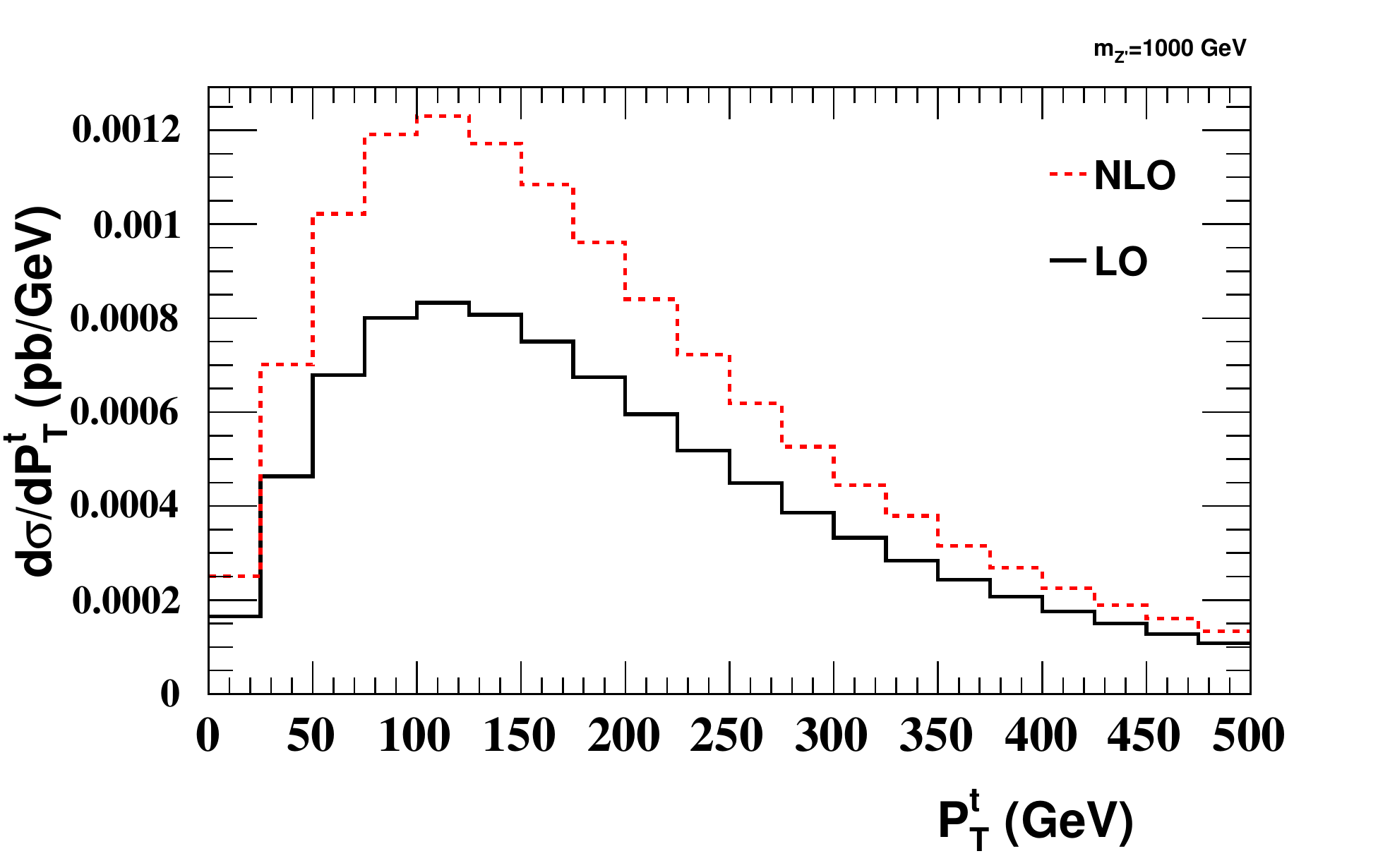}} \scalebox{0.4}{\includegraphics{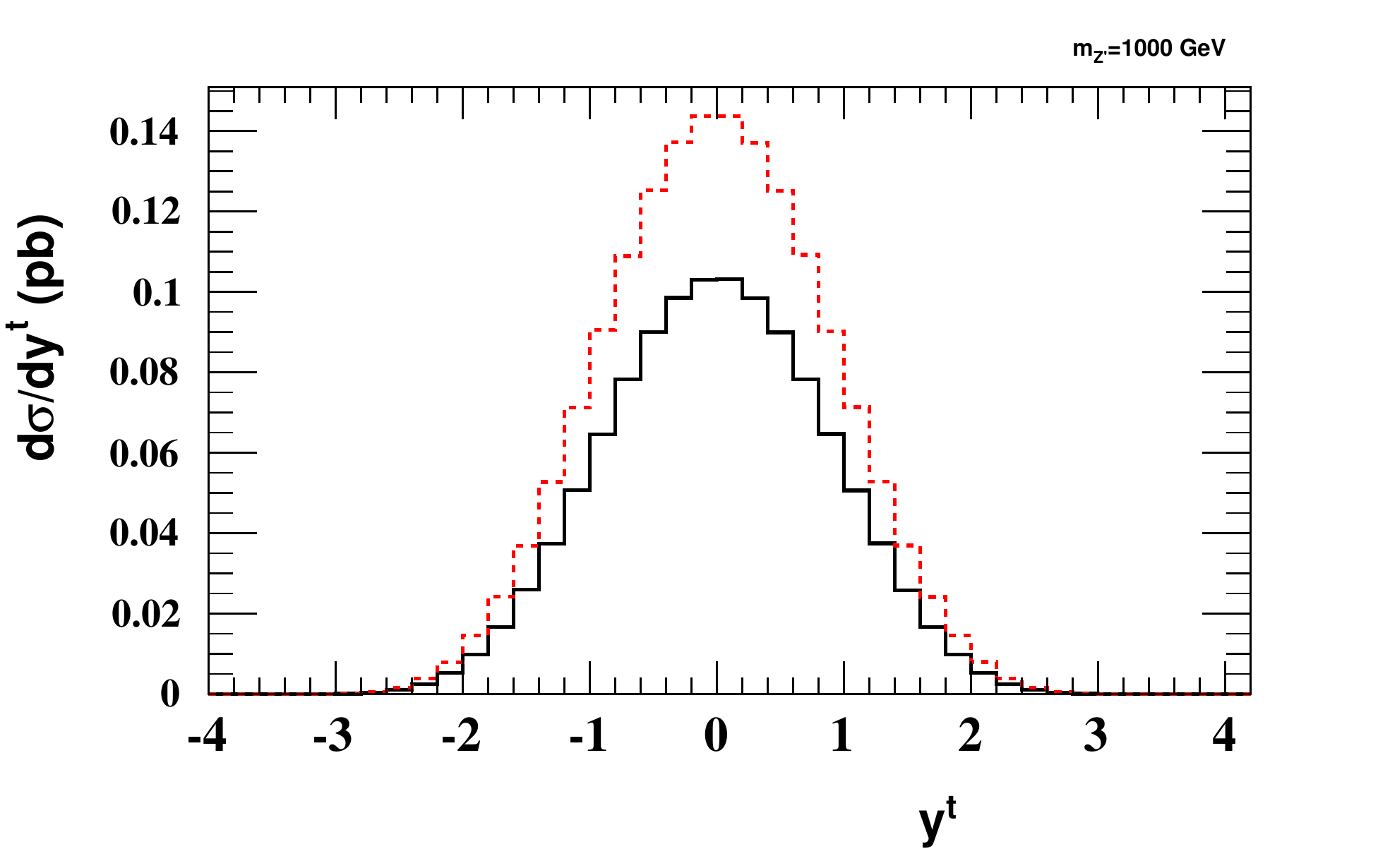}}\\
\scalebox{0.4}{\includegraphics{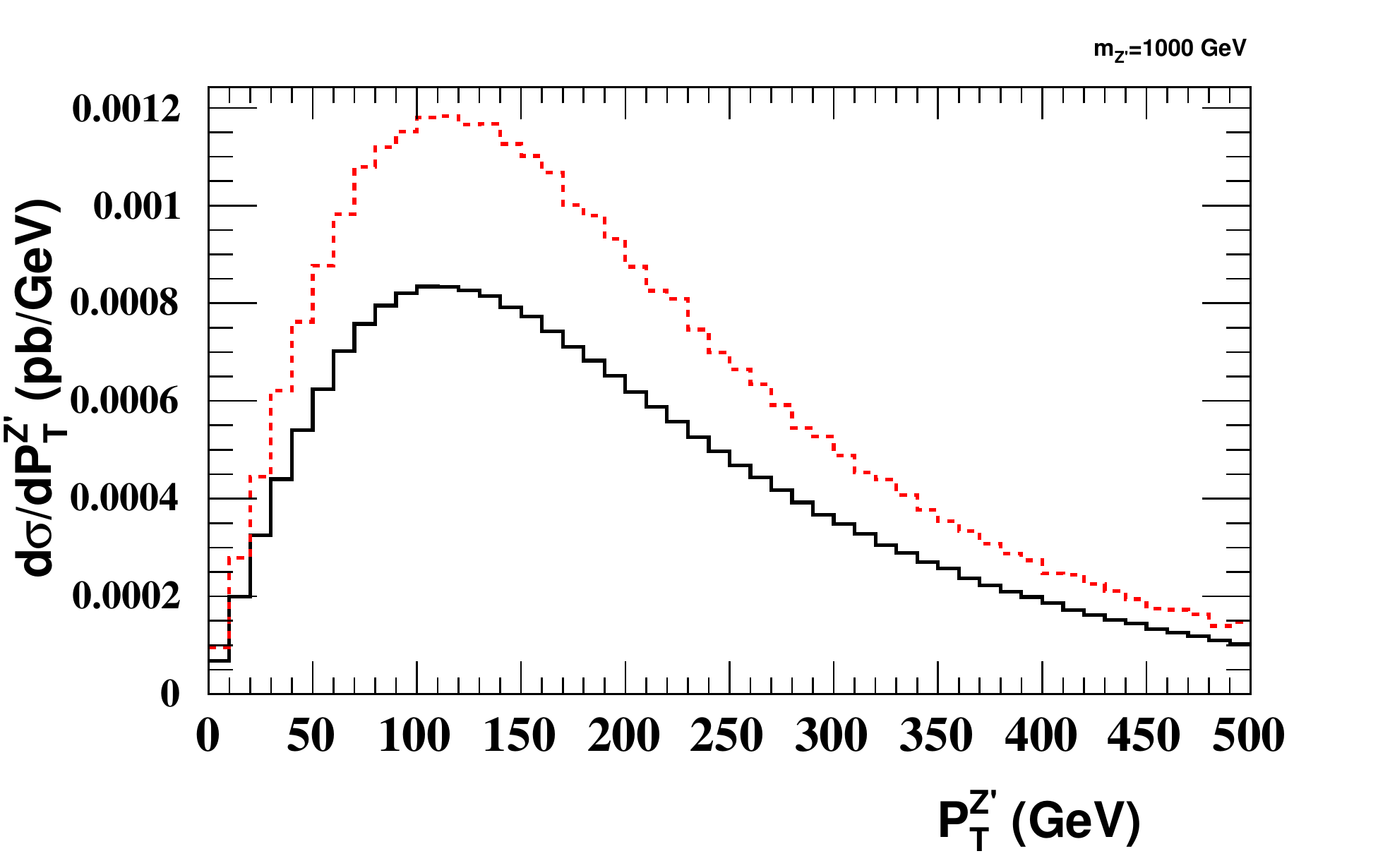}} \scalebox{0.4}{\includegraphics{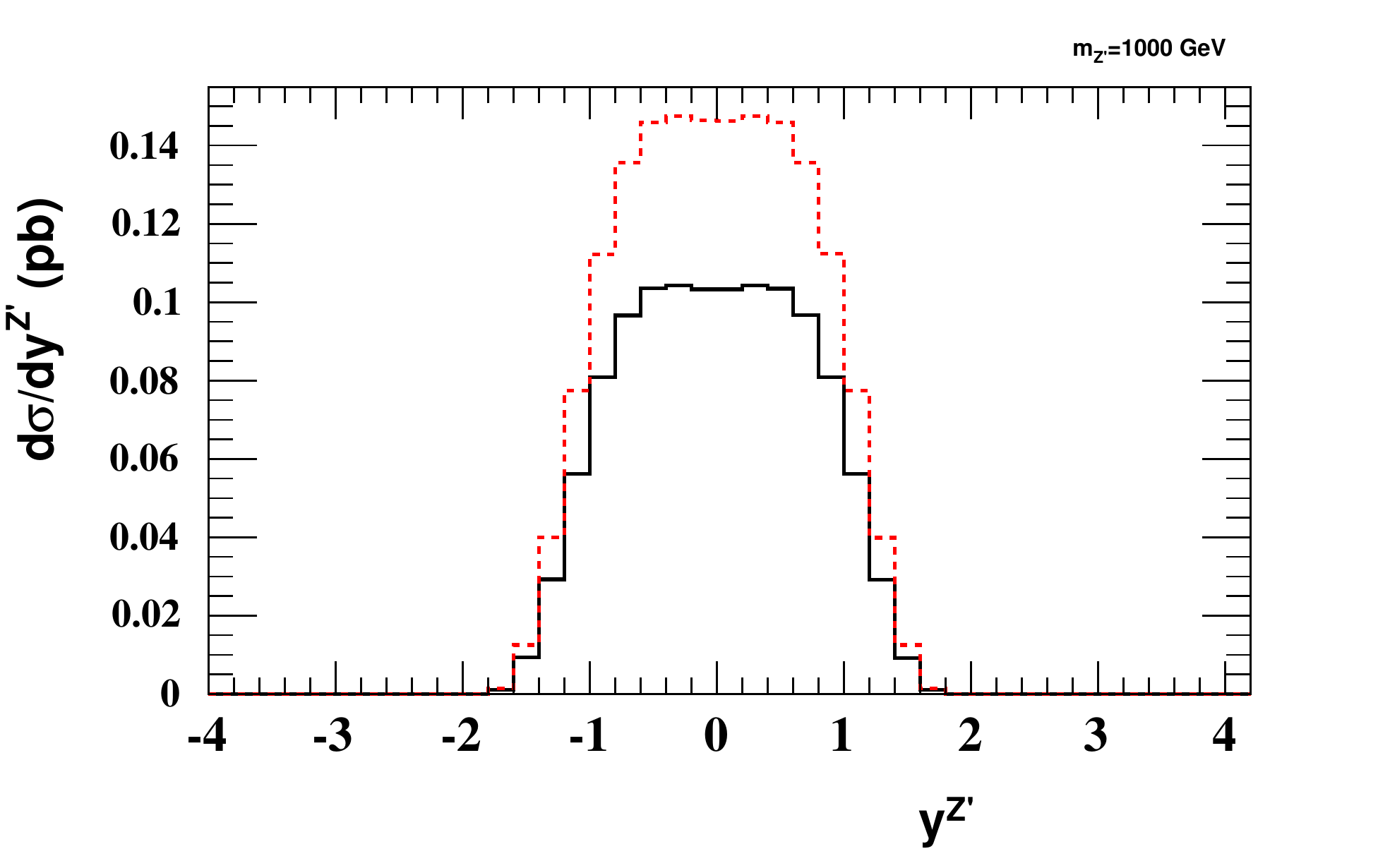}}\\
\caption{Transverse-momentum and rapidity distributions for the top quark and
 gauge boson in $tZ'$ production for $Z'$ masses of 200\,GeV and 1000\,GeV.}
\label{fig:zp200-kin}
\end{center}
\end{figure}

In figure~\ref{fig:zp200-kin}, we show the 
transverse-momentum and rapidity distributions for the top quark and gauge
 boson in  $tZ'$ production, for $Z'$ masses of 200\,GeV and 1000\,GeV respectively. One
sees that the top quark has a wide rapidity distribution peaking at zero, whereas 
the $Z'$ boson is preferentially produced in either the forward or backward 
direction. This is due to the fact that we are here considering the 
case of production via coupling of the top to an up quark. The latter is a 
valence quark, and thus carries more momentum on average than the initial 
state gluon. This results in a boosted final state, where the $Z'$ boson is 
preferentially emitted in the direction of the incoming up quark, from
helicity considerations. The double peak structure will be absent for couplings
between the top and second generation quarks, or for antitop production, as
there is then no valence quark in the initial state. 
Furthermore, the rapidity distributions become narrower as the gauge boson mass
increases, reflecting the fact that production becomes more central for heavier
final states. This is also reflected in the fact that the peaks of the 
transverse momentum distributions (for both the top quark and gauge boson) 
shift upwards as the gauge boson mass increases.\\

\begin{figure}
\begin{center}
\scalebox{0.4}{\includegraphics{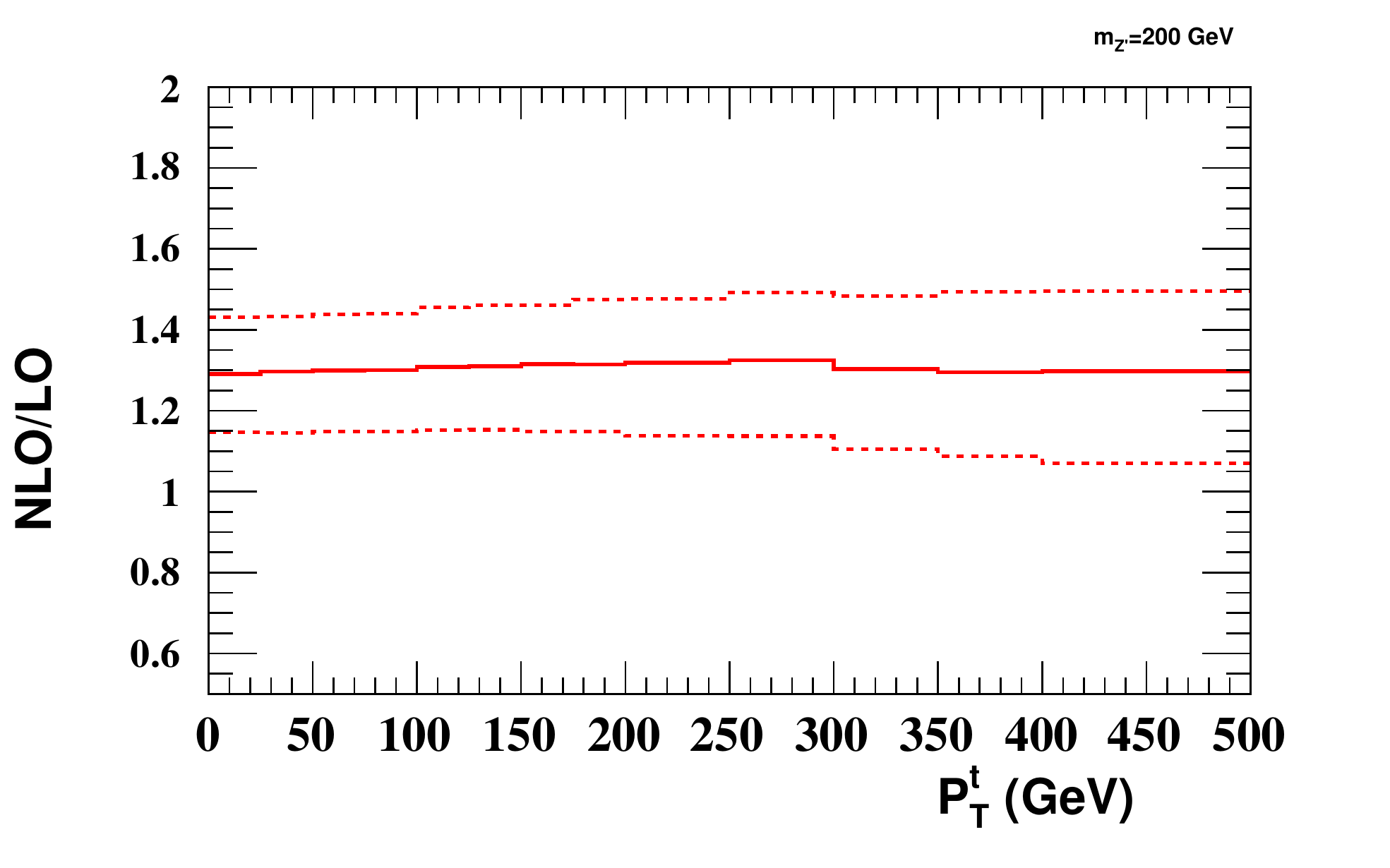}} \scalebox{0.4}{\includegraphics{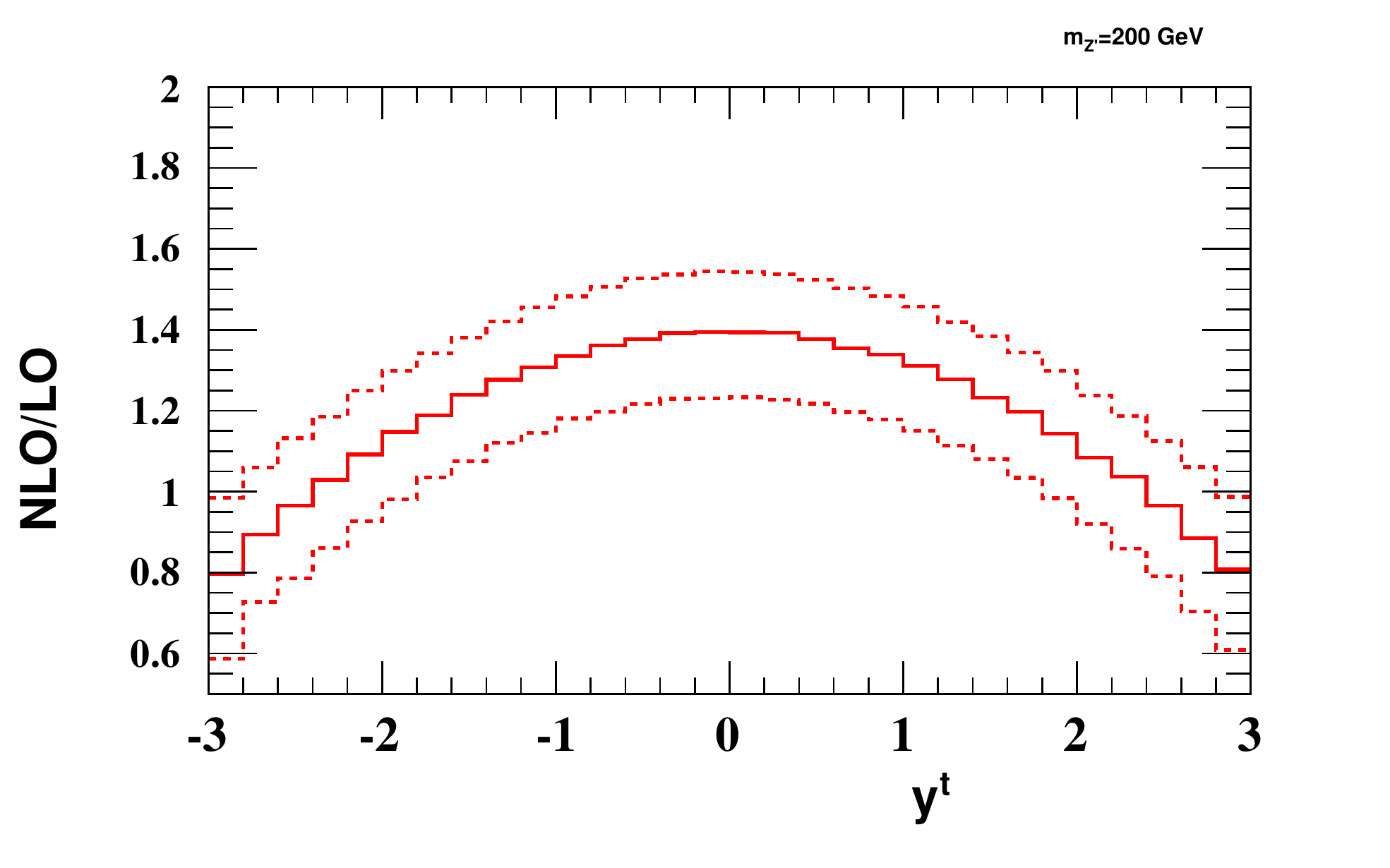}}\\
\scalebox{0.4}{\includegraphics{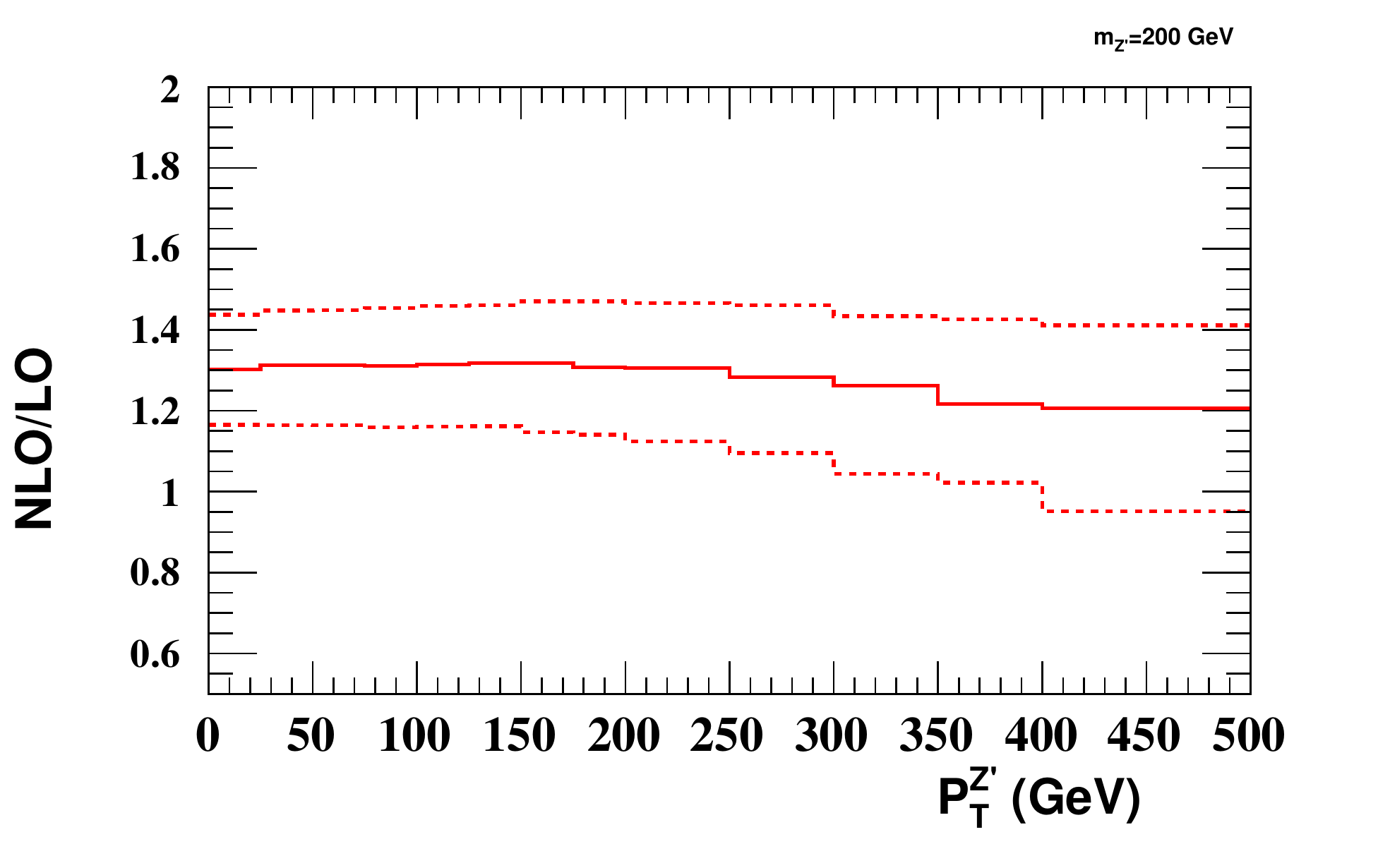}} \scalebox{0.4}{\includegraphics{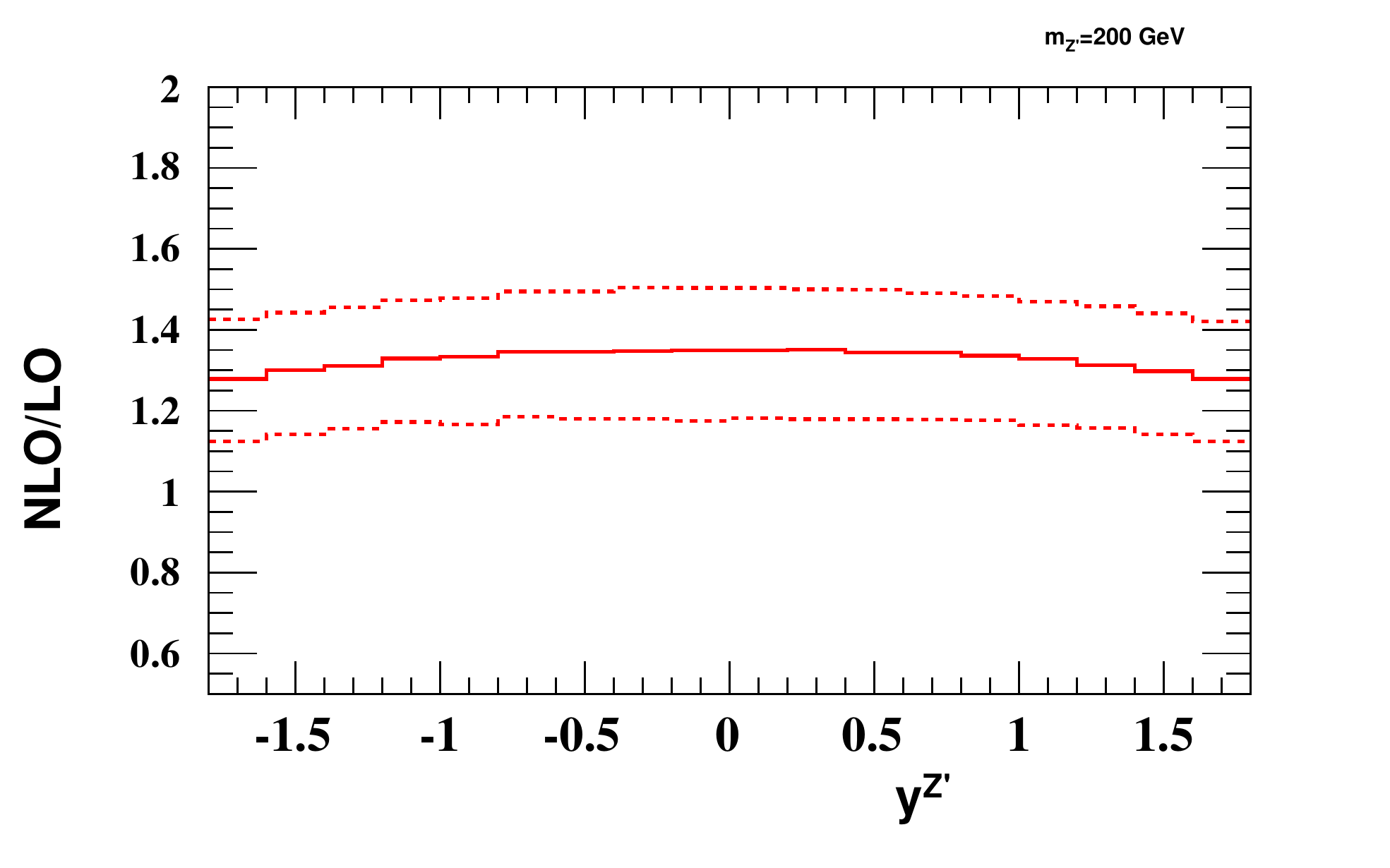}}\\
\scalebox{0.4}{\includegraphics{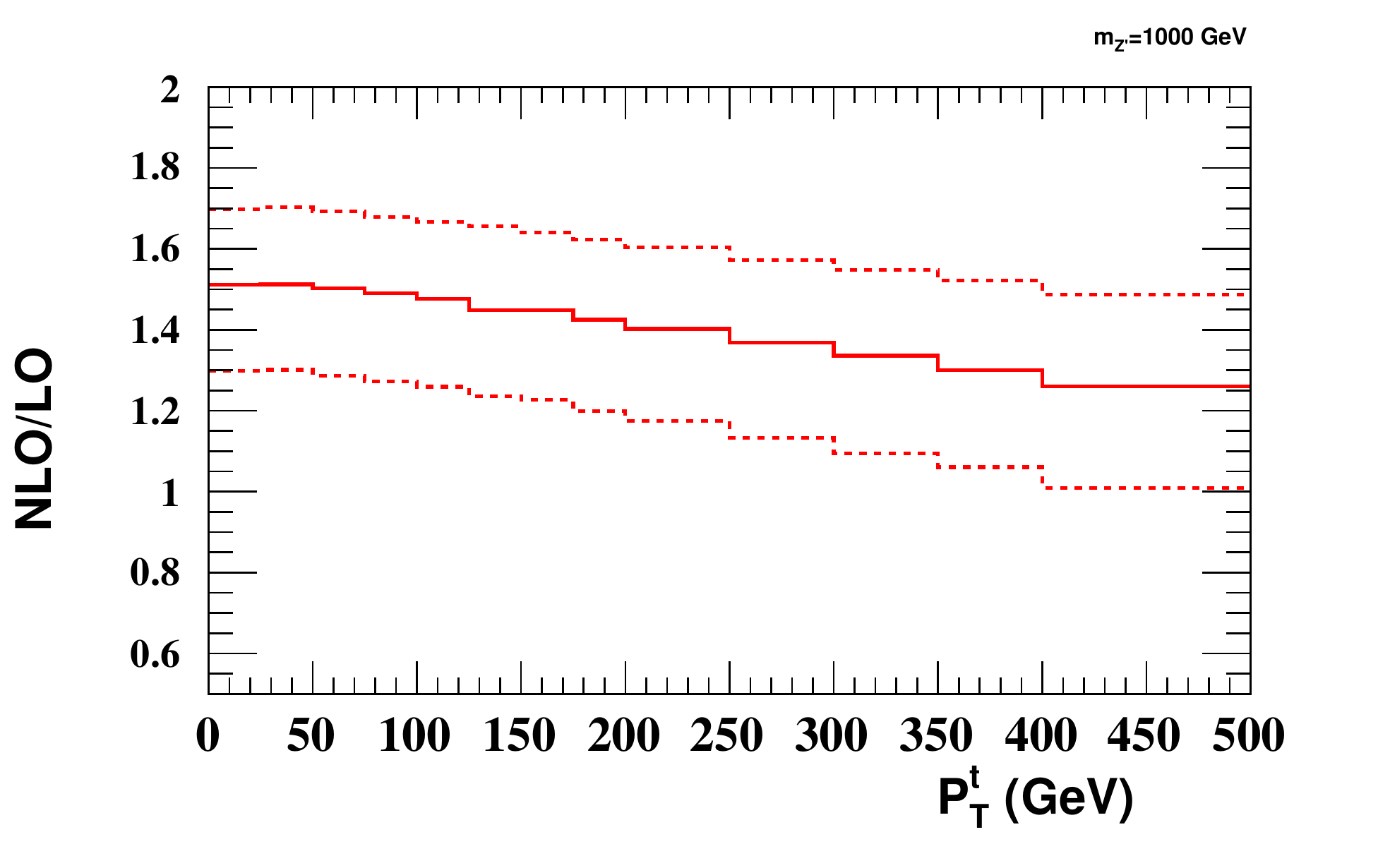}} \scalebox{0.4}{\includegraphics{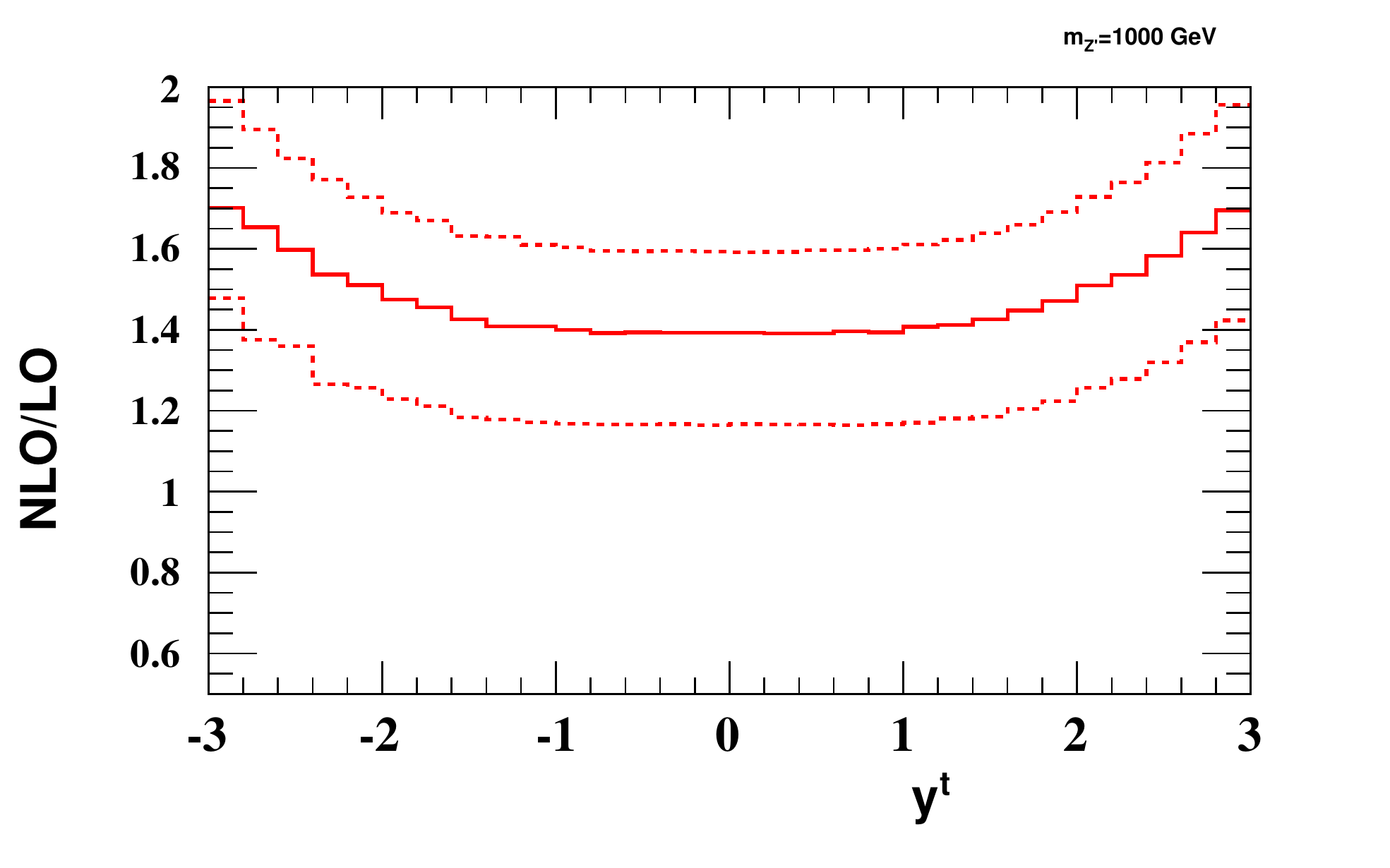}}\\
\scalebox{0.4}{\includegraphics{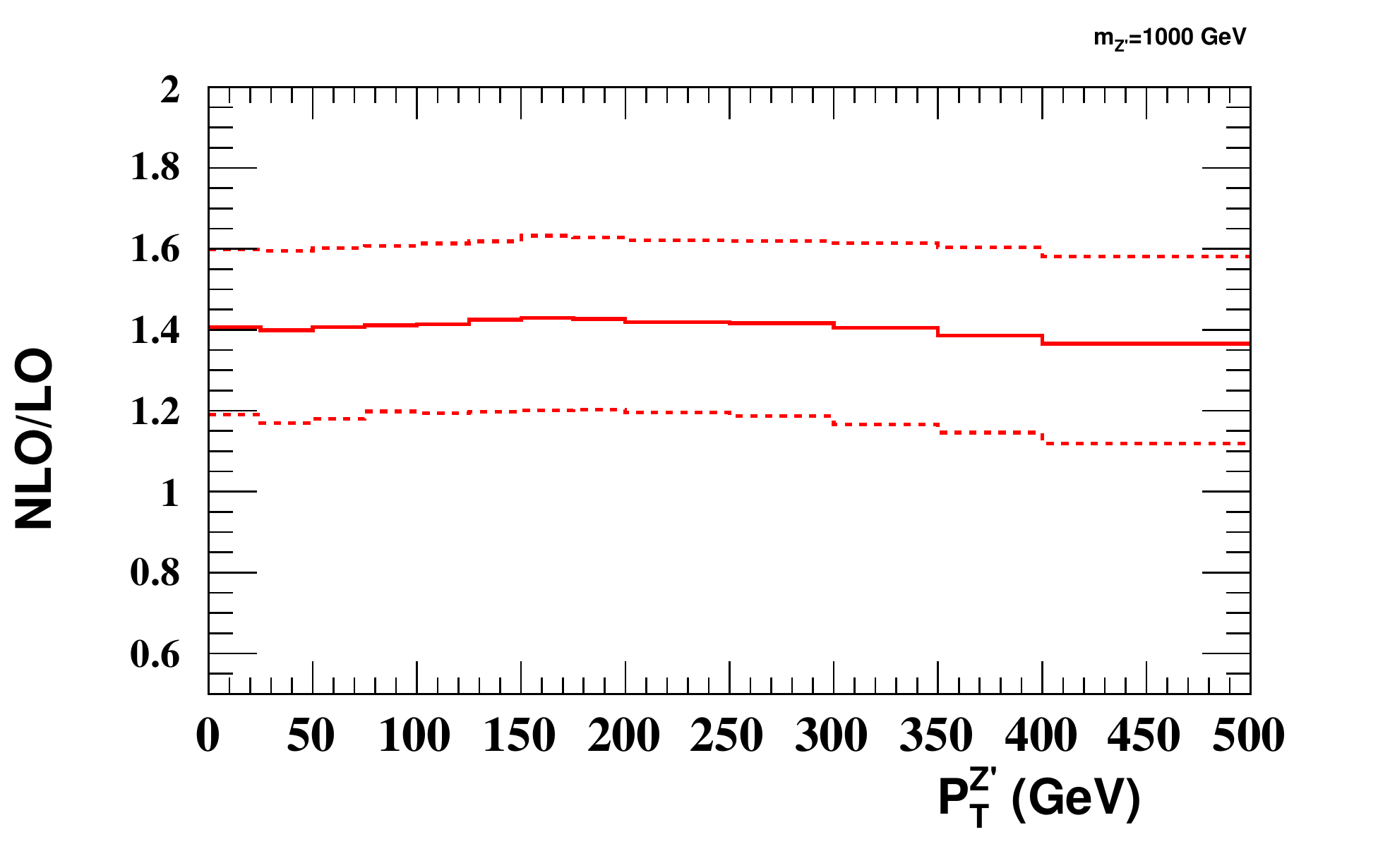}} \scalebox{0.4}{\includegraphics{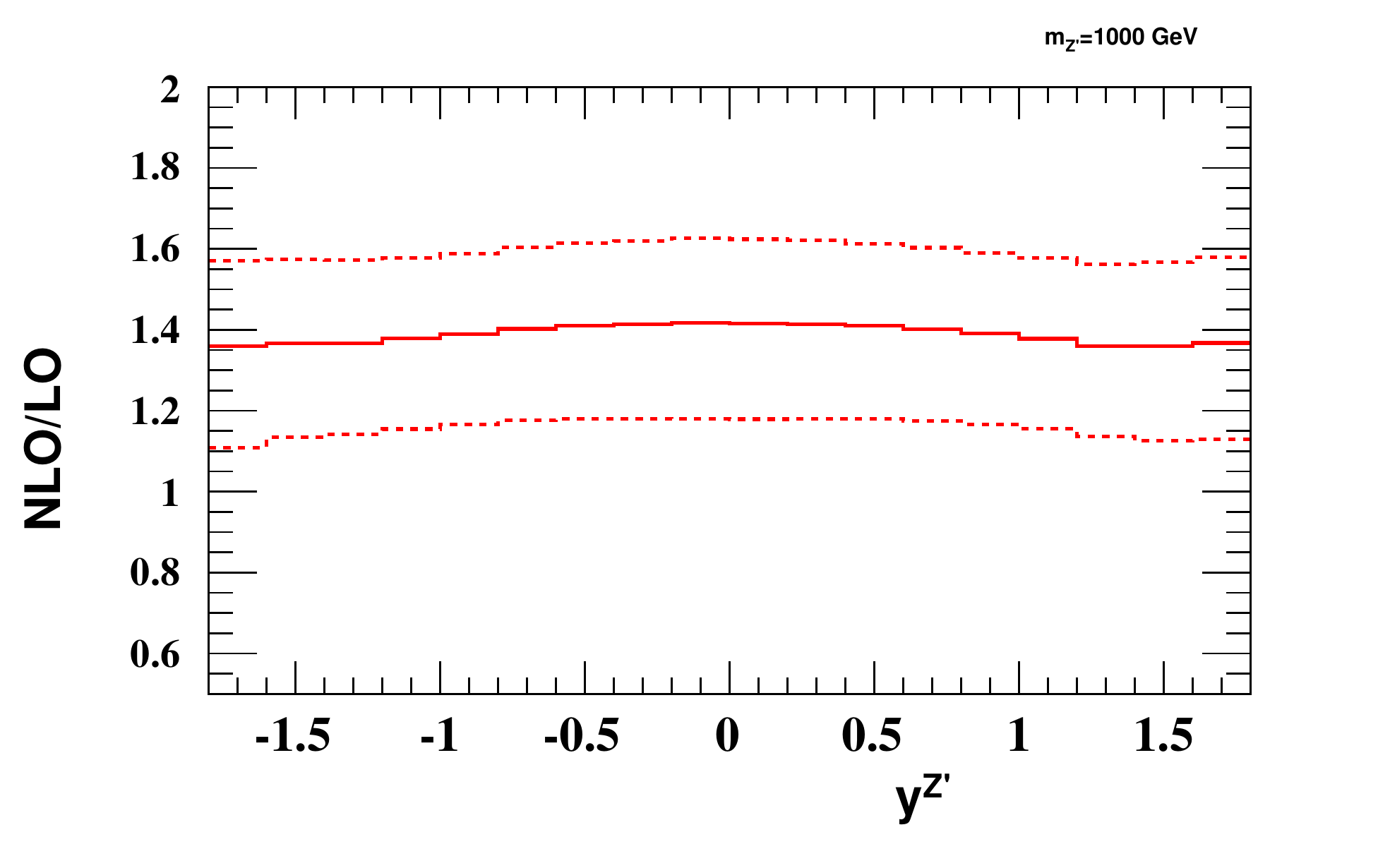}}\\
\caption{Ratio of the NLO and LO predictions for the transverse-momentum and rapidity distributions of
the top quark and gauge boson in $tZ'$ production with $Z'$ masses of 200\,GeV and 1000\,GeV. The scale uncertainties are shown as dashed lines.}
\label{fig:zp200-kin-ratio}
\end{center}
\end{figure}

\begin{figure}
\begin{center}
\scalebox{0.4}{\includegraphics{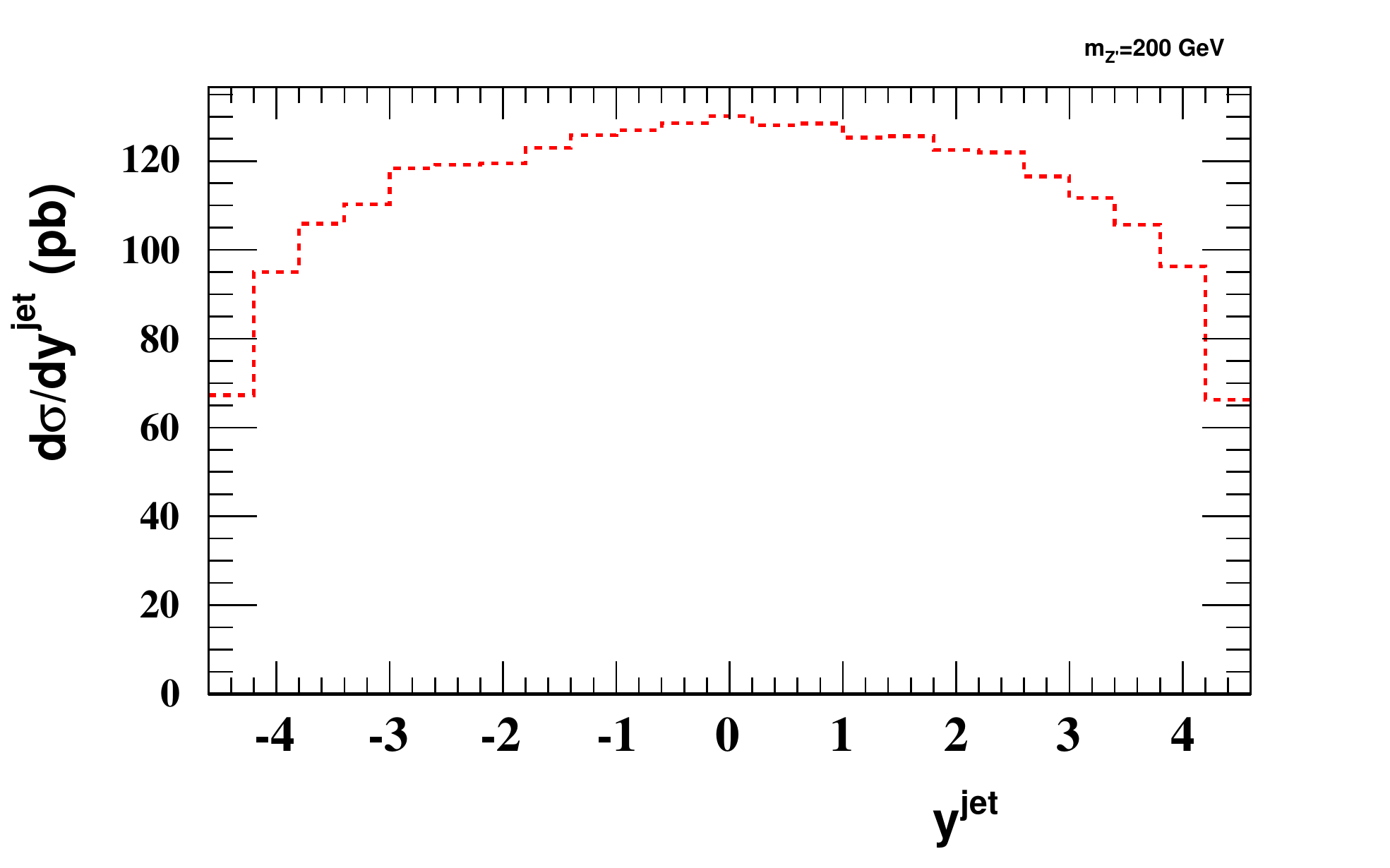}} \scalebox{0.4}{\includegraphics{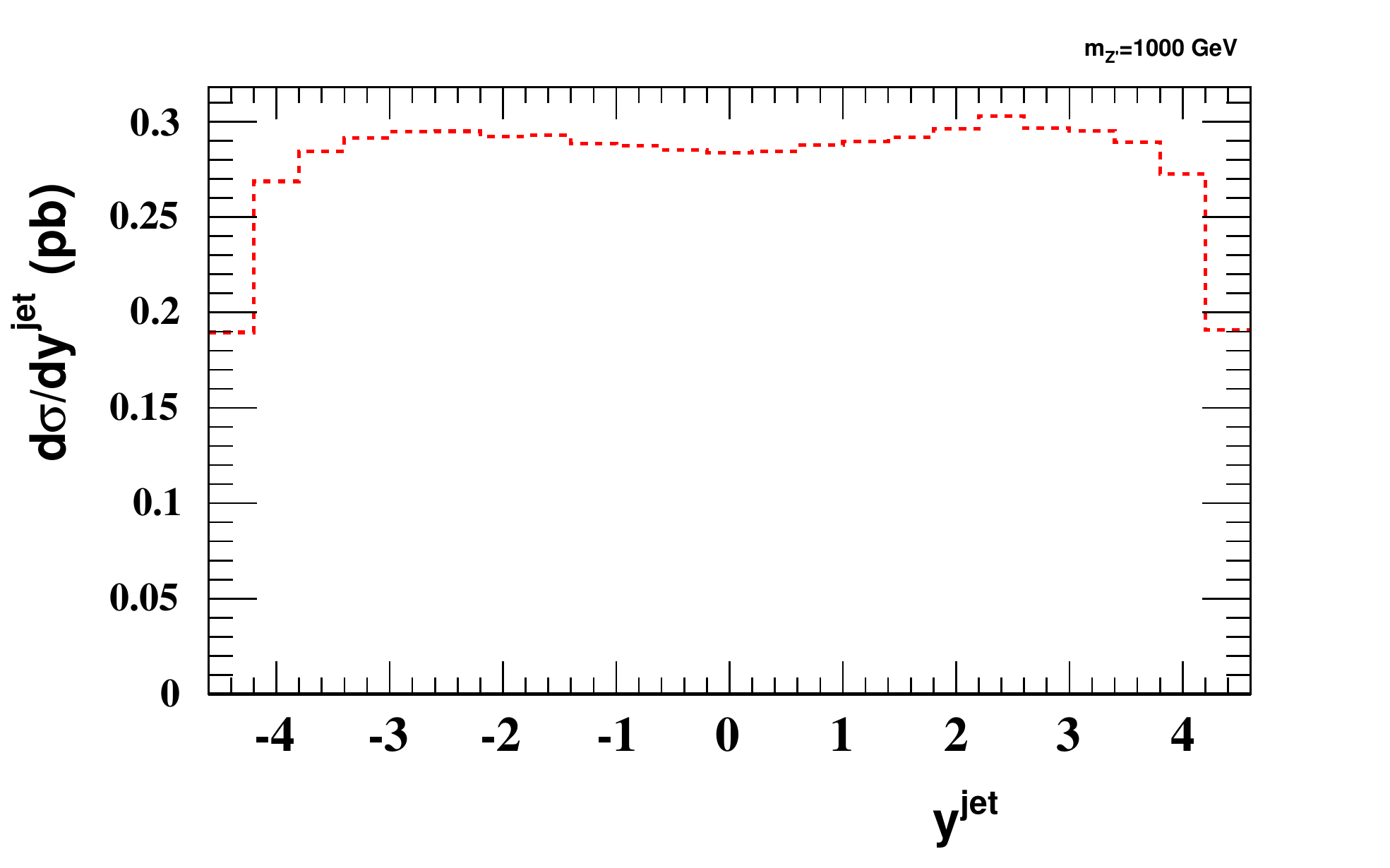}}\\
\scalebox{0.4}{\includegraphics{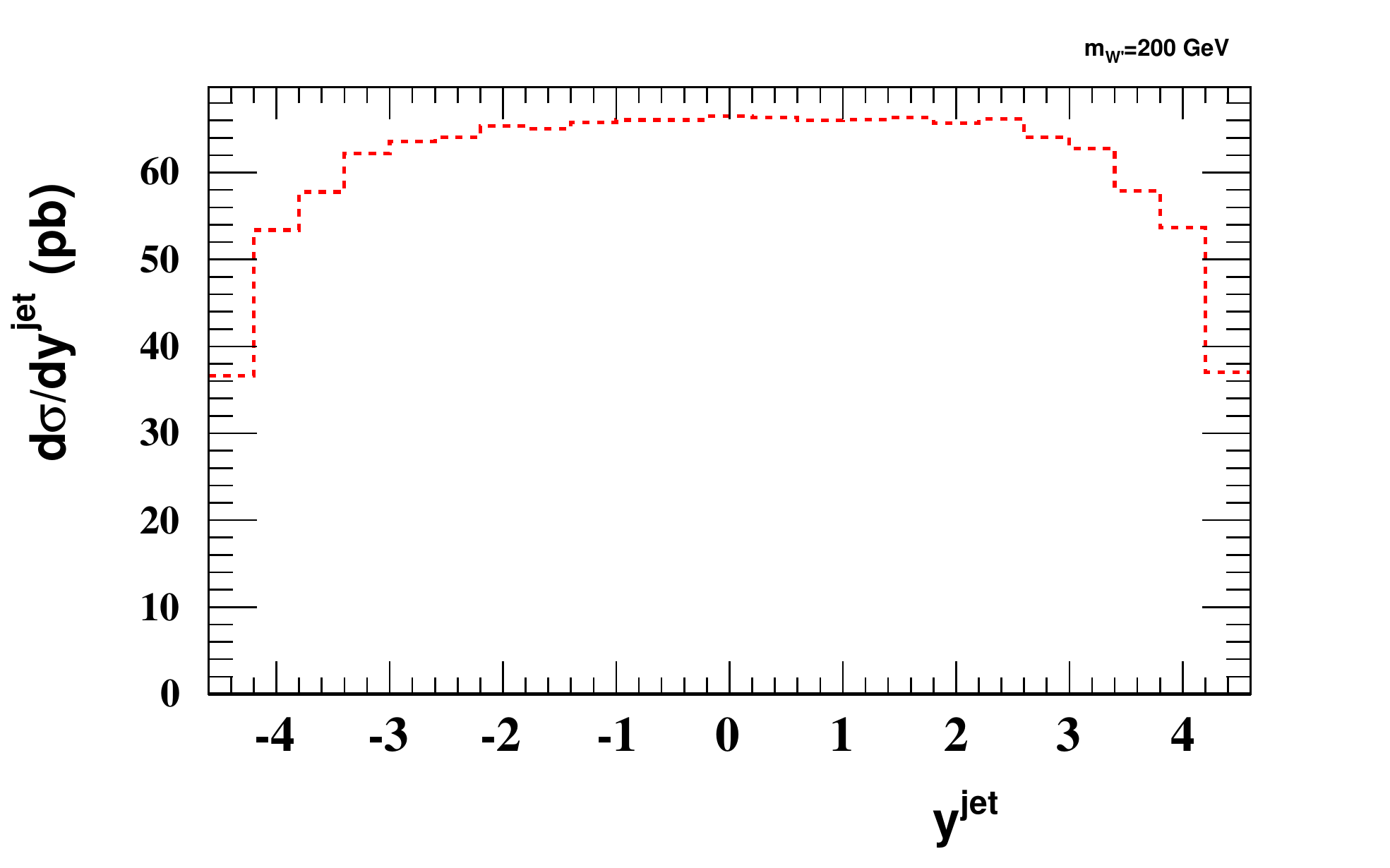}} \scalebox{0.4}{\includegraphics{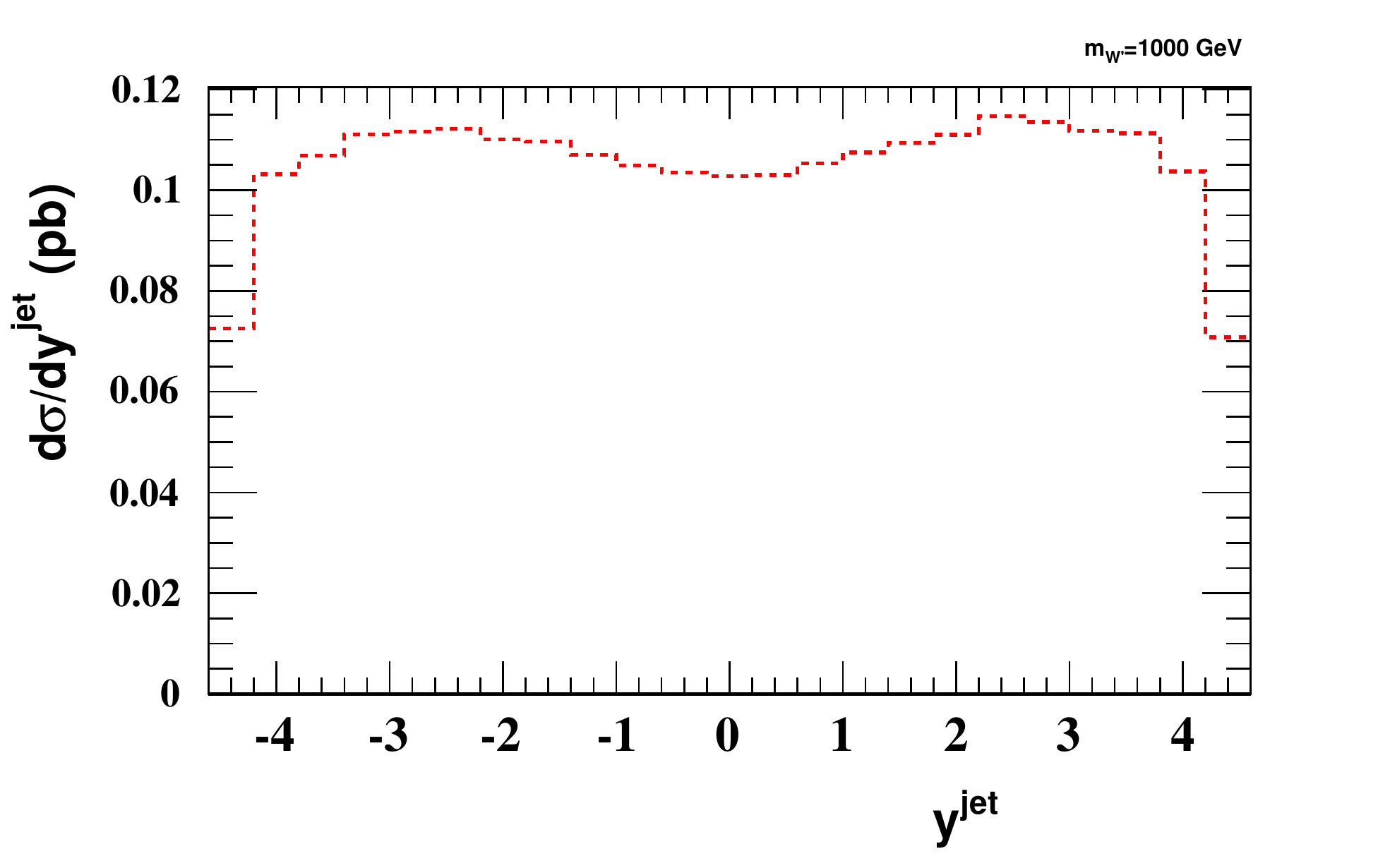}}\\
\caption{Rapidity distributions for the extra parton at NLO, for gauge boson
masses of 200\,GeV (left) and 1000\,GeV (right), in $Z'$ (upper) and $W'$ (lower)
production.}
\label{fig:extraparton}
\end{center}
\end{figure}

We show the ratio of the NLO and LO results for the above distributions 
for a 200 GeV gauge $Z'$ boson in figure~\ref{fig:zp200-kin-ratio}. One sees
that for lower gauge boson masses, NLO corrections have little impact on the
shape of the transverse momentum distributions of the top quark or $Z'$ boson
(although the tails of the NLO distributions are slightly softer, as expected).
This can be understood by the fact that QCD radiation is dominated by 
emissions which are collinear to the incoming particles: the 
heavy final state limits the phase space for hard gluon emission, and contains
a non-radiating colour-singlet particle and a heavy quark, such that there are
no final state collinear singularities. This expectation is confirmed in 
figure~\ref{fig:extraparton}, which shows the rapidity distribution of the
additional parton emitted at NLO. The distribution is wide and flat, and thus
characteristic of initial state radiation. This acts to make the rapidity
distributions of the $t$ and $Z'$ more strongly peaked at central values (i.e.
there is less energy available to the $tZ'$ pair), and thus to proportionally 
reduce the double peak structure in the $Z'$-boson rapidity distribution. Ratio
plots for a heavy gauge boson (1000 GeV) are also shown in 
figure~\ref{fig:zp200-kin-ratio}. Here one sees that the transverse momentum
of the top gets somewhat softer at NLO (more so than for lower gauge boson 
masses), and also its rapidity distribution is slightly less central. This
can be understood by noting that the top quark is more energetic on average
for heavier gauge boson masses, due to momentum conservation. It can therefore
emit harder radiation, such that the jet and top quark are produced slightly 
off-centre. This is borne out by the rapidity distribution of the extra parton
in figure~\ref{fig:extraparton}, which shows a shape difference with respect 
to the case of a 200 GeV boson. \\

\begin{figure}
\begin{center}
\scalebox{0.4}{\includegraphics{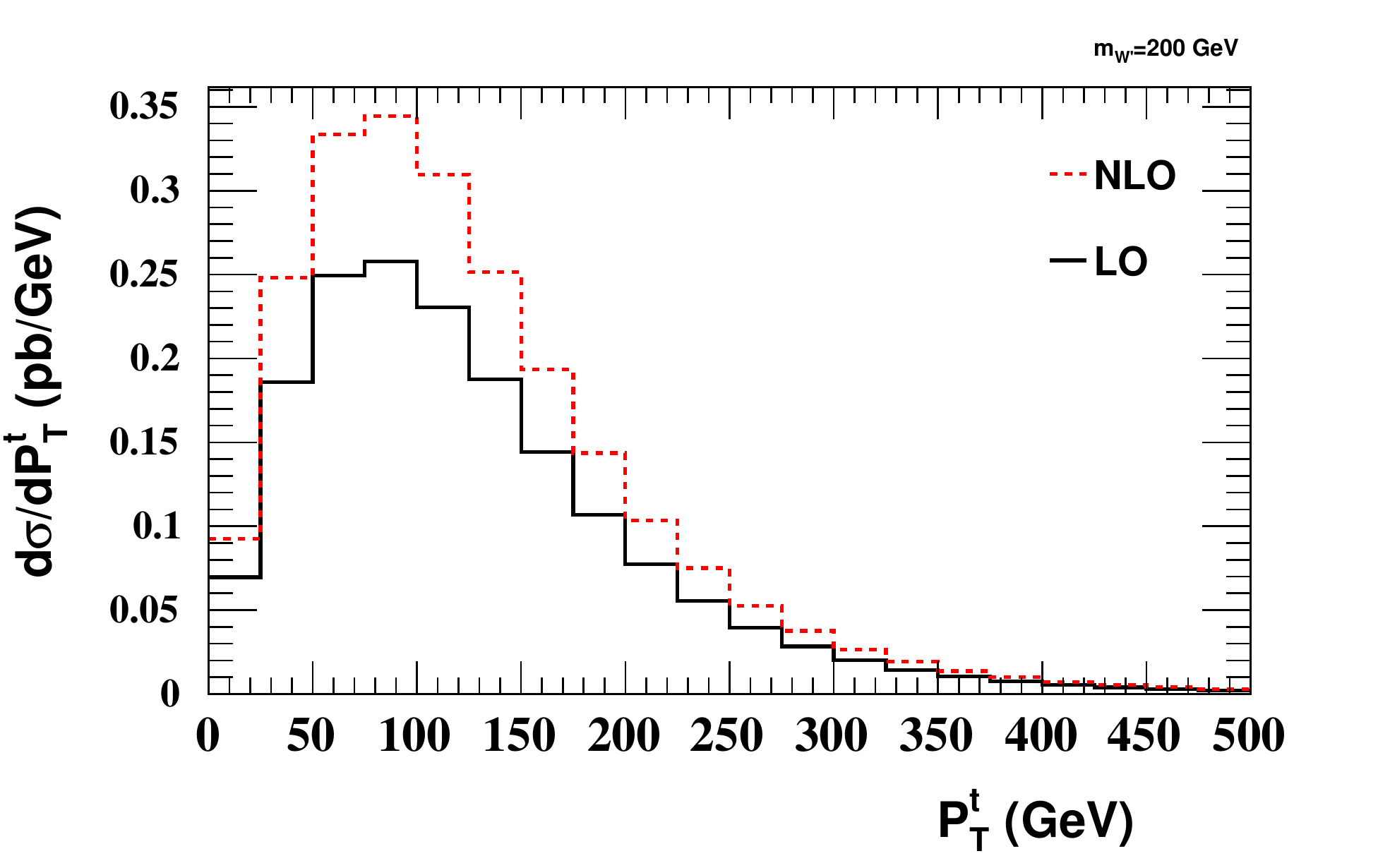}} \scalebox{0.4}{\includegraphics{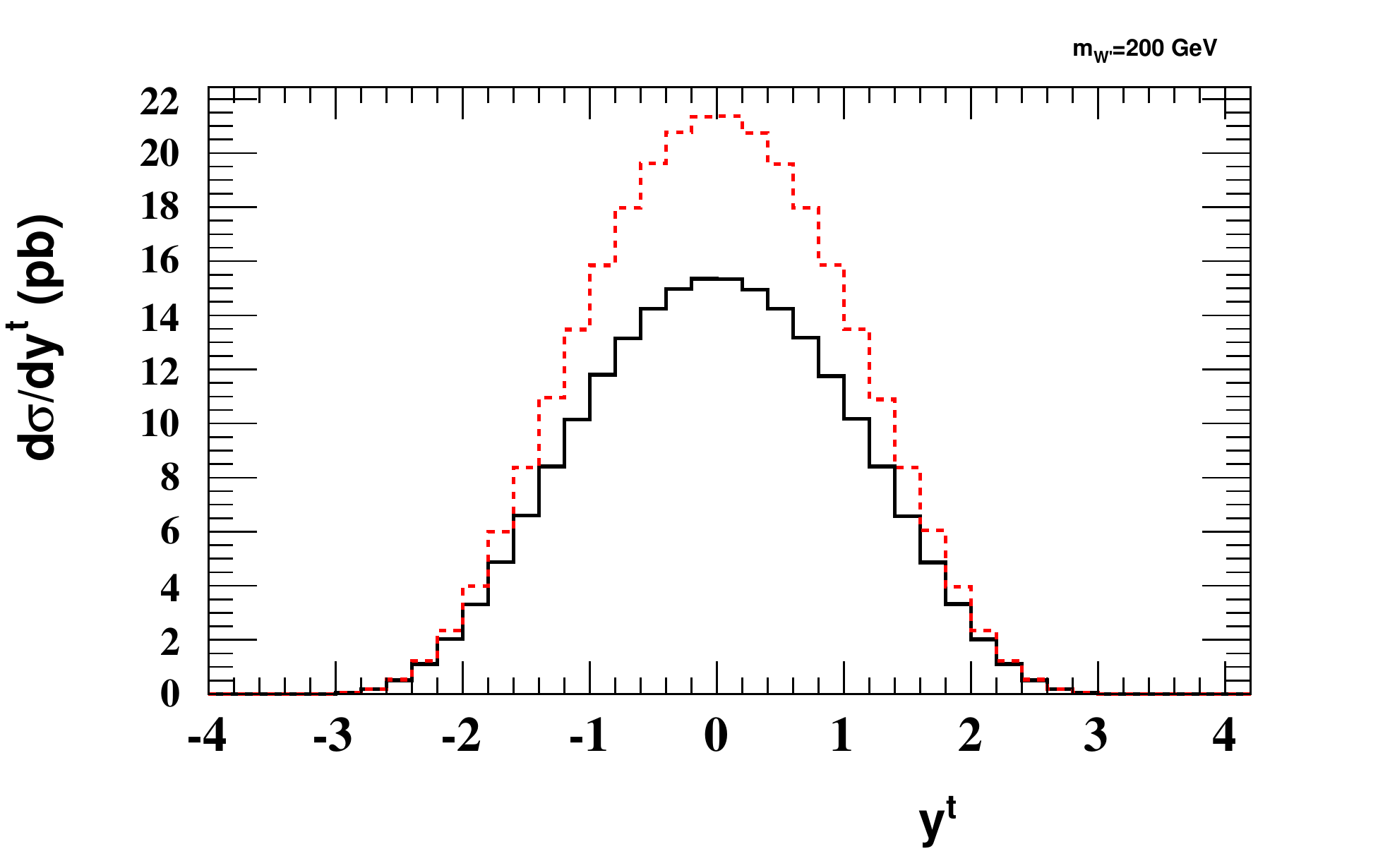}}\\
\scalebox{0.4}{\includegraphics{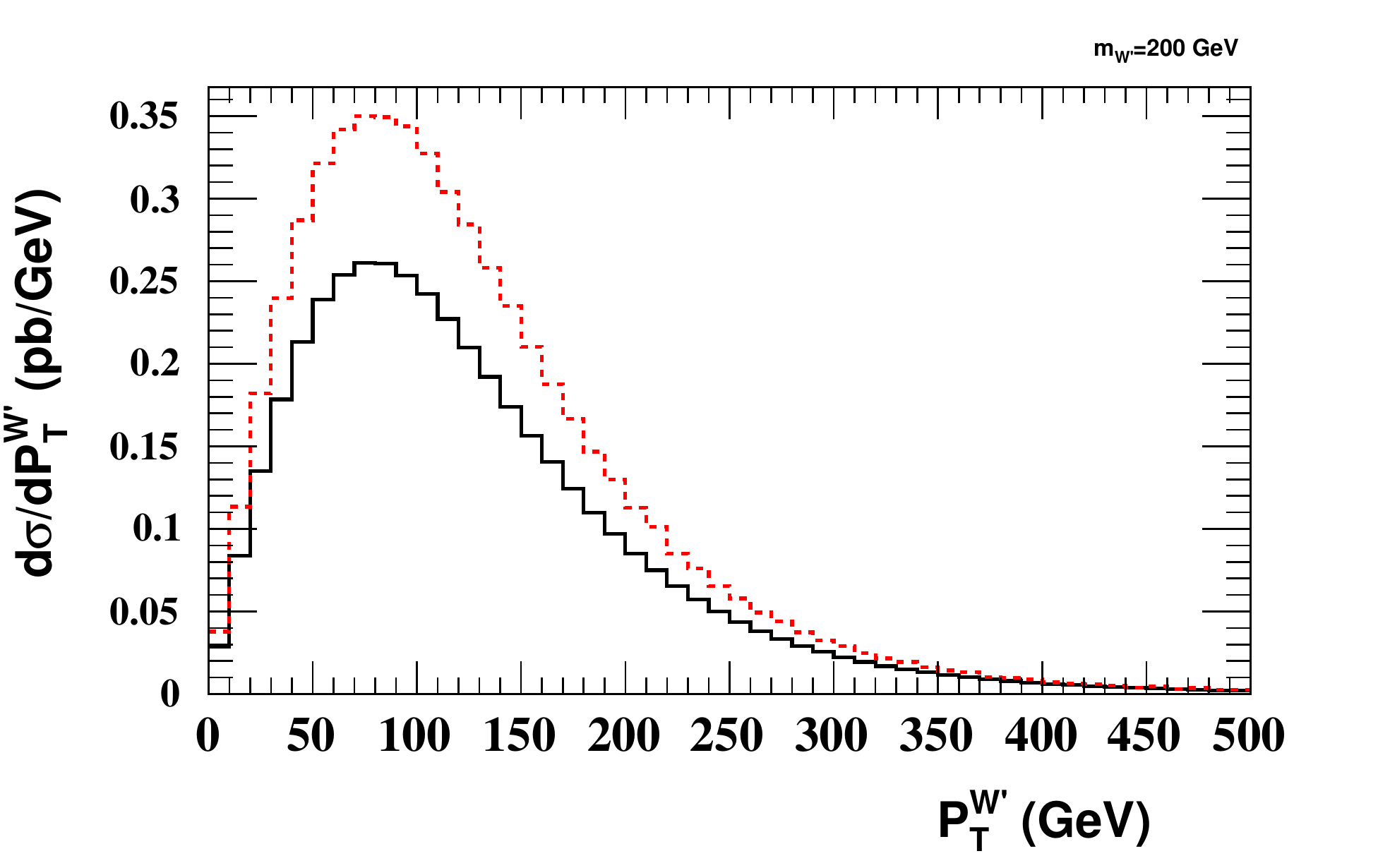}} \scalebox{0.4}{\includegraphics{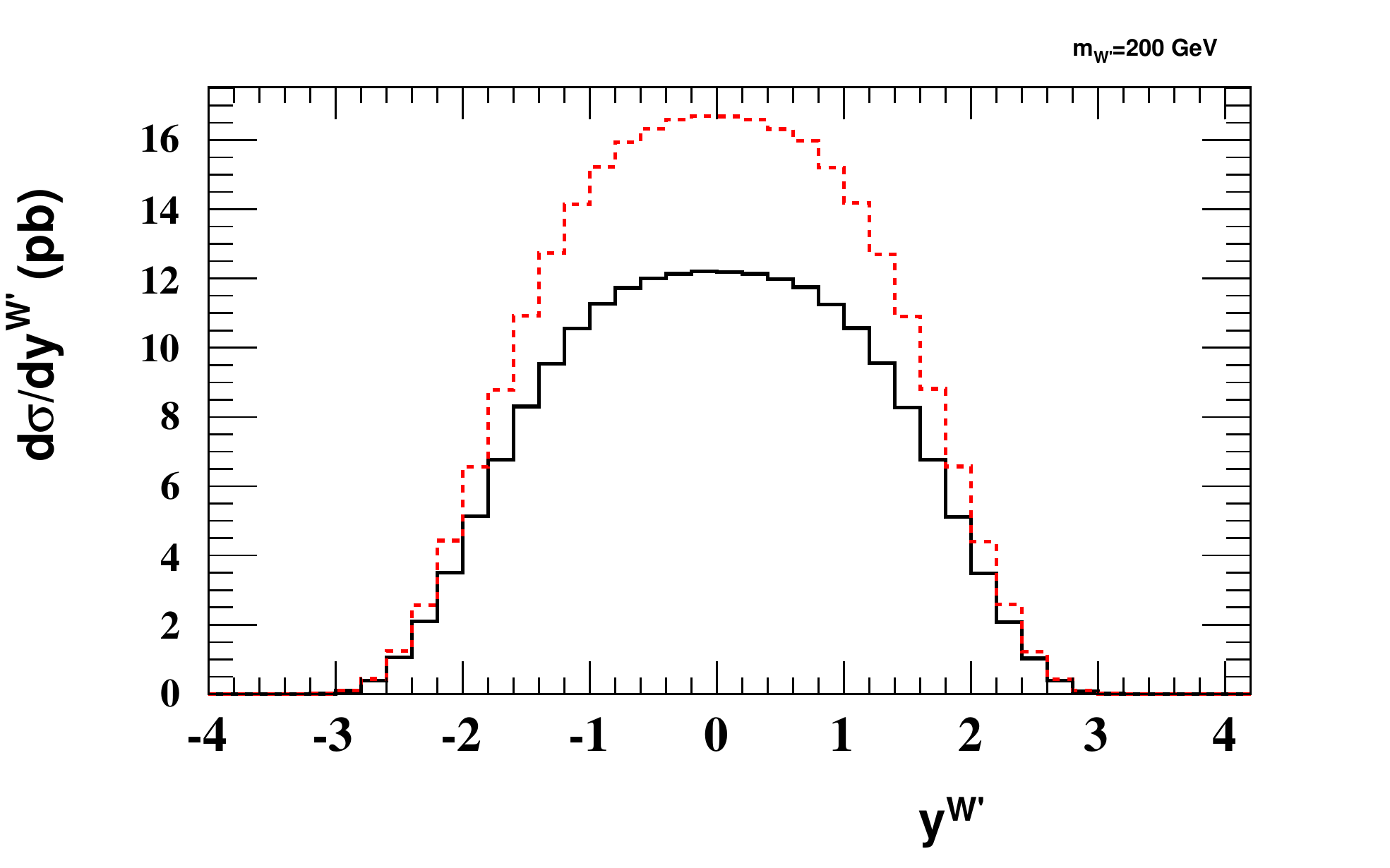}}\\
\scalebox{0.4}{\includegraphics{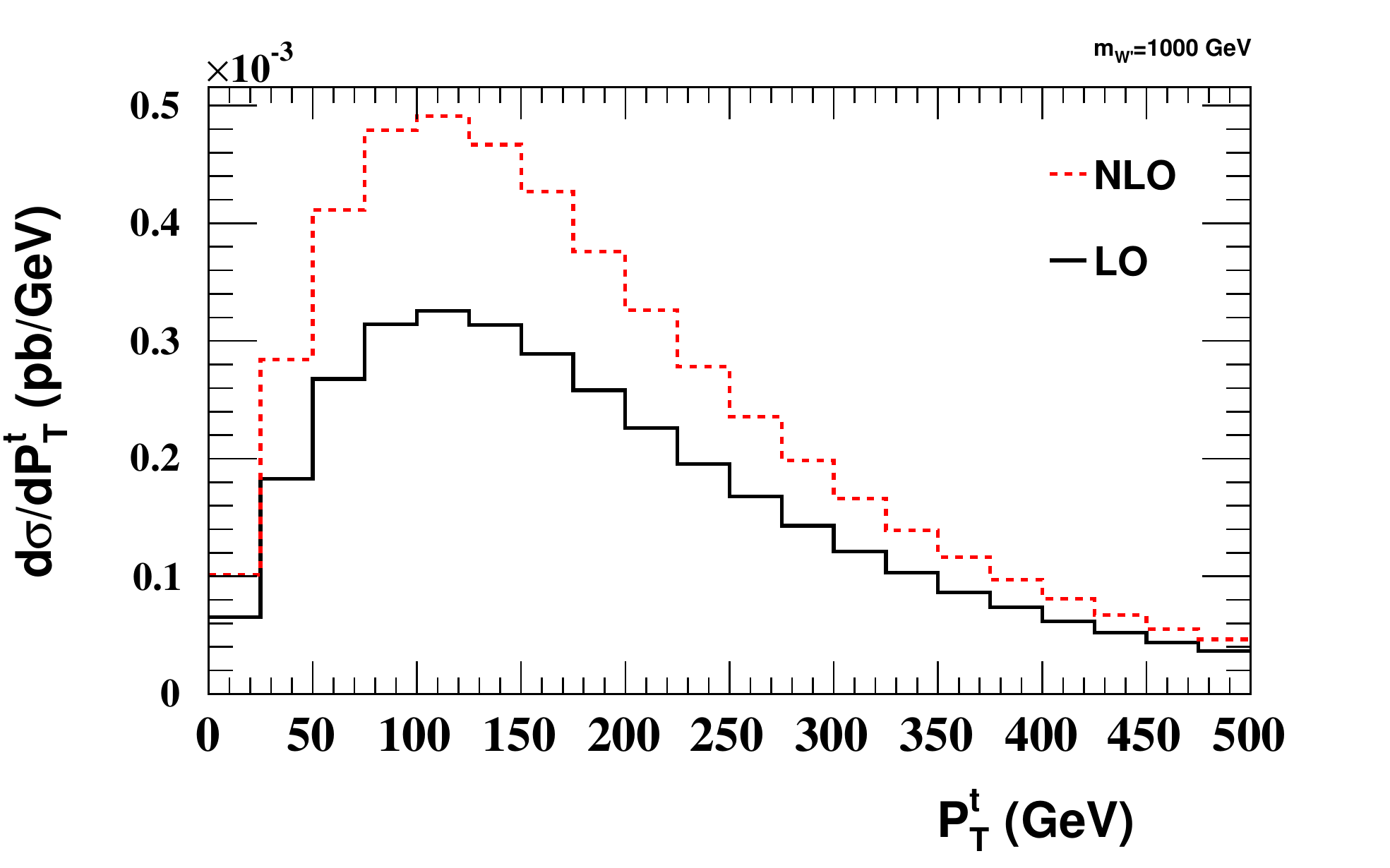}} \scalebox{0.4}{\includegraphics{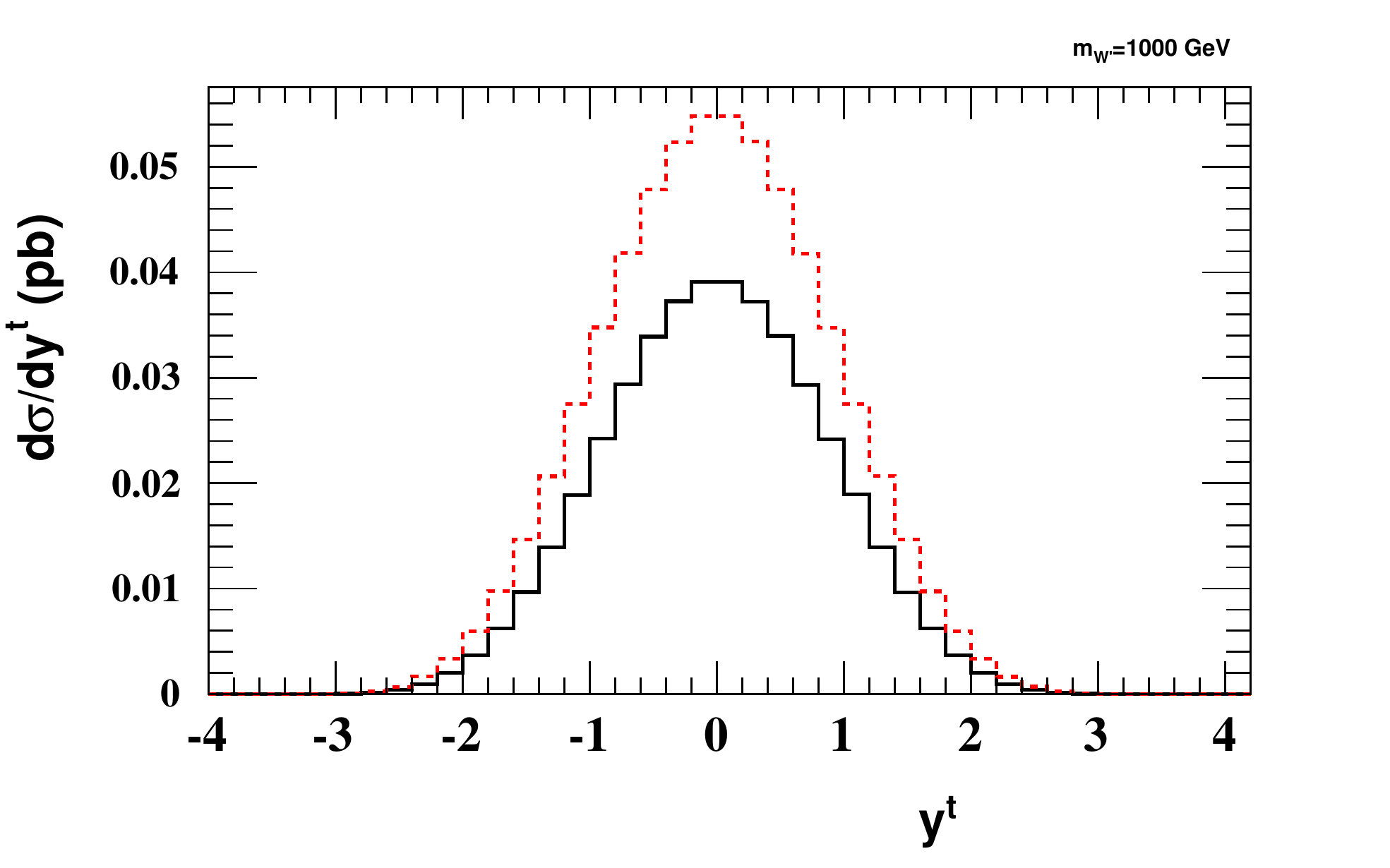}}\\
\scalebox{0.4}{\includegraphics{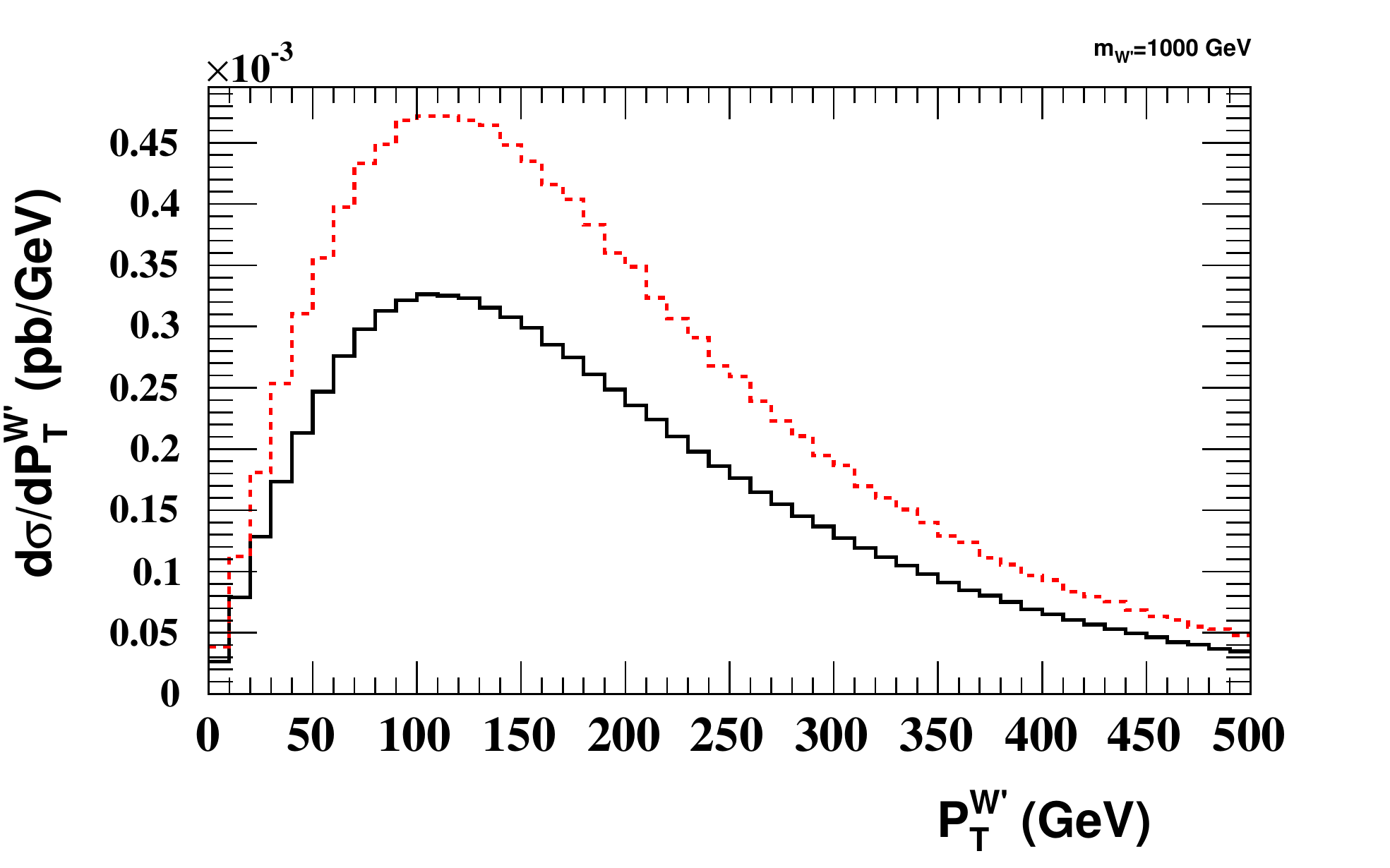}} \scalebox{0.4}{\includegraphics{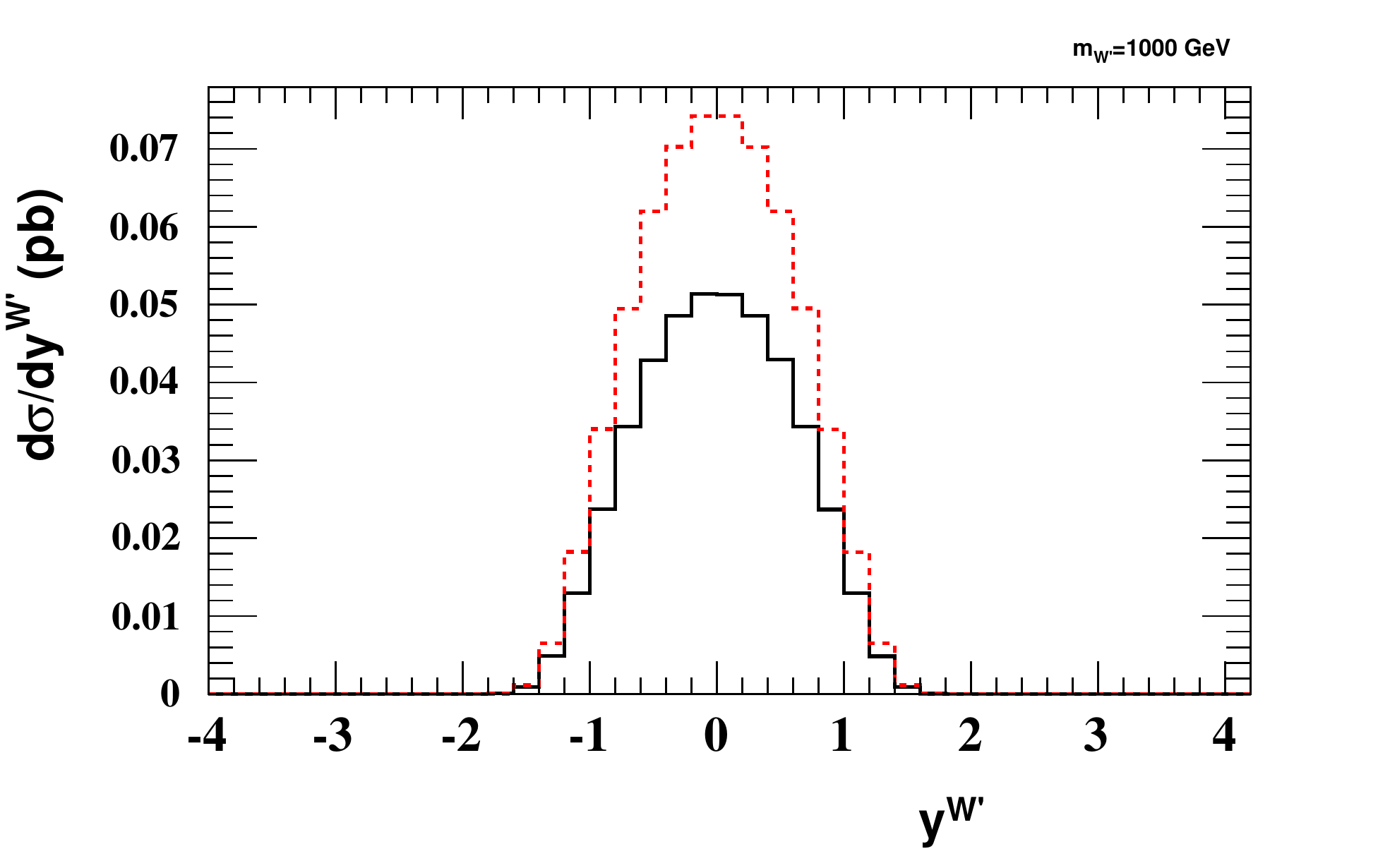}}\\
\caption{Transverse-momentum and rapidity distributions for the top quark and
gauge boson in $tW'$ production with $W'$ masses 200\,GeV and 1000\,GeV.}
\label{fig:wp200-kin}
\end{center}
\end{figure}

\begin{figure}
\begin{center}
\scalebox{0.4}{\includegraphics{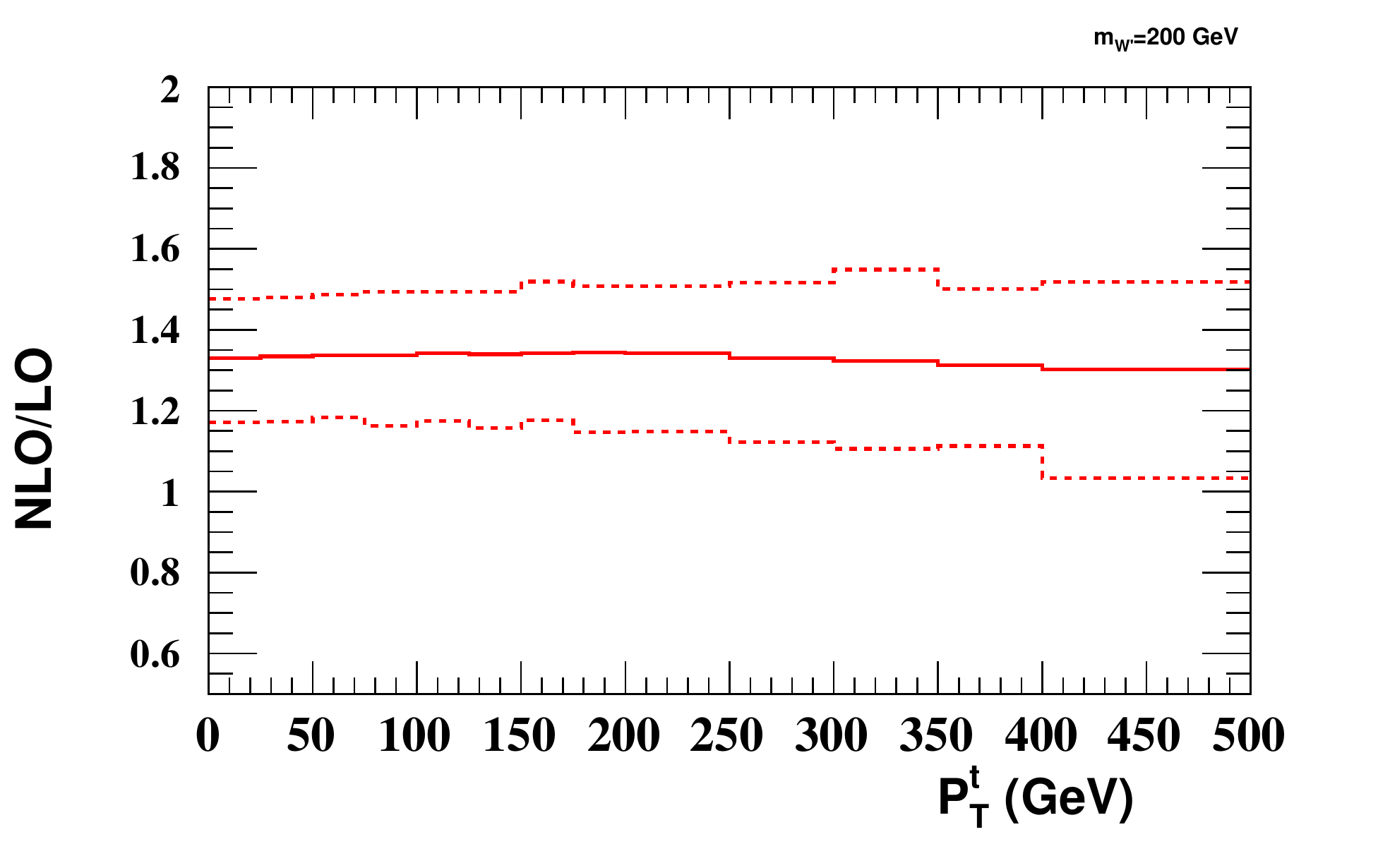}} \scalebox{0.4}{\includegraphics{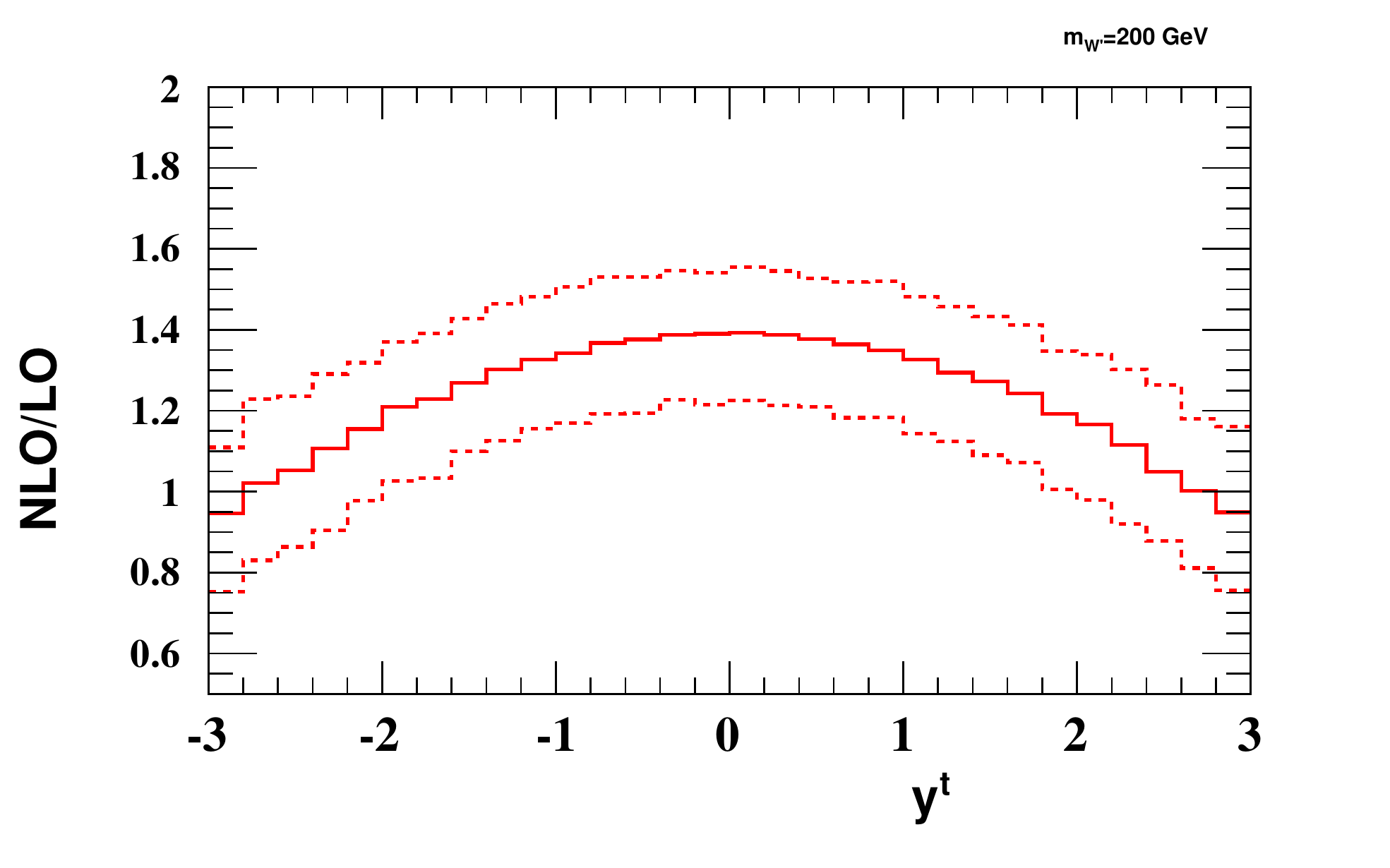}}\\
\scalebox{0.4}{\includegraphics{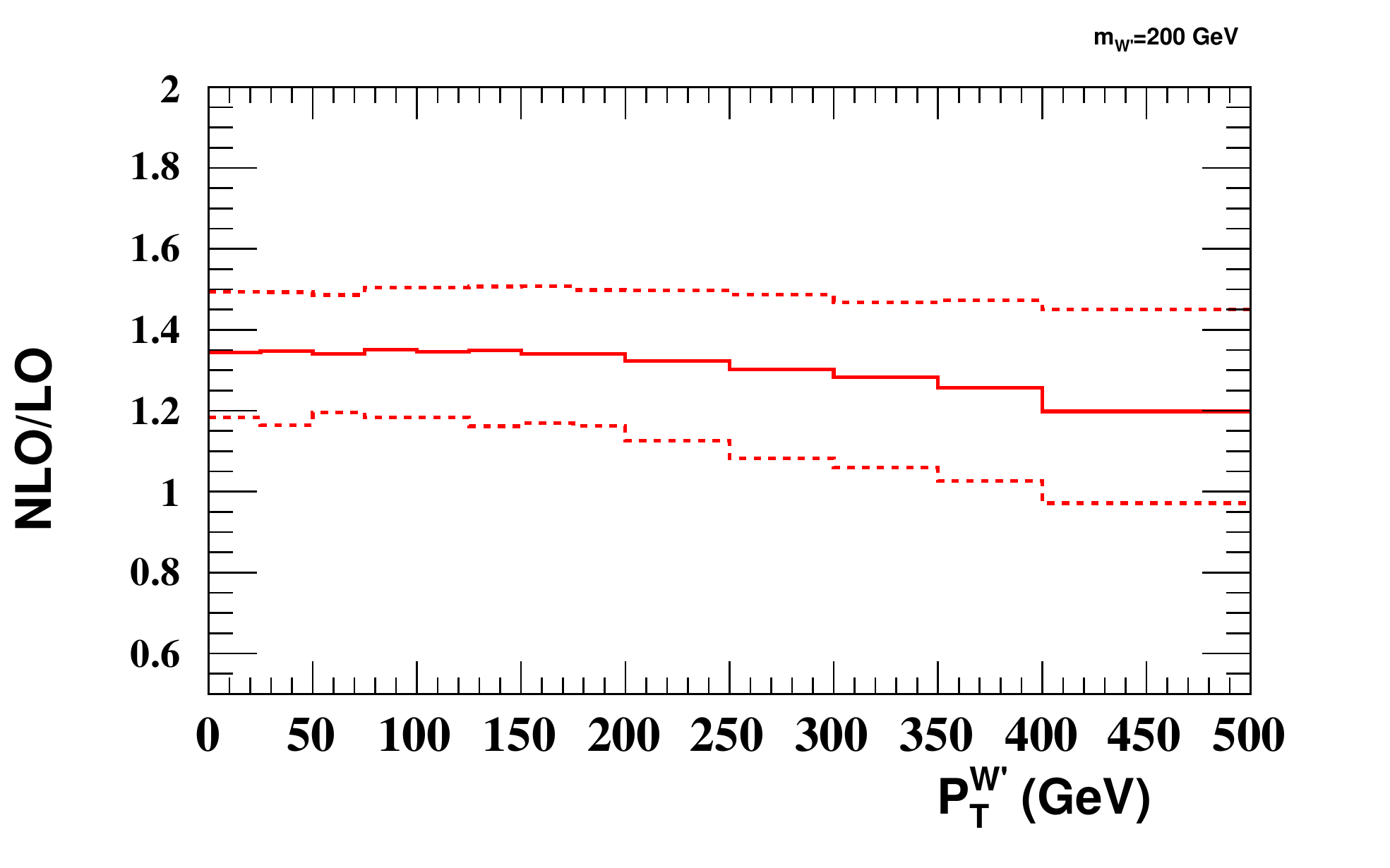}} \scalebox{0.4}{\includegraphics{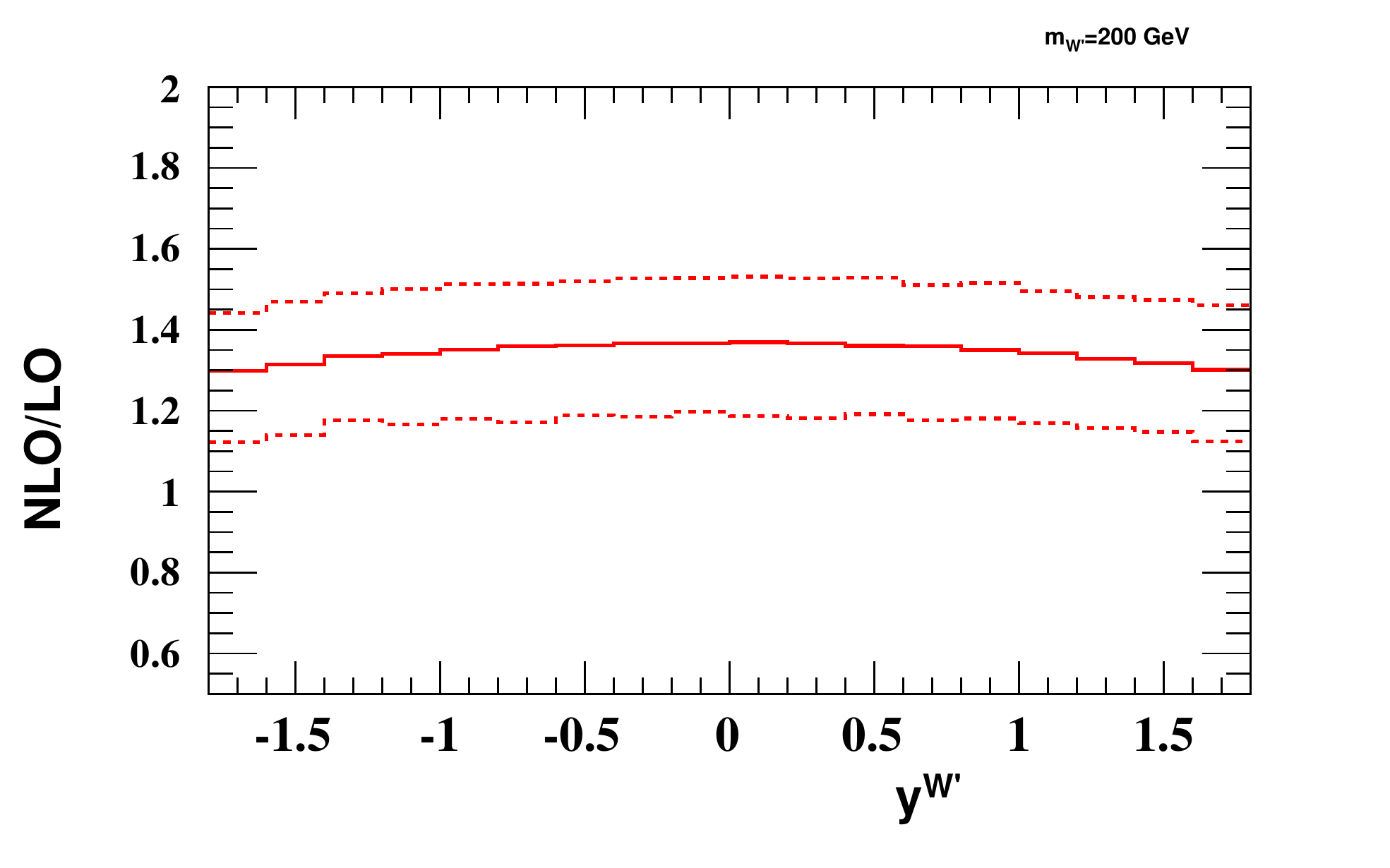}}\\
\scalebox{0.4}{\includegraphics{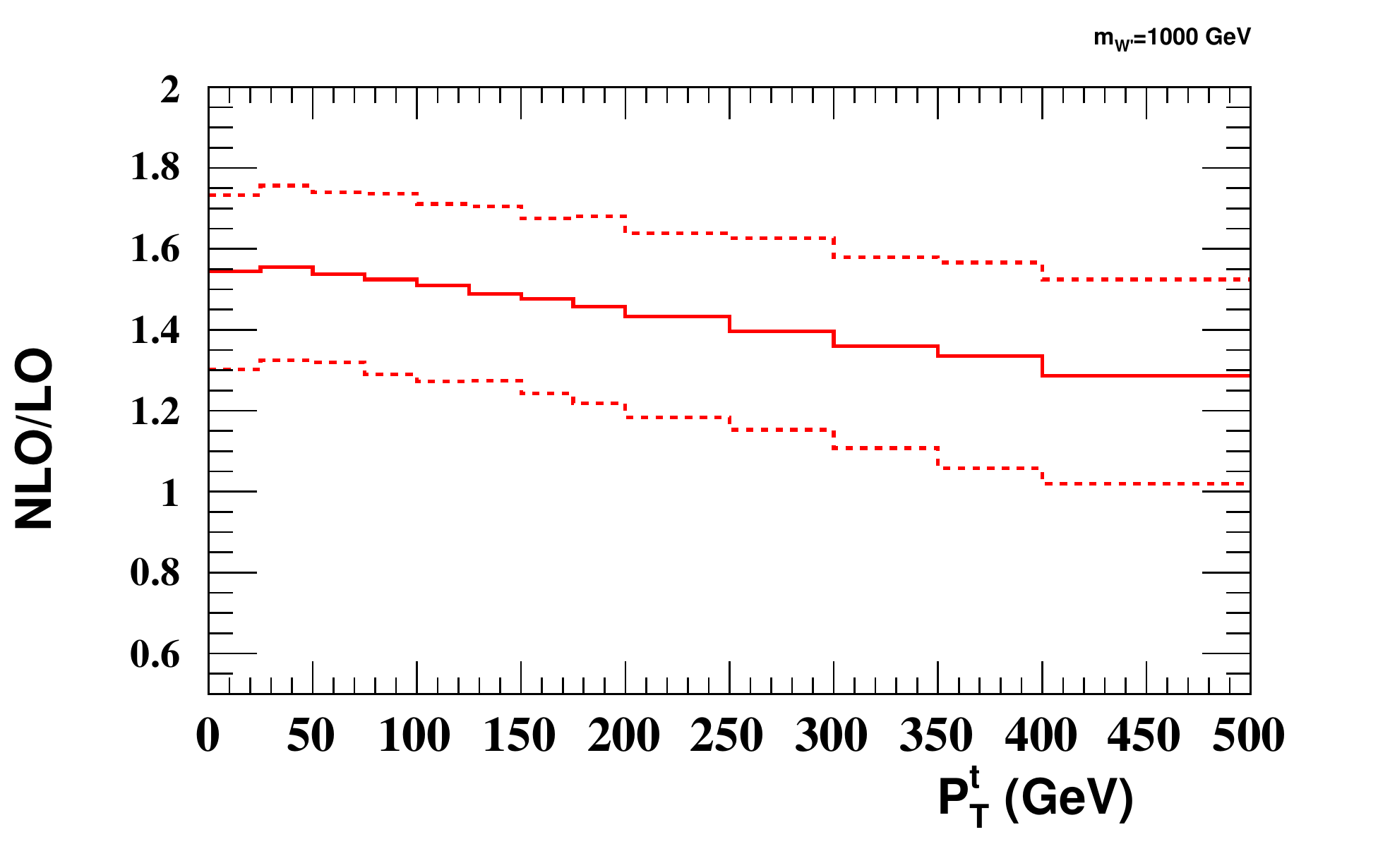}} \scalebox{0.4}{\includegraphics{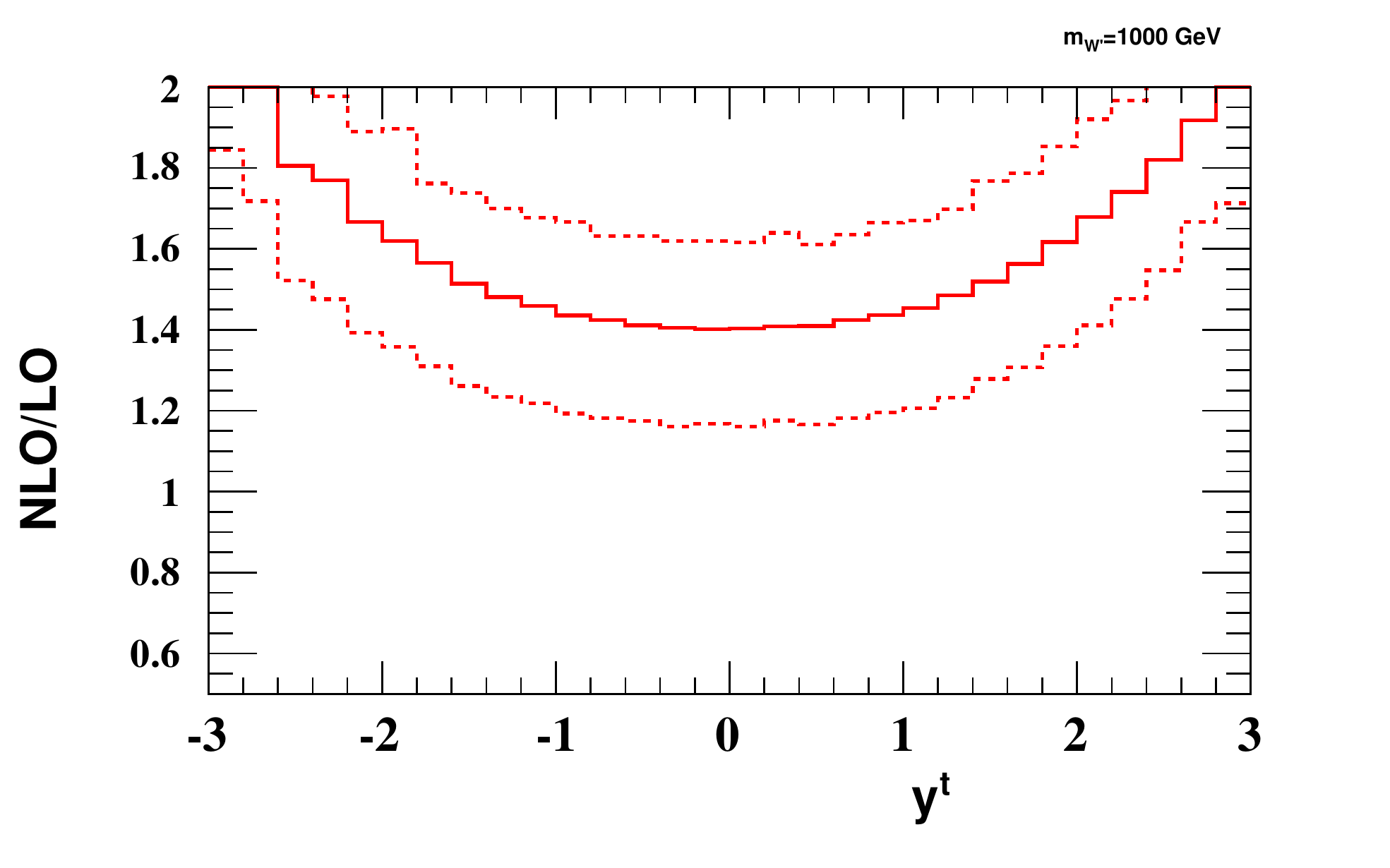}}\\
\scalebox{0.4}{\includegraphics{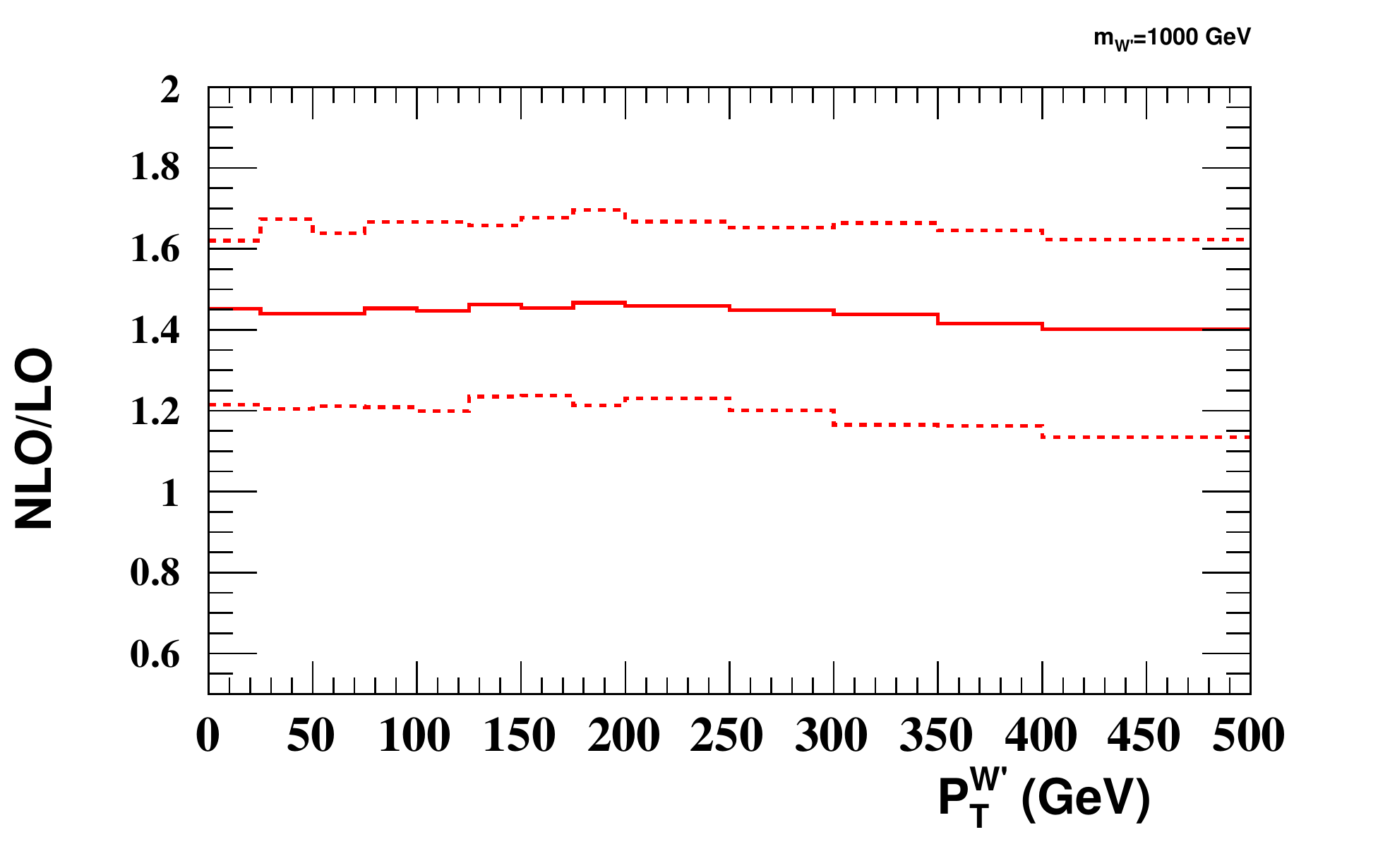}} \scalebox{0.4}{\includegraphics{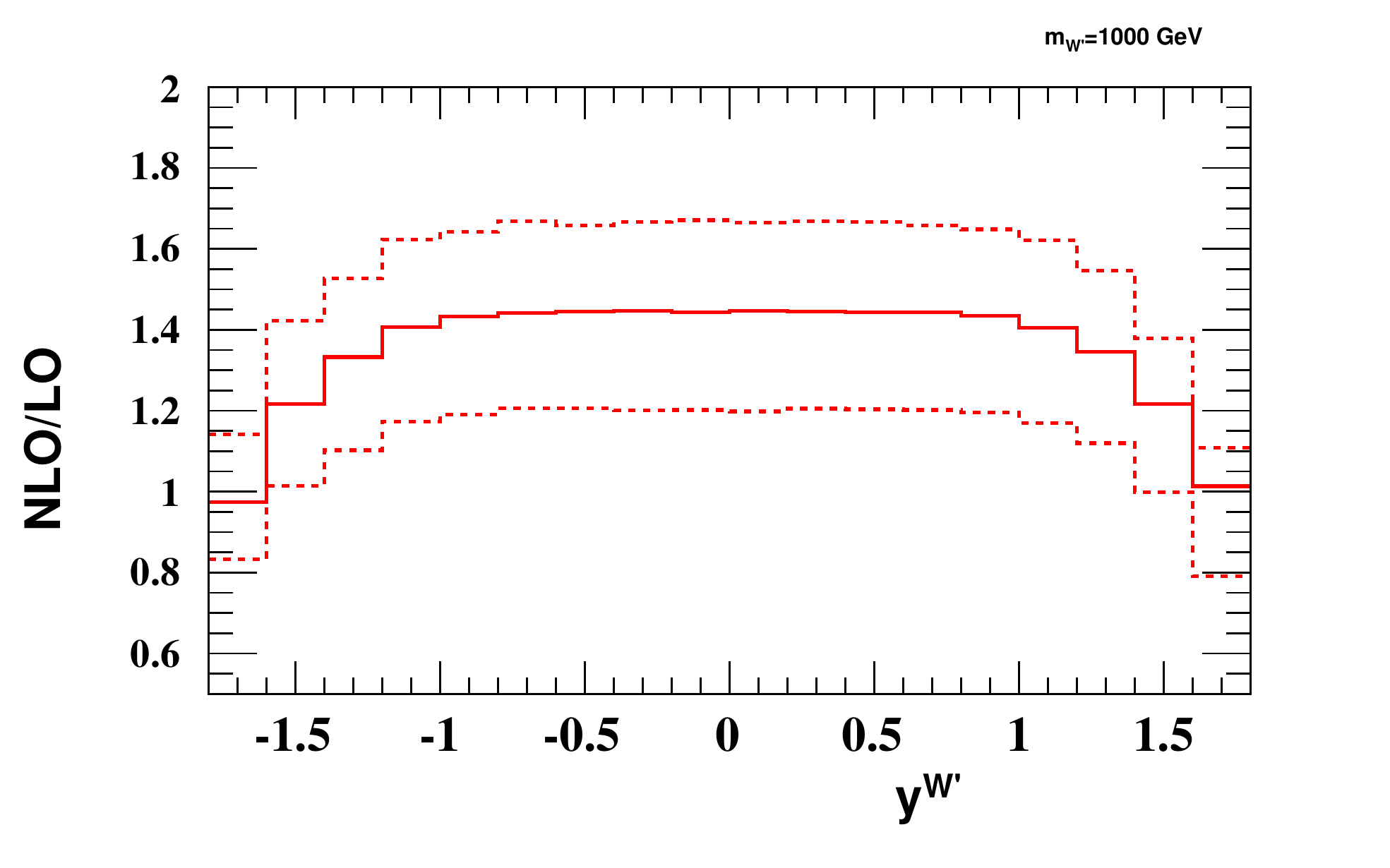}}\\
\caption{Ratio of NLO and LO predictions for the transverse-momentum and rapidity distributions for the top quark and
gauge boson in $tW'$ production for $W'$ masses of 200\,GeV and 1000\,GeV. The scale uncertainties are shown as dashed lines.}
\label{fig:wp200-kin-ratio}
\end{center}
\end{figure}

Analogous results for $tW'$ production are shown in 
figure~\ref{fig:wp200-kin}, again for gauge boson
masses of 200 GeV and 1000 GeV. Corresponding ratio plots are shown in
figure~\ref{fig:wp200-kin-ratio}. 
The results are broadly similar to the $Z'$
results apart from one qualitiative feature: the lack of a double peak in the
$W'$-boson rapidity distribution. This is due to the fact that the initial state 
contains a down rather than an up quark, which carries proportionally less
momentum, thus weakening the boost of the final state. A reasonable effect 
persists however, as can be seen by the fact that the $W'$ rapidity 
distribution is noticeably wider than that of the top quark. Again, real 
emission corrections are dominated by initial state radiation, which is born
out by the rapidity distribution of the extra parton in the lower panels
of figure~\ref{fig:extraparton}. There is again a difference in how rapidity 
distributions change at NLO in going from lower to higher gauge-boson masses,
due to the fact that final-state radiation from the top is harder on average
due to its recoiling against a highly massive boson.\\

In this and the previous section, we have examined the impact of NLO 
cross sections on total rates and distributions. A useful preliminary 
application of these results is to examine their impact on current searches
for top-associated $W'$ and $Z'$ production. This is the subject of the 
following section.

\section{Implications for $\mathbf{tW'}$ and $\mathbf{tZ'}$ searches}
\label{sec:imps}

In this section we discuss the implications of our results for experimental
searches. As a specific example, we consider the impact on results from a recent ATLAS search~\cite{ATLASCONF,ATLASPAPER} for $tW'$ with $W' \rightarrow td$, in order to illustrate the impact of the results in this paper. \\

Experimental searches for  $tZ'$ or $tW'$ production make use of LO Monte Carlo simulations, typically with events from the {\sc Madgraph}~\cite{Alwall:2011uj} matrix element generator. We note that the latest release (2.6) of {\tt Herwig++}~\cite{Arnold:2012fq} also implements relevant models for $tW'$ and $tZ'$.
 The cross-sections for the new-physics processes and kinematics of the top quark and associated heavy boson are generally derived from these LO codes. Correction of the cross section to the NLO values is straightforward using the results from Section \ref{sec:total} (see also the supplementary material for different coupling choices). Thus experimental upper limits on cross-sections correspond to stronger limits on the allowed couplings. Furthermore the reduced scale dependence at  NLO means that a limit on the coupling that fully accounts for the uncertainty on the cross section would be substantially improved. 
The kinematics of the top and associated heavy boson can also be corrected to account for the differing distributions at NLO. We stress that, since at NLO the top and heavy boson are produced more centrally, the acceptance would be higher for events produced at NLO than LO. It is thus the case that using LO samples to derive acceptance and set upper limits on cross sections is a conservative approach.\\

The charge asymmetry in the production rate of the top quarks from heavy boson decays can also be exploited in a search. The SM $t\bar{t}+$jet background is charge symmetric, so choosing to look at the anti-top+jet rather than top +jet mass spectrum can reduce the background by a factor of two while preserving the majority of signal. This is the case in the recent ATLAS search~\cite{ATLASPAPER} where windows in the top+jet mass vs anti-top+jet mass plane that give optimal sensitivity to signal are chosen.  We showed in Section~\ref{sec:total} that this asymmetry is quite stable in moving from  LO to NLO and have provided the necessary information to account for this effect. \\

\begin{figure}
\begin{center}
\scalebox{0.7}{\includegraphics{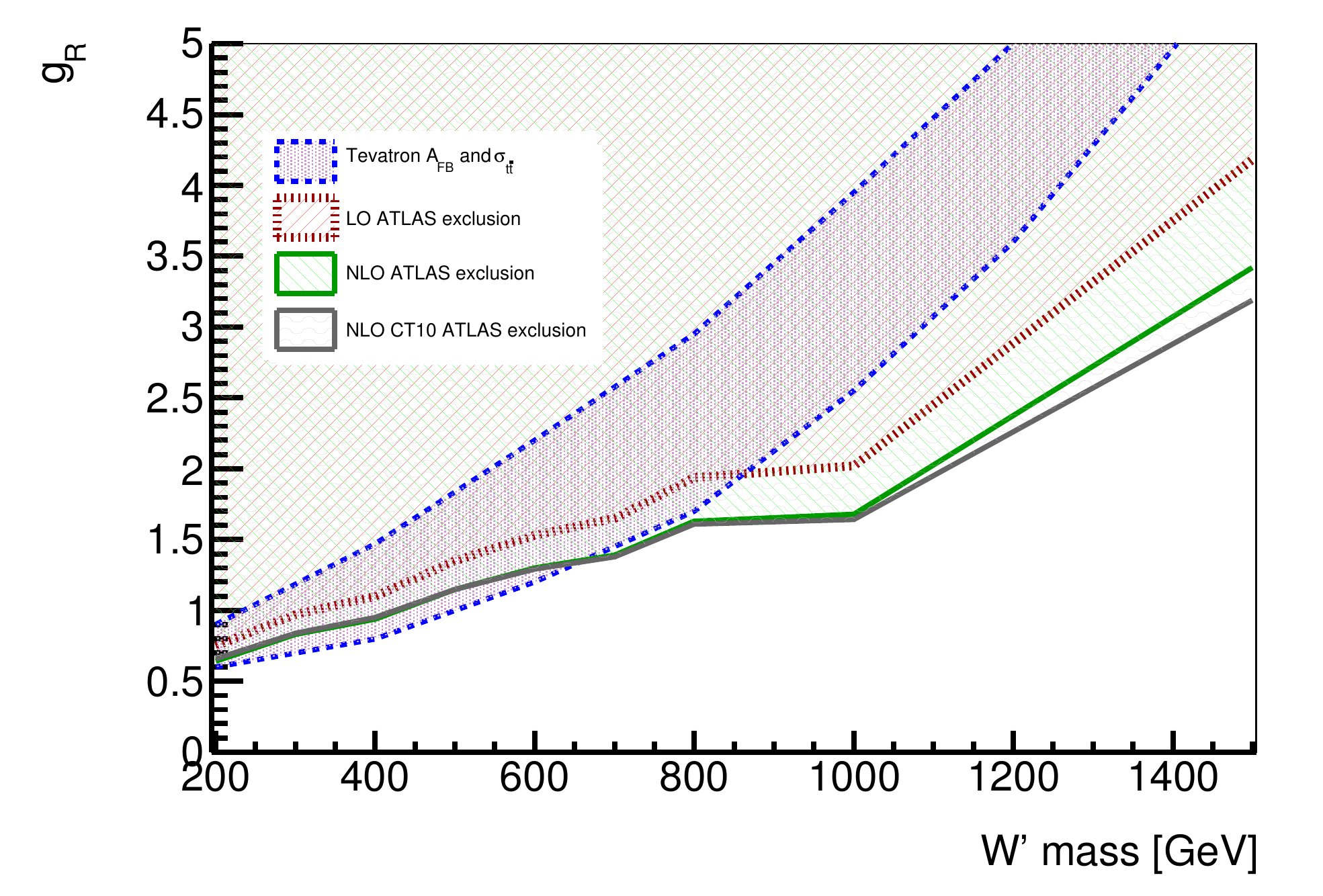}}
\end{center}
\caption{Example of the strengthening of constraints on the coupling $g_R$ when taking the NLO corrections to $tW'$ production into account. The LO exclusion curve is made using the calculations with the MSTW2008 LO  PDF set. The NLO curves are made using the calculations with the MSTW2008 LO and CT10 PDF sets. \label{fig:exclusion}}
\end{figure}

In order to demonstrate the impact of our NLO corrections, we first show in figure~\ref{fig:exclusion} a series of exclusion limits in the $(g_R,M_{W'})$ plane. The LO and NLO figures are obtained by taking the cross-section limit from the ATLAS search~\cite{ATLASCONF,ATLASPAPER}, and using our LO and NLO calculations respectively, with scale choices and other parameters as defined in section~\ref{sec:total}. Also shown is the region favoured by the Tevatron forward-backward asymmetry measurements~\cite{Aaltonen:2011kc,Abazov:2011rq}, here taken from~\cite{ATLASPAPER}. We see that NLO corrections significantly increase the constraints in parameter space, effectively cutting off most of the region that is still favoured by the Tevatron measurements. Whilst $W'$ models do not have to explain the forward-backward asymmetry, this was a significant motivation for considering them in the first place. Thus, the fact that only a sliver of asymmetry-favoured space remains is an important result, justifying the calculation of NLO corrections.\\

A similar conclusion has been reached by the ATLAS collaboration itself, who have used the results of the present paper in their most up-to-date analysis~\cite{ATLASPAPER}. Note that our NLO exclusion limits differ slightly from theirs, due to the fact that it has been calculated independently, using estimates of the cross-section limits presented in~\cite{ATLASCONF,ATLASPAPER}. Also, we have here used a default common factorisation and renormalisation scale of $(m_t+m_X)/2$, as opposed to the fixed scale of 200 GeV used by ATLAS for their LO limit~\cite{ATLASCONF}. We do not see a motivation for the latter scale for high gauge boson masses. Nevertheless, the results are broadly similar to those of~\cite{ATLASPAPER}. \\

The CMS collaboration have also presented bounds on $W'$ production~\cite{CMSWtd}. However, unlike the ATLAS analysis, this bound is on the combination of the resonant $W't$ process, and the $t$-channel exchange process. One could take this to be a conservative estimate on the resonant process, but this would give a much weaker bound on the $W't$ cross-section than would be obtained in practice. For this reason, we do not show the CMS result in figure~\ref{fig:exclusion}, so as not to introduce an unfair comparison. Another reason for not including this is a recent study~\cite{Endo:2012mi} showing that interference with SM $t\bar{t}+$ jet production is important for the CMS search, and significantly weakens the limit. The effect of such interference  is negligible for searches that focus on the resonant $W't$ process.\\

Note that our analysis in this section is only an approximate estimate of the exclusion limits implied by our NLO calculation. Firstly, and as stated above, we have not taken into account explicitly that the theoretical uncertainty of the NLO cross-section (due predominantly to scale variation) is significantly reduced relative to LO. Secondly, we have not accounted for the change in kinematic distributions, which would affect the acceptance. There is also an important interplay between these direct searches and additional constraints such as those discussed in the introduction.
A more realistic analysis can be found in the ATLAS paper~\cite{ATLASPAPER}, but we think it still worth presenting an independent exclusion plot here. Firstly, it is interesting to note that a rough estimate gives results which are reasonably similar to the ATLAS analysis. Secondly, we use a more physically motivated scale choice, which can be much different to the fixed LO scale of 200 GeV used by~\cite{Gresham:2011dg} and the ATLAS analysis, at high gauge boson masses. It is reassuring to see that this does not lead to a large deviation from their results.\\

In this section, we have illustrated the impact of our NLO results in the well-studied case (both theoretically and experimentally) in which the top couples to the first generation quarks only. It is worth stressing that our NLO calculation would also be useful in scenarios with alternative coupling choices, where their impact may be larger. 



\section{Conclusion}
\label{sec:conclusion}

In this paper, we have calculated NLO QCD corrections to the associated
production of $W'$ and $Z'$ boson production with a top quark. This production
mode is being actively searched for by both the ATLAS and CMS collaborations,
and higher-order corrections are important in order to set tight bounds on
couplings and gauge boson masses, due to both the size of NLO effects, and the
requisite reduction in theoretical uncertainty. As we have seen, for models in
which the top quark couples to first generation quarks, the uncertainty is
dominated by scale variation, whereas the PDF uncertainty is small, due to
the fact that high-$x$ valence quark distributions are well-constrained in 
global fits. One thus achieves a significant decrease in theoretical 
uncertainty in going to NLO, as can be seen directly in the total cross-section
plots of figures~\ref{fig:zp} and~\ref{fig:wp}.\\

Although we have focussed on the particular Lagrangian of 
eq.~(\ref{couplings}), our analysis is more general in that one can 
straightforwardly combine the results collected in the main body 
of this paper and the supplementary material in order to generate total cross sections for models
which have non-trivial mixing between quark generations or a mixture of
left- and right-handed couplings. 
Thus, our results can be directly and 
generally applied in current experimental analyses, which may differ somewhat
in the particular models which are being considered. Our aim has been to 
provide as complete a set of numbers as possible for use in experimental
searches, including cross sections at a centre-of-mass energy of 8\,TeV,
which can be found in the supplementary material. Results for alternative
parameter choices, if desired, may be requested by contacting the authors.\\

As well as total cross-sections, we also considered transverse momentum and
rapidity distributions of the top quark and gauge bosons. We find that NLO
corrections lead, predictably, to a softening of transverse momentum 
distributions, which effect increases with the gauge boson mass. Rapidity
distributions are more affected even at low gauge boson masses, due to the fact
that real emission corrections are dominated by initial state radiation. The
top quark becomes more (less) central at low (high) gauge boson masses. \\

In section~\ref{sec:imps}, we briefly examined the implications of our results
for current searches, and estimated how coupling and mass limits would change
in the recently presented ATLAS analysis of~\cite{ATLASPAPER}. We find
that coupling bounds are strengthened by around 20\%, and that the region favoured by the Tevatron forward-backward asymmetry is all but excluded. The latter is a significant result, given that this asymmetry is a major motivation for considering $W'$ and $Z'$ models in the first place.
Our analysis is an approximate estimate, and could be enhanced by taking into account the fact that the theoretical uncertainty diminishes when NLO corrections are implemented, and that kinematic distributions change (affecting the acceptance of signal events). \\

To summarise, we have provided NLO QCD corrections for use in experimental 
searches for $Z'$ and $W'$ bosons in association with a top quark. The 
hunt for such particles is ongoing.

\section*{Acknowledgements} 

This work was supported in part by grant DE-FG02-92ER40704 from the U.S. Dept. 
of Energy. We thank Daniel Whiteson for discussions and comments on the manuscript and Steven Worm for discussions regarding the CMS limits. JA thanks Moira I. Gresham, Ian-Woo Kim and Kathryn M.  Zurek for useful 
discussions about the $tW'$ and $tZ'$ models. 
JF is supported by the UK Science and Technology Facilities Council (STFC). 
CDW thanks Stefano Frixione for useful conversations, and also David Miller for
discussions. He is supported by the STFC Postdoctoral Fellowship ``Collider 
Physics at the LHC''.

\appendix
\clearpage
\section{Supplementary material}
\label{sec:supp}
In this supplementary material, we collect various additional results from our NLO $W'$ and 
$Z'$ calculation, which are of further use for current (and future) 
experimental analyses. As in the main paper, we adopt the Lagrangians
\begin{align}
{\cal L}_{W'}=\frac{P_{tD}}{\sqrt{2}}\bar{D}\gamma^\mu g_RP_Rt
W'+{\rm h.c.};\notag\\
{\cal L}_{Z'}=\frac{Q_{tU}}{\sqrt{2}}\bar{U}\gamma^\mu g_RP_Rt
Z'+{\rm h.c.},
\label{couplings}
\end{align}
where $P_{tD}$ and $Q_{tU}$ are suitable mixing matrices. The main paper 
focuses on the cases $P_{td}=Q_{tU}=1$ (all other $P_{ij}=Q_{ij}=0$). Here we
consider other possible coupling choices, and also how to combine these in 
order to obtain total NLO cross-sections in general models. For completeness,
we also give results for $t$ and $\bar{t}$ production separately, beginning
in the following section.

\subsection{Cross-section results at 7 TeV separated by top charge}
\label{app:charge}

This section contains the total cross sections for $t$ and $\bar{t}$ production separately. These are for cases where the $Z'$ ($W'$) couples to $tu$ ($td$). For couplings to second and third generation quarks, no charge asymmetry is expected due to the equality $q(x)=\bar{q}(x)$ for sea quark distributions\footnote{Modern PDF analyses typically allow $s(x)\neq\bar{s}(x)$~\cite{Olness:2003wz,Lai:2007dq,Martin:2009iq}. However, this would result in a negligible charge asymmetry.}.\\

Tables~\ref{tab:lo-zp-plus} and \ref{tab:lo-zp-minus} give the cross sections at LO for $tZ'$ production where the $t$ is specifically a top quark and anti-top quark respectively. The NLO cross sections for the same process are given
in Tables~\ref{tab:nlo-zp-plus} and \ref{tab:nlo-zp-minus}. The equivalent cross sections for $tW'$ production are given in Tables \ref{tab:lo-wp-plus}-\ref{tab:nlo-wp-minus}.

\begin{table}
\begin{center}
\begin{tabular}{|c||r@{.}ll|r@{.}lll|}
\hline
$M(Z')$\,(GeV) & \multicolumn{3}{|c|}{$\sigma^{\mathrm{CTEQ6L1}}_{\mathrm{born}} (\bar{t}Z')$\,(pb)}& \multicolumn{4}{|c|}{$\sigma^{\mathrm{MSTW\ 2008\ LO}}_{\mathrm{born}} (\bar{t}Z')$\,(pb)}\\ 
\hline 
200 & 6&51 & $^{+1.42}_{-1.08}$& 8&04 & $^{+1.85}_{-1.40}$ & $^{+0.29}_{-0.30}$\\ 
300 & 1&89 & $^{+0.44}_{-0.33}$& 2&41 & $^{+0.59}_{-0.44}$ & $^{+0.10}_{-0.10}$\\ 
400 & 0&66 & $^{+0.16}_{-0.12}$& 0&87 & $^{+0.23}_{-0.17}$ & $^{+0.04}_{-0.04}$\\ 
500 & 0&262 & $^{+0.066}_{-0.049}$& 0&353 & $^{+0.096}_{-0.070}$ & $^{+0.019}_{-0.019}$\\ 
600 & 0&113 & $^{+0.029}_{-0.022}$& 0&156 & $^{+0.044}_{-0.032}$ & $^{+0.010}_{-0.010}$\\ 
700 & 0&052 & $^{+0.014}_{-0.010}$& 0&073 & $^{+0.021}_{-0.015}$ & $^{+0.005}_{-0.005}$\\ 
800 & 0&0251 & $^{+0.0069}_{-0.0050}$& 0&0363 & $^{+0.0109}_{-0.0077}$ & $^{+0.0028}_{-0.0028}$\\
\hline 
\end{tabular} 

\end{center}
\caption{Leading order cross sections for $pp \rightarrow \bar{t}Z'$ at the 7 TeV LHC. \label{tab:lo-zp-plus}}
\end{table}

\begin{table}
\begin{center}
\begin{tabular}{|c||r@{.}ll|r@{.}lll|}
\hline
$M(Z')$\,(GeV) & \multicolumn{3}{|c|}{$\sigma^{\mathrm{CTEQ6L1}}_{\mathrm{born}} (tZ')$\,(pb)}& \multicolumn{4}{|c|}{$\sigma^{\mathrm{MSTW\ 2008\ LO}}_{\mathrm{born}} (tZ')$\,(pb)}\\ 
\hline 
200 & 71&3 & $^{+14.9}_{-11.4}$& 71&2 & $^{+15.5}_{-11.8}$ & $^{+ 0.9}_{- 1.0}$\\ 
300 & 24&74 & $^{+ 5.49}_{- 4.17}$& 24&85 & $^{+ 5.78}_{- 4.34}$ & $^{+ 0.34}_{- 0.35}$\\ 
400 & 10&14 & $^{+ 2.36}_{- 1.77}$& 10&27 & $^{+ 2.51}_{- 1.87}$ & $^{+ 0.16}_{- 0.15}$\\ 
500 & 4&62 & $^{+1.12}_{-0.83}$& 4&73 & $^{+1.21}_{-0.89}$ & $^{+0.08}_{-0.08}$\\ 
600 & 2&27 & $^{+0.57}_{-0.42}$& 2&35 & $^{+0.62}_{-0.45}$ & $^{+0.04}_{-0.04}$\\ 
700 & 1&181 & $^{+0.302}_{-0.223}$& 1&232 & $^{+0.336}_{-0.245}$ & $^{+0.026}_{-0.024}$\\ 
800 & 0&641 & $^{+0.168}_{-0.123}$& 0&676 & $^{+0.190}_{-0.137}$ & $^{+0.016}_{-0.015}$\\
\hline 
\end{tabular} 

\end{center}
\caption{Leading order cross sections for $pp \rightarrow tZ'$ at the 7 TeV LHC. \label{tab:lo-zp-minus}}
\end{table}

\begin{table}
\begin{center}
\begin{tabular}{|c||r@{.}lll|r@{.}lll|}
\hline
$M(Z')$\,(GeV) & \multicolumn{4}{|c|}{$\sigma^{\mathrm{CT10}}_{\mathrm{nlo}} (\bar{t}Z')$\,(pb)}& \multicolumn{4}{|c|}{$\sigma^{\mathrm{MSTW\ 2008\ NLO}}_{\mathrm{nlo}} (\bar{t}Z')$\,(pb)}\\ 
\hline 
200 & 9&17 & $^{+0.60}_{-0.66}$ & $^{+0.80}_{-0.89}$& 10&31 & $^{+ 0.70}_{- 0.77}$ & $^{+ 0.35}_{- 0.32}$\\ 
300 & 2&74 & $^{+0.18}_{-0.21}$ & $^{+0.29}_{-0.31}$& 3&09 & $^{+0.21}_{-0.24}$ & $^{+0.13}_{-0.12}$\\ 
400 & 0&99 & $^{+0.07}_{-0.08}$ & $^{+0.13}_{-0.13}$& 1&12 & $^{+0.08}_{-0.09}$ & $^{+0.06}_{-0.05}$\\ 
500 & 0&402 & $^{+0.028}_{-0.033}$ & $^{+0.061}_{-0.057}$& 0&457 & $^{+0.033}_{-0.039}$ & $^{+0.027}_{-0.025}$\\ 
600 & 0&178 & $^{+0.013}_{-0.015}$ & $^{+0.032}_{-0.028}$& 0&203 & $^{+0.016}_{-0.018}$ & $^{+0.014}_{-0.013}$\\ 
700 & 0&085 & $^{+0.007}_{-0.008}$ & $^{+0.018}_{-0.015}$& 0&096 & $^{+0.008}_{-0.009}$ & $^{+0.008}_{-0.007}$\\ 
800 & 0&0422 & $^{+0.0034}_{-0.0040}$ & $^{+0.0100}_{-0.0082}$& 0&0479 & $^{+0.0040}_{-0.0046}$ & $^{+0.0042}_{-0.0039}$\\
\hline 
\end{tabular} 

\end{center}
\caption{Next-to-leading order cross sections for $pp \rightarrow \bar{t}Z'$ at the 7 TeV LHC. \label{tab:nlo-zp-plus}}
\end{table}

\begin{table}
\begin{center}
\begin{tabular}{|c||r@{.}lll|r@{.}lll|}
\hline
$M(Z')$\,(GeV) & \multicolumn{4}{|c|}{$\sigma^{\mathrm{CT10}}_{\mathrm{nlo}} (tZ')$\,(pb)}& \multicolumn{4}{|c|}{$\sigma^{\mathrm{MSTW\ 2008\ NLO}}_{\mathrm{nlo}} (tZ')$\,(pb)}\\ 
\hline 
200 & 92&0 & $^{+ 6.1}_{- 6.5}$ & $^{+ 2.4}_{- 2.6}$& 93&1 & $^{+ 6.4}_{- 6.7}$ & $^{+ 1.1}_{- 0.7}$\\ 
300 & 32&55 & $^{+ 2.13}_{- 2.38}$ & $^{+ 0.99}_{- 1.01}$& 32&97 & $^{+ 2.23}_{- 2.47}$ & $^{+ 0.37}_{- 0.26}$\\ 
400 & 13&60 & $^{+ 0.90}_{- 1.04}$ & $^{+ 0.51}_{- 0.49}$& 13&78 & $^{+ 0.94}_{- 1.07}$ & $^{+ 0.17}_{- 0.13}$\\ 
500 & 6&33 & $^{+0.43}_{-0.50}$ & $^{+0.29}_{-0.27}$& 6&40 & $^{+0.45}_{-0.52}$ & $^{+0.09}_{-0.07}$\\ 
600 & 3&17 & $^{+0.22}_{-0.26}$ & $^{+0.18}_{-0.16}$& 3&21 & $^{+0.23}_{-0.27}$ & $^{+0.06}_{-0.05}$\\ 
700 & 1&682 & $^{+0.122}_{-0.145}$ & $^{+0.111}_{-0.100}$& 1&698 & $^{+0.127}_{-0.149}$ & $^{+0.034}_{-0.029}$\\ 
800 & 0&932 & $^{+0.070}_{-0.083}$ & $^{+0.072}_{-0.064}$& 0&938 & $^{+0.073}_{-0.085}$ & $^{+0.022}_{-0.019}$\\
\hline 
\end{tabular} 

\end{center}
\caption{Next-to-leading order cross sections for $pp \rightarrow tZ'$ at the 7 TeV LHC. \label{tab:nlo-zp-minus}}
\end{table}

\begin{table}
\begin{center}
\begin{tabular}{|c||r@{.}ll|r@{.}lll|}
\hline
$M(W'^{+})$\,(GeV) & \multicolumn{3}{|c|}{$\sigma^{\mathrm{CTEQ6L1}}_{\mathrm{born}} (\bar{t}W'^{+})$\,(pb)}& \multicolumn{4}{|c|}{$\sigma^{\mathrm{MSTW\ 2008\ LO}}_{\mathrm{born}} (\bar{t}W'^{+})$\,(pb)}\\ 
\hline 
200 & 8&40 & $^{+1.88}_{-1.42}$& 9&58 & $^{+2.27}_{-1.70}$ & $^{+0.39}_{-0.33}$\\ 
300 & 2&49 & $^{+0.59}_{-0.44}$& 2&88 & $^{+0.73}_{-0.54}$ & $^{+0.13}_{-0.11}$\\ 
400 & 0&88 & $^{+0.22}_{-0.16}$& 1&03 & $^{+0.28}_{-0.20}$ & $^{+0.06}_{-0.05}$\\ 
500 & 0&350 & $^{+0.091}_{-0.067}$& 0&414 & $^{+0.116}_{-0.084}$ & $^{+0.025}_{-0.020}$\\ 
600 & 0&151 & $^{+0.040}_{-0.030}$& 0&180 & $^{+0.053}_{-0.038}$ & $^{+0.012}_{-0.010}$\\ 
700 & 0&0692 & $^{+0.019}_{-0.014}$& 0&083 & $^{+0.025}_{-0.018}$ & $^{+0.006}_{-0.005}$\\ 
800 & 0&0333 & $^{+0.0094}_{-0.0068}$& 0&0403 & $^{+0.0126}_{-0.0089}$ & $^{+0.0034}_{-0.0027}$\\
\hline 
\end{tabular} 

\end{center}
\caption{Leading order cross sections for $pp \rightarrow \bar{t}W'^{+}$ at the 7 TeV LHC. \label{tab:lo-wp-plus}}
\end{table}

\begin{table}
\begin{center}
\begin{tabular}{|c||r@{.}ll|r@{.}lll|}
\hline
$M(W'^{-})$\,(GeV) & \multicolumn{3}{|c|}{$\sigma^{\mathrm{CTEQ6L1}}_{\mathrm{born}} (tW'^{-})$\,(pb)}& \multicolumn{4}{|c|}{$\sigma^{\mathrm{MSTW\ 2008\ LO}}_{\mathrm{born}} (tW'^{-})$\,(pb)}\\ 
\hline 
200 & 33&4 & $^{+ 7.3}_{- 5.5}$& 33&1 & $^{+ 7.6}_{- 5.7}$ & $^{+ 0.6}_{- 0.6}$\\ 
300 & 11&11 & $^{+ 2.56}_{- 1.93}$& 11&04 & $^{+ 2.68}_{- 2.00}$ & $^{+ 0.22}_{- 0.21}$\\ 
400 & 4&381 & $^{+1.06}_{-0.79}$& 4&38 & $^{+1.12}_{-0.82}$ & $^{+0.10}_{-0.09}$\\ 
500 & 1&93 & $^{+0.48}_{-0.36}$& 1&94 & $^{+0.52}_{-0.38}$ & $^{+0.05}_{-0.04}$\\ 
600 & 0&916 & $^{+0.236}_{-0.174}$& 0&928 & $^{+0.256}_{-0.186}$ & $^{+0.025}_{-0.023}$\\ 
700 & 0&461 & $^{+0.122}_{-0.089}$& 0&470 & $^{+0.134}_{-0.097}$ & $^{+0.014}_{-0.013}$\\ 
800 & 0&243 & $^{+0.066}_{-0.048}$& 0&249 & $^{+0.073}_{-0.052}$ & $^{+0.008}_{-0.007}$\\
\hline 
\end{tabular} 

\end{center}
\caption{Leading order cross sections for $pp \rightarrow tW'^{-}$ at the 7 TeV LHC. \label{tab:lo-wp-minus}}
\end{table}

\begin{table}
\begin{center}
\begin{tabular}{|c||r@{.}lll|r@{.}lll|}
\hline
$M(W'^{+})$\,(GeV) & \multicolumn{4}{|c|}{$\sigma^{\mathrm{CT10}}_{\mathrm{nlo}} (\bar{t}W'^{+})$\,(pb)}& \multicolumn{4}{|c|}{$\sigma^{\mathrm{MSTW\ 2008\ NLO}}_{\mathrm{nlo}} (\bar{t}W'^{+})$\,(pb)}\\ 
\hline 
200 & 11&0 & $^{+ 0.7}_{- 0.8}$ & $^{+ 1.0}_{- 1.3}$& 12&0 & $^{+ 0.8}_{- 0.9}$ & $^{+ 0.4}_{- 0.4}$\\ 
300 & 3&30 & $^{+0.22}_{-0.26}$ & $^{+0.37}_{-0.44}$& 3&59 & $^{+0.25}_{-0.29}$ & $^{+0.15}_{-0.15}$\\ 
400 & 1&19 & $^{+0.08}_{-0.10}$ & $^{+0.16}_{-0.18}$& 1&29 & $^{+0.09}_{-0.11}$ & $^{+0.06}_{-0.06}$\\ 
500 & 0&483 & $^{+0.035}_{-0.041}$ & $^{+0.076}_{-0.084}$& 0&516 & $^{+0.039}_{-0.046}$ & $^{+0.029}_{-0.029}$\\ 
600 & 0&213 & $^{+0.016}_{-0.019}$ & $^{+0.039}_{-0.041}$& 0&224 & $^{+0.018}_{-0.021}$ & $^{+0.014}_{-0.015}$\\ 
700 & 0&100 & $^{+0.008}_{-0.009}$ & $^{+0.021}_{-0.022}$& 0&103 & $^{+0.009}_{-0.010}$ & $^{+0.008}_{-0.008}$\\ 
800 & 0&049 & $^{+0.004}_{-0.005}$ & $^{+0.012}_{-0.012}$& 0&050 & $^{+0.004}_{-0.005}$ & $^{+0.004}_{-0.004}$\\
\hline 
\end{tabular} 

\end{center}
\caption{Next-to-leading order cross sections for $pp \rightarrow \bar{t}W'^{+}$ at the 7 TeV LHC. \label{tab:nlo-wp-plus}}
\end{table}

\begin{table}
\begin{center}
\begin{tabular}{|c||r@{.}lll|r@{.}lll|}
\hline
$M(W'^{-})$\,(GeV) & \multicolumn{4}{|c|}{$\sigma^{\mathrm{CT10}}_{\mathrm{nlo}} (tW'^{-})$\,(pb)}& \multicolumn{4}{|c|}{$\sigma^{\mathrm{MSTW\ 2008\ NLO}}_{\mathrm{nlo}} (tW'^{-})$\,(pb)}\\ 
\hline 
200 & 44&6 & $^{+ 3.0}_{- 3.2}$ & $^{+ 2.1}_{- 2.0}$& 45&0 & $^{+ 3.1}_{- 3.4}$ & $^{+ 0.6}_{- 0.6}$\\ 
300 & 15&16 & $^{+ 1.00}_{- 1.14}$ & $^{+ 0.81}_{- 0.77}$& 15&22 & $^{+ 1.05}_{- 1.18}$ & $^{+ 0.23}_{- 0.23}$\\ 
400 & 6&12 & $^{+0.41}_{-0.48}$ & $^{+0.38}_{-0.35}$& 6&10 & $^{+0.43}_{-0.49}$ & $^{+0.10}_{-0.11}$\\ 
500 & 2&76 & $^{+0.19}_{-0.23}$ & $^{+0.20}_{-0.18}$& 2&73 & $^{+0.20}_{-0.23}$ & $^{+0.05}_{-0.06}$\\ 
600 & 1&341 & $^{+0.097}_{-0.115}$ & $^{+0.113}_{-0.099}$& 1&320 & $^{+0.099}_{-0.116}$ & $^{+0.030}_{-0.031}$\\ 
700 & 0&692 & $^{+0.052}_{-0.061}$ & $^{+0.067}_{-0.058}$& 0&676 & $^{+0.053}_{-0.062}$ & $^{+0.018}_{-0.018}$\\ 
800 & 0&374 & $^{+0.029}_{-0.034}$ & $^{+0.041}_{-0.035}$& 0&361 & $^{+0.029}_{-0.034}$ & $^{+0.011}_{-0.011}$\\
\hline 
\end{tabular} 

\end{center}
\caption{Next-to-leading order cross sections for $pp \rightarrow tW'^{-}$ at the 7 TeV LHC. \label{tab:nlo-wp-minus}}
\end{table}

\subsection{Cross-section results for other partonic couplings}
\label{app:othercouplings}

In the main paper, we have explicitly analysed the Lagrangian of
eq~(\ref{couplings}), with the only non-zero couplings being those between
the third and first quark generations ($P_{td}=Q_{tu}=1$). 
In this appendix, we collect total cross-sections for alternative coupling 
choices, before discussing how these results can be combined in the general
case.\\

In tables~\ref{tab:lo-wp-t2}-\ref{tab:nlo-wp-tb} we present total LO and NLO 
cross-sections for $tW'$ production for the choices $P_{ts}=1$ and, separately,
$P_{tb}=1$ (all other $P_{tD}$ zero in each case). Similarly, the total 
cross-section for $tZ'$ production at LO and NLO is shown in 
tables~\ref{tab:lo-zp-t2} and~\ref{tab:nlo-zp-t2} for $Q_{tc}=1$ (all other 
$Q_{tU}$ zero). Some comments are in order regarding a couple of features noticable in the results. Firstly, the $K$-factor for both $W'$ and $Z'$ production decreases with increasing gauge boson mass for the MSTW parton choice. This is a reflection of the fact that the initial state involves only gluons and sea quark (or heavy flavour) distributions. As the final state gets heavier, these distributions are probed at higher $x$ values, where they fall off rapidly. This effect is most pronounced for $Z'$ production with a $tc$ coupling, where the $K$ factor becomes less than one at the highest gauge boson masses. Secondly, the $K$ factor results show much more variation when CTEQ6L1 and CT10 partons are used, including wider disagreements with the MSTW results. We put this down to the fact that CTEQ substantially revised their treatment of heavy flavour effects from CTEQ6.5 onwards~\cite{Tung:2006tb}. One may consequently question the validity of cross-section results obtained using CTEQ6L1 partons and involving initial state heavy quarks, particularly at high $x$ values (applicable to high gauge boson masses). Nevertheless, we include these here given that CTEQ6L1 partons remain in use by experimental collaborations. \\

In general, there may be non-zero values for all relevant $P_{tD}$ and 
$Q_{tU}$, and the total cross-section for such a model is not simply related 
to the sum of the cross-sections for the individual choices $P_{tD}=1$, 
$Q_{tU}=1$. This is due to the presence of partonic channels appearing in the
real emission corrections at NLO, such as $gg\rightarrow tW'\bar{D}$ and
$q\bar{q}\rightarrow tW'\bar{D}$ (where $\bar{D}$ is any down-type antiquark),
which involve a sum over possible decays of an off-shell intermediate antitop. 
However, we have explicitly checked that the contribution of such channels is
numerically extremely small (due principally to the damping effect of 
initial state sea quark and gluon distributions at large $x$). A very good
approximate prescription for obtaining the total cross-section
for any model is then as follows.\\

For $tW'$ production, let $\sigma^{tW'}_{tD}$ be the total cross-section
for $P_{tD}=1$ (all other $P_{tj}=0$), and $g_R=1$ in eq.~(\ref{couplings}).
The total cross-section in a model with arbitrary left- and right-handed 
couplings $g_L$ and $g_R$ is then given by
\begin{equation}
\sigma\simeq \left(|g_L|^2+|g_R|^2\right)\left[|P_{td}|^2\sigma^{tW'}_{td}
+|P_{ts}|^2\sigma^{tW'}_{ts}+|P_{tb}|^2\sigma^{tW'}_{tb}\right].
\label{totcoupl1}
\end{equation}
Similarly, for $tZ'$ production one has
\begin{equation}
\sigma\simeq \left(|g_L|^2+|g_R|^2\right)\left[|Q_{tu}|^2\sigma^{tZ'}_{tu}
+|Q_{tc}|^2\sigma^{tZ'}_{tc}\right],
\label{totcoupl2}
\end{equation}
where $\sigma^{tZ'}_{tU}$ is the total cross-section for $tZ'$ production with
$Q_{tU}=1$ (all other $Q_{tj}=0$). Note that eqs.~(\ref{totcoupl1}) 
and~(\ref{totcoupl2}) are exact at LO, and are broken at NLO only by the
additional partonic subchannels described above. This happens at the 
sub-percent level, and is certainly within the theoretical uncertainty as
described by scale and PDF variation. 

\begin{table}
\begin{center}
\begin{tabular}{|c||r@{.}ll|r@{.}lll|}
\hline
$M(W')$\,(GeV) & \multicolumn{3}{|c|}{$\sigma^{\mathrm{CTEQ6L1}}_{\mathrm{born}} (tW'^{-}+\bar{t}W^{+})$\,(pb)}& \multicolumn{4}{|c|}{$\sigma^{\mathrm{MSTW\ 2008\ LO}}_{\mathrm{born}} (tW'^{-}+\bar{t}W^{-})$\,(pb)}\\ 
\hline 
200 & 9&2 & $^{+1.9}_{-1.4}$& 10&3 & $^{+ 2.2}_{- 1.7}$ & $^{+ 0.6}_{- 0.6}$\\ 
300 & 2&63 & $^{+0.58}_{-0.44}$& 3&01 & $^{+0.70}_{-0.53}$ & $^{+0.21}_{-0.22}$\\ 
400 & 0&91 & $^{+0.21}_{-0.16}$& 1&06 & $^{+0.26}_{-0.20}$ & $^{+0.08}_{-0.09}$\\ 
500 & 0&35 & $^{+0.09}_{-0.06}$& 0&42 & $^{+0.11}_{-0.08}$ & $^{+0.04}_{-0.04}$\\ 
600 & 0&151 & $^{+0.038}_{-0.028}$& 0&183 & $^{+0.049}_{-0.036}$ & $^{+0.018}_{-0.018}$\\ 
700 & 0&068 & $^{+0.018}_{-0.013}$& 0&085 & $^{+0.024}_{-0.017}$ & $^{+0.010}_{-0.009}$\\ 
800 & 0&033 & $^{+0.009}_{-0.006}$& 0&041 & $^{+0.012}_{-0.009}$ & $^{+0.005}_{-0.005}$\\ 
900 & 0&0162 & $^{+0.0044}_{-0.0032}$& 0&0209 & $^{+0.0062}_{-0.0044}$ & $^{+0.0029}_{-0.0026}$\\ 
1000 & 0&0083 & $^{+0.0023}_{-0.0017}$& 0&0109 & $^{+0.0033}_{-0.0024}$ & $^{+0.0017}_{-0.0015}$\\ 
1100 & 0&0044 & $^{+0.0012}_{-0.0009}$& 0&0059 & $^{+0.0018}_{-0.0013}$ & $^{+0.0010}_{-0.0008}$\\ 
1200 & 0&0023 & $^{+0.0007}_{-0.0005}$& 0&0032 & $^{+0.0010}_{-0.0007}$ & $^{+0.0006}_{-0.0005}$\\ 
1300 & 0&00128 & $^{+0.00037}_{-0.00027}$& 0&00180 & $^{+0.00059}_{-0.00041}$ & $^{+0.00036}_{-0.00028}$\\ 
1400 & 0&00071 & $^{+0.00021}_{-0.00015}$& 0&00103 & $^{+0.00034}_{-0.00024}$ & $^{+0.00022}_{-0.00017}$\\ 
1500 & 0&00040 & $^{+0.00012}_{-0.00009}$& 0&00059 & $^{+0.00020}_{-0.00014}$ & $^{+0.00014}_{-0.00010}$\\ 
1600 & 0&00023 & $^{+0.00007}_{-0.00005}$& 0&00035 & $^{+0.00012}_{-0.00008}$ & $^{+0.00009}_{-0.00006}$\\ 
1700 & 0&000129 & $^{+0.000039}_{-0.000028}$& 0&000203 & $^{+0.000071}_{-0.000049}$ & $^{+0.000056}_{-0.000040}$\\ 
1800 & 0&000074 & $^{+0.000023}_{-0.000016}$& 0&000120 & $^{+0.000043}_{-0.000029}$ & $^{+0.000035}_{-0.000025}$\\ 
1900 & 0&000043 & $^{+0.000013}_{-0.000010}$& 0&000072 & $^{+0.000026}_{-0.000018}$ & $^{+0.000023}_{-0.000016}$\\ 
2000 & 0&000025 & $^{+0.000008}_{-0.000006}$& 0&000043 & $^{+0.000016}_{-0.000011}$ & $^{+0.000015}_{-0.000010}$\\ 
\hline 
\end{tabular} 

\end{center}
\caption{The sum of the leading order cross sections for $pp \rightarrow tW'^-$ and $pp \rightarrow \bar{t}W^+$  at the 7 TeV LHC. Where $W'$ couples to $ts$. \label{tab:lo-wp-t2}}
\end{table}

\begin{table}
\begin{center}
\begin{tabular}{|c||r@{.}lll|r@{.}lll|}
\hline
$M(W')$\,(GeV) & \multicolumn{4}{|c|}{$\sigma^{\mathrm{CT10}}_{\mathrm{nlo}} (tW'^{-}+\bar{t}W'^{+})$\,(pb)}& \multicolumn{4}{|c|}{$\sigma^{\mathrm{MSTW\ 2008\ NLO}}_{\mathrm{nlo}} (tW'^-+\bar{t}W'^+)$\,(pb)}\\ 
\hline 
200 & 13&8 & $^{+ 0.9}_{- 1.0}$ & $^{+ 3.1}_{- 1.8}$& 14&4 & $^{+ 0.9}_{- 1.0}$ & $^{+ 0.9}_{- 0.8}$\\ 
300 & 3&99 & $^{+0.25}_{-0.29}$ & $^{+0.99}_{-0.60}$& 4&22 & $^{+0.27}_{-0.31}$ & $^{+0.30}_{-0.28}$\\ 
400 & 1&40 & $^{+0.09}_{-0.11}$ & $^{+0.38}_{-0.24}$& 1&49 & $^{+0.10}_{-0.12}$ & $^{+0.12}_{-0.12}$\\ 
500 & 0&55 & $^{+0.04}_{-0.04}$ & $^{+0.17}_{-0.11}$& 0&60 & $^{+0.04}_{-0.05}$ & $^{+0.05}_{-0.05}$\\ 
600 & 0&240 & $^{+0.017}_{-0.020}$ & $^{+0.081}_{-0.052}$& 0&259 & $^{+0.019}_{-0.022}$ & $^{+0.026}_{-0.026}$\\ 
700 & 0&111 & $^{+0.008}_{-0.010}$ & $^{+0.042}_{-0.027}$& 0&120 & $^{+0.009}_{-0.011}$ & $^{+0.014}_{-0.013}$\\ 
800 & 0&054 & $^{+0.004}_{-0.005}$ & $^{+0.023}_{-0.015}$& 0&058 & $^{+0.005}_{-0.006}$ & $^{+0.007}_{-0.007}$\\ 
900 & 0&0276 & $^{+0.0022}_{-0.0026}$ & $^{+0.0129}_{-0.0081}$& 0&0293 & $^{+0.0025}_{-0.0029}$ & $^{+0.0041}_{-0.0038}$\\ 
1000 & 0&0145 & $^{+0.0012}_{-0.0014}$ & $^{+0.0076}_{-0.0046}$& 0&0153 & $^{+0.0014}_{-0.0016}$ & $^{+0.0023}_{-0.0021}$\\ 
1100 & 0&0079 & $^{+0.0007}_{-0.0008}$ & $^{+0.0047}_{-0.0027}$& 0&0081 & $^{+0.0008}_{-0.0009}$ & $^{+0.0014}_{-0.0012}$\\ 
1200 & 0&0044 & $^{+0.0004}_{-0.0005}$ & $^{+0.0029}_{-0.0016}$& 0&0044 & $^{+0.0004}_{-0.0005}$ & $^{+0.0008}_{-0.0007}$\\ 
1300 & 0&00247 & $^{+0.00024}_{-0.00027}$ & $^{+0.00184}_{-0.00101}$& 0&00245 & $^{+0.00025}_{-0.00028}$ & $^{+0.00049}_{-0.00041}$\\ 
1400 & 0&00142 & $^{+0.00014}_{-0.00016}$ & $^{+0.00119}_{-0.00062}$& 0&00138 & $^{+0.00014}_{-0.00016}$ & $^{+0.00030}_{-0.00024}$\\ 
1500 & 0&00083 & $^{+0.00009}_{-0.00010}$ & $^{+0.00078}_{-0.00039}$& 0&00078 & $^{+0.00009}_{-0.00009}$ & $^{+0.00019}_{-0.00015}$\\ 
1600 & 0&00049 & $^{+0.00005}_{-0.00006}$ & $^{+0.00052}_{-0.00025}$& 0&00045 & $^{+0.00005}_{-0.00006}$ & $^{+0.00012}_{-0.00009}$\\ 
1700 & 0&000293 & $^{+0.000033}_{-0.000035}$ & $^{+0.000352}_{-0.000158}$& 0&000259 & $^{+0.000030}_{-0.000033}$ & $^{+0.000073}_{-0.000053}$\\ 
1800 & 0&000177 & $^{+0.000020}_{-0.000022}$ & $^{+0.000240}_{-0.000102}$& 0&000150 & $^{+0.000018}_{-0.000019}$ & $^{+0.000046}_{-0.000032}$\\ 
1900 & 0&000107 & $^{+0.000013}_{-0.000014}$ & $^{+0.000165}_{-0.000066}$& 0&000088 & $^{+0.000011}_{-0.000012}$ & $^{+0.000029}_{-0.000020}$\\ 
2000 & 0&000066 & $^{+0.000008}_{-0.000009}$ & $^{+0.000114}_{-0.000043}$& 0&000051 & $^{+0.000007}_{-0.000007}$ & $^{+0.000018}_{-0.000012}$\\ 
\hline 
\end{tabular} 

\end{center}
\caption{The sum of the next-to-leading order cross sections for $pp \rightarrow tW'^-$ and  $pp \rightarrow \bar{t}W^+$  at the 7 TeV LHC. Where $W'$ couples to $ts$. \label{tab:nlo-wp-t2}}
\end{table}

\begin{table}
\begin{center}
\begin{tabular}{|c||r@{.}ll|r@{.}lll|}
\hline
$M(W')$\,(GeV) & \multicolumn{3}{|c|}{$\sigma^{\mathrm{CTEQ6L1}}_{\mathrm{born}} (tW'^-+\bar{t}W^+)$\,(pb)}& \multicolumn{4}{|c|}{$\sigma^{\mathrm{MSTW\ 2008\ LO}}_{\mathrm{born}} (tW'^-+\bar{t}W^+)$\,(pb)}\\ 
\hline 
200 & 3&22 & $^{+0.37}_{-0.33}$& 4&08 & $^{+0.51}_{-0.47}$ & $^{+0.17}_{-0.17}$\\ 
300 & 0&88 & $^{+0.11}_{-0.10}$& 1&17 & $^{+0.17}_{-0.15}$ & $^{+0.06}_{-0.06}$\\ 
400 & 0&293 & $^{+0.042}_{-0.037}$& 0&406 & $^{+0.068}_{-0.057}$ & $^{+0.023}_{-0.023}$\\ 
500 & 0&110 & $^{+0.017}_{-0.015}$& 0&159 & $^{+0.029}_{-0.024}$ & $^{+0.010}_{-0.010}$\\ 
600 & 0&0451 & $^{+0.008}_{-0.006}$& 0&068 & $^{+0.013}_{-0.011}$ & $^{+0.005}_{-0.005}$\\ 
700 & 0&0198 & $^{+0.0035}_{-0.0029}$& 0&0310 & $^{+0.0065}_{-0.0052}$ & $^{+0.0026}_{-0.0025}$\\ 
800 & 0&0092 & $^{+0.0017}_{-0.0014}$& 0&0149 & $^{+0.0033}_{-0.0026}$ & $^{+0.0014}_{-0.0013}$\\ 
900 & 0&0044 & $^{+0.0009}_{-0.0007}$& 0&0075 & $^{+0.0017}_{-0.0014}$ & $^{+0.0008}_{-0.0007}$\\ 
1000 & 0&00221 & $^{+0.00044}_{-0.00035}$& 0&00386 & $^{+0.00094}_{-0.00072}$ & $^{+0.00044}_{-0.00042}$\\ 
1100 & 0&00113 & $^{+0.00023}_{-0.00018}$& 0&00205 & $^{+0.00052}_{-0.00039}$ & $^{+0.00026}_{-0.00024}$\\ 
1200 & 0&00060 & $^{+0.00013}_{-0.00010}$& 0&00112 & $^{+0.00029}_{-0.00022}$ & $^{+0.00015}_{-0.00014}$\\ 
1300 & 0&00032 & $^{+0.00007}_{-0.00005}$& 0&00062 & $^{+0.00017}_{-0.00012}$ & $^{+0.00009}_{-0.00008}$\\ 
1400 & 0&000174 & $^{+0.000038}_{-0.000030}$& 0&000348 & $^{+0.000097}_{-0.000072}$ & $^{+0.000056}_{-0.000051}$\\ 
1500 & 0&000096 & $^{+0.000021}_{-0.000017}$& 0&000198 & $^{+0.000057}_{-0.000042}$ & $^{+0.000034}_{-0.000031}$\\ 
1600 & 0&000054 & $^{+0.000012}_{-0.000010}$& 0&000114 & $^{+0.000034}_{-0.000025}$ & $^{+0.000021}_{-0.000019}$\\ 
1700 & 0&000031 & $^{+0.000007}_{-0.000005}$& 0&000066 & $^{+0.000020}_{-0.000015}$ & $^{+0.000013}_{-0.000012}$\\ 
1800 & 0&0000174 & $^{+0.0000041}_{-0.0000031}$& 0&0000389 & $^{+0.0000121}_{-0.0000087}$ & $^{+0.0000083}_{-0.0000073}$\\ 
1900 & 0&0000100 & $^{+0.0000024}_{-0.0000018}$& 0&0000229 & $^{+0.0000073}_{-0.0000052}$ & $^{+0.0000052}_{-0.0000045}$\\ 
2000 & 0&0000058 & $^{+0.0000014}_{-0.0000011}$& 0&0000136 & $^{+0.0000044}_{-0.0000031}$ & $^{+0.0000033}_{-0.0000028}$\\ 
\hline 
\end{tabular} 

\end{center}
\caption{The sum of the leading order cross sections for $pp \rightarrow tW'^-$ and  $pp \rightarrow \bar{t}W^+$ at the 7 TeV LHC. Where $W'$ couples to $tb$. \label{tab:llo-wp-tb}}
\end{table}

\begin{table}
\begin{center}
\begin{tabular}{|c||r@{.}lll|r@{.}lll|}
\hline
$M(W')$\,(GeV) & \multicolumn{4}{|c|}{$\sigma^{\mathrm{CT10}}_{\mathrm{nlo}} (tW'^- + \bar{t}W'^+)$\,(pb)}& \multicolumn{4}{|c|}{$\sigma^{\mathrm{MSTW\ 2008\ NLO}}_{\mathrm{nlo}} (tW'^-+\bar{t}W'^+)$\,(pb)}\\ 
\hline 
200 & 4&61 & $^{+0.29}_{-0.25}$ & $^{+0.57}_{-0.49}$& 4&98 & $^{+0.32}_{-0.27}$ & $^{+0.22}_{-0.29}$\\ 
300 & 1&30 & $^{+0.08}_{-0.06}$ & $^{+0.21}_{-0.18}$& 1&40 & $^{+0.08}_{-0.06}$ & $^{+0.08}_{-0.10}$\\ 
400 & 0&452 & $^{+0.024}_{-0.019}$ & $^{+0.089}_{-0.074}$& 0&480 & $^{+0.026}_{-0.021}$ & $^{+0.032}_{-0.039}$\\ 
500 & 0&178 & $^{+0.009}_{-0.008}$ & $^{+0.042}_{-0.034}$& 0&187 & $^{+0.010}_{-0.009}$ & $^{+0.015}_{-0.017}$\\ 
600 & 0&077 & $^{+0.004}_{-0.004}$ & $^{+0.022}_{-0.017}$& 0&078 & $^{+0.004}_{-0.004}$ & $^{+0.007}_{-0.008}$\\ 
700 & 0&0355 & $^{+0.0016}_{-0.0019}$ & $^{+0.0118}_{-0.0087}$& 0&0363 & $^{+0.0018}_{-0.0021}$ & $^{+0.0037}_{-0.0041}$\\ 
800 & 0&0173 & $^{+0.0008}_{-0.0010}$ & $^{+0.0066}_{-0.0047}$& 0&0174 & $^{+0.0009}_{-0.0011}$ & $^{+0.0020}_{-0.0022}$\\ 
900 & 0&0088 & $^{+0.0004}_{-0.0005}$ & $^{+0.0039}_{-0.0026}$& 0&0087 & $^{+0.0005}_{-0.0006}$ & $^{+0.0011}_{-0.0012}$\\ 
1000 & 0&00464 & $^{+0.00025}_{-0.00030}$ & $^{+0.00233}_{-0.00151}$& 0&00448 & $^{+0.00026}_{-0.00031}$ & $^{+0.00062}_{-0.00066}$\\ 
1100 & 0&00252 & $^{+0.00014}_{-0.00017}$ & $^{+0.00143}_{-0.00089}$& 0&00238 & $^{+0.00014}_{-0.00017}$ & $^{+0.00036}_{-0.00038}$\\ 
1200 & 0&00140 & $^{+0.00008}_{-0.00010}$ & $^{+0.00090}_{-0.00053}$& 0&00129 & $^{+0.00008}_{-0.00010}$ & $^{+0.00021}_{-0.00022}$\\ 
1300 & 0&00079 & $^{+0.00005}_{-0.00006}$ & $^{+0.00057}_{-0.00033}$& 0&00071 & $^{+0.00005}_{-0.00006}$ & $^{+0.00013}_{-0.00013}$\\ 
1400 & 0&000458 & $^{+0.000030}_{-0.000036}$ & $^{+0.000369}_{-0.000204}$& 0&000400 & $^{+0.000028}_{-0.000034}$ & $^{+0.000075}_{-0.000077}$\\ 
1500 & 0&000269 & $^{+0.000018}_{-0.000022}$ & $^{+0.000242}_{-0.000128}$& 0&000227 & $^{+0.000017}_{-0.000020}$ & $^{+0.000046}_{-0.000046}$\\ 
1600 & 0&000159 & $^{+0.000011}_{-0.000014}$ & $^{+0.000160}_{-0.000081}$& 0&000131 & $^{+0.000010}_{-0.000012}$ & $^{+0.000028}_{-0.000028}$\\ 
1700 & 0&000096 & $^{+0.000007}_{-0.000009}$ & $^{+0.000107}_{-0.000052}$& 0&000076 & $^{+0.000006}_{-0.000007}$ & $^{+0.000017}_{-0.000017}$\\ 
1800 & 0&000058 & $^{+0.000005}_{-0.000006}$ & $^{+0.000072}_{-0.000034}$& 0&000044 & $^{+0.000004}_{-0.000004}$ & $^{+0.000011}_{-0.000011}$\\ 
1900 & 0&0000352 & $^{+0.0000029}_{-0.0000035}$ & $^{+0.0000485}_{-0.0000219}$& 0&0000260 & $^{+0.0000023}_{-0.0000026}$ & $^{+0.0000067}_{-0.0000065}$\\ 
2000 & 0&0000215 & $^{+0.0000019}_{-0.0000022}$ & $^{+0.0000330}_{-0.0000142}$& 0&0000153 & $^{+0.0000014}_{-0.0000016}$ & $^{+0.0000042}_{-0.0000040}$\\ 
\hline 
\end{tabular} 

\end{center}
\caption{The sum of the next-to-leading order cross sections for $pp \rightarrow tW'^-$ and  $pp \rightarrow \bar{t}W^+$ at the 7 TeV LHC. Where $W'$ couples to $tb$. \label{tab:nlo-wp-tb}}
\end{table}

\begin{table}
\begin{center}
\begin{tabular}{|c||r@{.}ll|r@{.}lll|}
\hline
$M(Z')$\,(GeV) & \multicolumn{3}{|c|}{$\sigma^{\mathrm{CTEQ6L1}}_{\mathrm{born}} (tZ'+\bar{t}Z')$\,(pb)}& \multicolumn{4}{|c|}{$\sigma^{\mathrm{MSTW\ 2008\ LO}}_{\mathrm{born}} (tZ'+\bar{t}Z')$\,(pb)}\\ 
\hline 
200 & 5&31 & $^{+0.91}_{-0.74}$& 6&89 & $^{+1.26}_{-1.02}$ & $^{+0.30}_{-0.30}$\\ 
300 & 1&46 & $^{+0.27}_{-0.22}$& 2&00 & $^{+0.41}_{-0.32}$ & $^{+0.10}_{-0.10}$\\ 
400 & 0&49 & $^{+0.10}_{-0.08}$& 0&70 & $^{+0.16}_{-0.12}$ & $^{+0.04}_{-0.04}$\\ 
500 & 0&183 & $^{+0.039}_{-0.030}$& 0&279 & $^{+0.065}_{-0.050}$ & $^{+0.019}_{-0.019}$\\ 
600 & 0&075 & $^{+0.017}_{-0.013}$& 0&121 & $^{+0.030}_{-0.022}$ & $^{+0.009}_{-0.009}$\\ 
700 & 0&0331 & $^{+0.008}_{-0.006}$& 0&056 & $^{+0.014}_{-0.011}$ & $^{+0.005}_{-0.005}$\\ 
800 & 0&0154 & $^{+0.0036}_{-0.0027}$& 0&0273 & $^{+0.0073}_{-0.0054}$ & $^{+0.0027}_{-0.0026}$\\ 
900 & 0&0074 & $^{+0.0018}_{-0.0013}$& 0&0139 & $^{+0.0038}_{-0.0028}$ & $^{+0.0015}_{-0.0015}$\\ 
1000 & 0&0037 & $^{+0.0009}_{-0.0007}$& 0&0073 & $^{+0.0021}_{-0.0015}$ & $^{+0.0009}_{-0.0008}$\\ 
1100 & 0&0019 & $^{+0.0005}_{-0.0004}$& 0&0039 & $^{+0.0012}_{-0.0008}$ & $^{+0.0005}_{-0.0005}$\\ 
1200 & 0&00101 & $^{+0.00026}_{-0.00019}$& 0&00216 & $^{+0.00065}_{-0.00047}$ & $^{+0.00031}_{-0.00029}$\\ 
1300 & 0&00054 & $^{+0.00014}_{-0.00010}$& 0&00121 & $^{+0.00038}_{-0.00027}$ & $^{+0.00019}_{-0.00018}$\\ 
1400 & 0&00030 & $^{+0.00008}_{-0.00006}$& 0&00069 & $^{+0.00022}_{-0.00016}$ & $^{+0.00012}_{-0.00011}$\\ 
1500 & 0&000163 & $^{+0.000043}_{-0.000032}$& 0&000400 & $^{+0.000130}_{-0.000091}$ & $^{+0.000074}_{-0.000068}$\\ 
1600 & 0&000091 & $^{+0.000025}_{-0.000018}$& 0&000234 & $^{+0.000078}_{-0.000054}$ & $^{+0.000047}_{-0.000042}$\\ 
1700 & 0&000052 & $^{+0.000014}_{-0.000010}$& 0&000138 & $^{+0.000047}_{-0.000032}$ & $^{+0.000030}_{-0.000026}$\\ 
1800 & 0&000030 & $^{+0.000008}_{-0.000006}$& 0&000082 & $^{+0.000028}_{-0.000020}$ & $^{+0.000019}_{-0.000017}$\\ 
1900 & 0&0000169 & $^{+0.0000047}_{-0.0000034}$& 0&0000490 & $^{+0.0000173}_{-0.0000119}$ & $^{+0.0000121}_{-0.0000106}$\\ 
2000 & 0&0000098 & $^{+0.0000027}_{-0.0000020}$& 0&0000294 & $^{+0.0000106}_{-0.0000072}$ & $^{+0.0000077}_{-0.0000067}$\\ 
\hline 
\end{tabular} 

\end{center}
\caption{The sum of the leading order cross sections for $pp \rightarrow \bar{t}Z'$ and  $pp \rightarrow tZ'$ at the 7 TeV LHC. Where $Z'$ couples to $tc$. \label{tab:lo-zp-t2}}
\end{table}

\begin{table}
\begin{center}
\begin{tabular}{|c||r@{.}lll|r@{.}lll|}
\hline
$M(Z')$\,(GeV) & \multicolumn{4}{|c|}{$\sigma^{\mathrm{CT10}}_{\mathrm{nlo}} (tZ'+\bar{t}Z')$\,(pb)}& \multicolumn{4}{|c|}{$\sigma^{\mathrm{MSTW\ 2008\ NLO}}_{\mathrm{nlo}} (tZ'+\bar{t}Z')$\,(pb)}\\ 
\hline 
200 & 7&92 & $^{+0.44}_{-0.45}$ & $^{+0.98}_{-0.86}$& 8&41 & $^{+0.49}_{-0.50}$ & $^{+0.40}_{-0.51}$\\ 
300 & 2&26 & $^{+0.12}_{-0.13}$ & $^{+0.36}_{-0.31}$& 2&38 & $^{+0.13}_{-0.15}$ & $^{+0.14}_{-0.17}$\\ 
400 & 0&78 & $^{+0.04}_{-0.05}$ & $^{+0.16}_{-0.13}$& 0&82 & $^{+0.05}_{-0.05}$ & $^{+0.06}_{-0.07}$\\ 
500 & 0&310 & $^{+0.018}_{-0.021}$ & $^{+0.075}_{-0.062}$& 0&321 & $^{+0.019}_{-0.023}$ & $^{+0.026}_{-0.031}$\\ 
600 & 0&134 & $^{+0.008}_{-0.009}$ & $^{+0.039}_{-0.031}$& 0&137 & $^{+0.009}_{-0.010}$ & $^{+0.013}_{-0.015}$\\ 
700 & 0&062 & $^{+0.004}_{-0.005}$ & $^{+0.021}_{-0.016}$& 0&063 & $^{+0.004}_{-0.005}$ & $^{+0.007}_{-0.008}$\\ 
800 & 0&0306 & $^{+0.0020}_{-0.0024}$ & $^{+0.0120}_{-0.0087}$& 0&0303 & $^{+0.0021}_{-0.0025}$ & $^{+0.0036}_{-0.0041}$\\ 
900 & 0&0157 & $^{+0.0011}_{-0.0013}$ & $^{+0.0071}_{-0.0049}$& 0&0152 & $^{+0.0011}_{-0.0013}$ & $^{+0.0020}_{-0.0022}$\\ 
1000 & 0&0083 & $^{+0.0006}_{-0.0007}$ & $^{+0.0043}_{-0.0029}$& 0&0079 & $^{+0.0006}_{-0.0007}$ & $^{+0.0012}_{-0.0013}$\\ 
1100 & 0&00454 & $^{+0.00034}_{-0.00040}$ & $^{+0.00266}_{-0.00169}$& 0&00422 & $^{+0.00034}_{-0.00039}$ & $^{+0.00067}_{-0.00072}$\\ 
1200 & 0&00254 & $^{+0.00020}_{-0.00023}$ & $^{+0.00168}_{-0.00103}$& 0&00230 & $^{+0.00019}_{-0.00022}$ & $^{+0.00040}_{-0.00042}$\\ 
1300 & 0&00145 & $^{+0.00012}_{-0.00014}$ & $^{+0.00108}_{-0.00063}$& 0&00128 & $^{+0.00011}_{-0.00013}$ & $^{+0.00024}_{-0.00025}$\\ 
1400 & 0&00085 & $^{+0.00007}_{-0.00008}$ & $^{+0.00070}_{-0.00040}$& 0&00072 & $^{+0.00007}_{-0.00007}$ & $^{+0.00015}_{-0.00015}$\\ 
1500 & 0&00050 & $^{+0.00005}_{-0.00005}$ & $^{+0.00046}_{-0.00025}$& 0&00041 & $^{+0.00004}_{-0.00004}$ & $^{+0.00009}_{-0.00009}$\\ 
1600 & 0&000298 & $^{+0.000028}_{-0.000032}$ & $^{+0.000308}_{-0.000161}$& 0&000239 & $^{+0.000024}_{-0.000026}$ & $^{+0.000055}_{-0.000056}$\\ 
1700 & 0&000180 & $^{+0.000018}_{-0.000020}$ & $^{+0.000207}_{-0.000104}$& 0&000139 & $^{+0.000014}_{-0.000016}$ & $^{+0.000034}_{-0.000034}$\\ 
1800 & 0&000110 & $^{+0.000011}_{-0.000012}$ & $^{+0.000140}_{-0.000067}$& 0&000082 & $^{+0.000009}_{-0.000010}$ & $^{+0.000021}_{-0.000021}$\\ 
1900 & 0&000068 & $^{+0.000007}_{-0.000008}$ & $^{+0.000096}_{-0.000044}$& 0&000048 & $^{+0.000005}_{-0.000006}$ & $^{+0.000013}_{-0.000013}$\\ 
2000 & 0&0000417 & $^{+0.0000046}_{-0.0000050}$ & $^{+0.0000657}_{-0.0000290}$& 0&0000287 & $^{+0.0000033}_{-0.0000035}$ & $^{+0.0000084}_{-0.0000081}$\\ 
\hline 
\end{tabular} 

\end{center}
\caption{The sum of the next-to-leading order cross sections for $pp \rightarrow \bar{t}Z'$ and  $pp \rightarrow tZ'$ at the 7 TeV LHC. Where $Z'$ couples to $tc$. \label{tab:nlo-zp-t2}}
\end{table}

\clearpage
\subsection{Cross-section results at 8 TeV}
\label{app:8TeV}

In this section, we present cross-section results at the LHC, for a centre of
mass energy of 8 TeV. These can then be directly applied to current ATLAS and
CMS analyses, as in the case of the 7 TeV results. \\

Total LO cross-sections for top or anti-top production in association with a
$Z'$ boson can be found in table~\ref{tab:lo-zp-8}. Individual results for
$t$ or $\bar{t}$ production are in tables~\ref{tab:lo-zp-plus-8} 
and~\ref{tab:lo-zp-minus-8}. The corresponding LO results for $W'$ production
can be found in tables~\ref{tab:lo-wp-8}-\ref{tab:lo-wp-minus-8}, and NLO
results for all the above in 
tables~\ref{tab:nlo-zp-8}-\ref{tab:nlo-wp-minus-8}.\\

For completeness, we also present results for different choices of the 
partonic couplings $P_{tD}$ and $Q_{tU}$ (similarly to 
appendix~\ref{app:othercouplings}), in 
tables~\ref{tab:lo-zp-t2-8}-\ref{tab:nlo-wp-tb-8}. These results can be 
combined to yield (very good) approximate cross-sections for any model 
encapsulated by eq.~(\ref{couplings}), according to the prescription of 
eqs.~(\ref{totcoupl1}) and~(\ref{totcoupl2}).

\begin{table}
\begin{center}
\begin{tabular}{|c||r@{.}ll|r@{.}lll|}
\hline
$M(Z')$\,(GeV) & \multicolumn{3}{|c|}{$\sigma^{\mathrm{CTEQ6L1}}_{\mathrm{born}} (tZ'+\bar{t}Z')$\,(pb)}& \multicolumn{4}{|c|}{$\sigma^{\mathrm{MSTW\ 2008\ LO}}_{\mathrm{born}} (tZ'+\bar{t}Z')$\,(pb)}\\ 
\hline 
200 & 104&7 & $^{+ 20.7}_{- 16.1}$& 106&2 & $^{+ 22.0}_{- 17.0}$ & $^{+  1.1}_{-  1.4}$\\ 
300 & 36&98 & $^{+ 7.80}_{- 5.99}$& 37&66 & $^{+ 8.33}_{- 6.33}$ & $^{+ 0.47}_{- 0.51}$\\ 
400 & 15&47 & $^{+ 3.43}_{- 2.61}$& 15&84 & $^{+ 3.68}_{- 2.77}$ & $^{+ 0.22}_{- 0.23}$\\ 
500 & 7&218 & $^{+1.662}_{-1.254}$& 7&440 & $^{+1.804}_{-1.346}$ & $^{+0.116}_{-0.114}$\\ 
600 & 3&637 & $^{+0.864}_{-0.648}$& 3&775 & $^{+0.949}_{-0.703}$ & $^{+0.065}_{-0.063}$\\ 
700 & 1&941 & $^{+0.474}_{-0.354}$& 2&030 & $^{+0.526}_{-0.388}$ & $^{+0.038}_{-0.037}$\\ 
800 & 1&083 & $^{+0.271}_{-0.201}$& 1&142 & $^{+0.304}_{-0.223}$ & $^{+0.024}_{-0.023}$\\
\hline 
\end{tabular} 

\end{center}
\caption{The sum of the leading order cross sections for $pp \rightarrow \bar{t}Z'$ and  $pp \rightarrow tZ'$  at the 8 TeV LHC. \label{tab:lo-zp-8}}
\end{table}

\begin{table}
\begin{center}
\begin{tabular}{|c||r@{.}ll|r@{.}lll|}
\hline
$M(Z')$\,(GeV) & \multicolumn{3}{|c|}{$\sigma^{\mathrm{CTEQ6L1}}_{\mathrm{born}} (\bar{t}Z')$\,(pb)}& \multicolumn{4}{|c|}{$\sigma^{\mathrm{MSTW\ 2008\ LO}}_{\mathrm{born}} (\bar{t}Z')$\,(pb)}\\ 
\hline 
200 & 9&5 & $^{+2.0}_{-1.5}$& 11&6 & $^{+ 2.5}_{- 1.9}$ & $^{+ 0.4}_{- 0.4}$\\ 
300 & 2&89 & $^{+0.64}_{-0.48}$& 3&61 & $^{+0.84}_{-0.63}$ & $^{+0.14}_{-0.14}$\\ 
400 & 1&05 & $^{+0.24}_{-0.18}$& 1&35 & $^{+0.33}_{-0.25}$ & $^{+0.06}_{-0.06}$\\ 
500 & 0&433 & $^{+0.104}_{-0.078}$& 0&569 & $^{+0.146}_{-0.108}$ & $^{+0.028}_{-0.028}$\\ 
600 & 0&194 & $^{+0.048}_{-0.036}$& 0&261 & $^{+0.069}_{-0.051}$ & $^{+0.015}_{-0.014}$\\ 
700 & 0&093 & $^{+0.024}_{-0.018}$& 0&128 & $^{+0.035}_{-0.025}$ & $^{+0.008}_{-0.008}$\\ 
800 & 0&047 & $^{+0.012}_{-0.009}$& 0&066 & $^{+0.019}_{-0.013}$ & $^{+0.005}_{-0.004}$\\
\hline 
\end{tabular} 

\end{center}
\caption{Leading order cross sections for $pp \rightarrow \bar{t}Z'$ at the 8 TeV LHC. \label{tab:lo-zp-plus-8}}
\end{table}

\begin{table}
\begin{center}
\begin{tabular}{|c||r@{.}ll|r@{.}lll|}
\hline
$M(Z')$\,(GeV) & \multicolumn{3}{|c|}{$\sigma^{\mathrm{CTEQ6L1}}_{\mathrm{born}} (tZ')$\,(pb)}& \multicolumn{4}{|c|}{$\sigma^{\mathrm{MSTW\ 2008\ LO}}_{\mathrm{born}} (tZ')$\,(pb)}\\ 
\hline 
200 & 95&2 & $^{+18.8}_{-14.6}$& 94&6 & $^{+19.5}_{-15.1}$ & $^{+ 1.1}_{- 1.3}$\\ 
300 & 34&1 & $^{+ 7.2}_{- 5.5}$& 34&1 & $^{+ 7.5}_{- 5.7}$ & $^{+ 0.4}_{- 0.5}$\\ 
400 & 14&42 & $^{+ 3.18}_{- 2.42}$& 14&49 & $^{+ 3.35}_{- 2.53}$ & $^{+ 0.21}_{- 0.21}$\\ 
500 & 6&79 & $^{+1.56}_{-1.18}$& 6&87 & $^{+1.66}_{-1.24}$ & $^{+0.11}_{-0.10}$\\ 
600 & 3&44 & $^{+0.82}_{-0.61}$& 3&52 & $^{+0.88}_{-0.65}$ & $^{+0.06}_{-0.06}$\\ 
700 & 1&848 & $^{+0.450}_{-0.336}$& 1&903 & $^{+0.491}_{-0.362}$ & $^{+0.035}_{-0.033}$\\ 
800 & 1&036 & $^{+0.259}_{-0.192}$& 1&076 & $^{+0.286}_{-0.209}$ & $^{+0.022}_{-0.021}$\\
\hline 
\end{tabular} 

\end{center}
\caption{Leading order cross sections for $pp \rightarrow tZ'$ at the 8 TeV LHC. \label{tab:lo-zp-minus-8}}
\end{table}

\begin{table}
\begin{center}
\begin{tabular}{|c||r@{.}ll|r@{.}lll|}
\hline
$M(W')$\,(GeV) & \multicolumn{3}{|c|}{$\sigma^{\mathrm{CTEQ6L1}}_{\mathrm{born}} (tW'^-+\bar{t}W'^+)$\,(pb)}& \multicolumn{4}{|c|}{$\sigma^{\mathrm{MSTW\ 2008\ LO}}_{\mathrm{born}} (tW'^-+\bar{t}W'^+)$\,(pb)}\\ 
\hline 
200 & 57&7 & $^{+11.9}_{- 9.2}$& 58&7 & $^{+12.8}_{- 9.8}$ & $^{+ 0.9}_{- 0.9}$\\ 
300 & 19&43 & $^{+ 4.27}_{- 3.25}$& 19&81 & $^{+ 4.60}_{- 3.46}$ & $^{+ 0.33}_{- 0.31}$\\ 
400 & 7&79 & $^{+1.79}_{-1.35}$& 7&96 & $^{+1.94}_{-1.45}$ & $^{+0.15}_{-0.13}$\\ 
500 & 3&49 & $^{+0.84}_{-0.62}$& 3&58 & $^{+0.91}_{-0.67}$ & $^{+0.08}_{-0.07}$\\ 
600 & 1&692 & $^{+0.418}_{-0.310}$& 1&743 & $^{+0.460}_{-0.337}$ & $^{+0.040}_{-0.035}$\\ 
700 & 0&871 & $^{+0.221}_{-0.163}$& 0&902 & $^{+0.245}_{-0.179}$ & $^{+0.023}_{-0.020}$\\ 
800 & 0&470 & $^{+0.122}_{-0.090}$& 0&488 & $^{+0.137}_{-0.099}$ & $^{+0.014}_{-0.012}$\\
\hline 
\end{tabular} 

\end{center}
\caption{The sum of the leading order cross sections for $pp \rightarrow tW'^-$ and $pp \rightarrow \bar{t}W'^+$ at the 8 TeV LHC. \label{tab:lo-wp-8}}
\end{table}

\begin{table}
\begin{center}
\begin{tabular}{|c||r@{.}ll|r@{.}lll|}
\hline
$M(W'^{+})$\,(GeV) & \multicolumn{3}{|c|}{$\sigma^{\mathrm{CTEQ6L1}}_{\mathrm{born}} (\bar{t}W'^{+})$\,(pb)}& \multicolumn{4}{|c|}{$\sigma^{\mathrm{MSTW\ 2008\ LO}}_{\mathrm{born}} (\bar{t}W'^{+})$\,(pb)}\\ 
\hline 
200 & 12&2 & $^{+ 2.6}_{- 2.0}$& 13&7 & $^{+ 3.1}_{- 2.3}$ & $^{+ 0.5}_{- 0.5}$\\ 
300 & 3&76 & $^{+0.85}_{-0.64}$& 4&30 & $^{+1.03}_{-0.77}$ & $^{+0.19}_{-0.16}$\\ 
400 & 1&39 & $^{+0.33}_{-0.25}$& 1&61 & $^{+0.41}_{-0.30}$ & $^{+0.08}_{-0.07}$\\ 
500 & 0&575 & $^{+0.142}_{-0.106}$& 0&674 & $^{+0.179}_{-0.131}$ & $^{+0.037}_{-0.030}$\\ 
600 & 0&259 & $^{+0.066}_{-0.049}$& 0&306 & $^{+0.084}_{-0.061}$ & $^{+0.018}_{-0.015}$\\ 
700 & 0&124 & $^{+0.033}_{-0.024}$& 0&148 & $^{+0.042}_{-0.030}$ & $^{+0.010}_{-0.008}$\\ 
800 & 0&062 & $^{+0.017}_{-0.012}$& 0&075 & $^{+0.022}_{-0.016}$ & $^{+0.006}_{-0.004}$\\
\hline 
\end{tabular} 

\end{center}
\caption{Leading order cross sections for $pp \rightarrow \bar{t}W'^{+}$ at the 8 TeV LHC. \label{tab:lo-wp-plus-8}}
\end{table}

\begin{table}
\begin{center}
\begin{tabular}{|c||r@{.}ll|r@{.}lll|}
\hline
$M(W'^{-})$\,(GeV) & \multicolumn{3}{|c|}{$\sigma^{\mathrm{CTEQ6L1}}_{\mathrm{born}} (tW'^{-})$\,(pb)}& \multicolumn{4}{|c|}{$\sigma^{\mathrm{MSTW\ 2008\ LO}}_{\mathrm{born}} (tW'^{-})$\,(pb)}\\ 
\hline 
200 & 45&5 & $^{+ 9.3}_{- 7.2}$& 45&0 & $^{+ 9.7}_{- 7.4}$ & $^{+ 0.8}_{- 0.8}$\\ 
300 & 15&68 & $^{+ 3.42}_{- 2.61}$& 15&51 & $^{+ 3.57}_{- 2.69}$ & $^{+ 0.29}_{- 0.29}$\\ 
400 & 6&40 & $^{+1.47}_{-1.11}$& 6&35 & $^{+1.54}_{-1.15}$ & $^{+0.13}_{-0.13}$\\ 
500 & 2&91 & $^{+0.69}_{-0.52}$& 2&91 & $^{+0.73}_{-0.54}$ & $^{+0.07}_{-0.06}$\\ 
600 & 1&433 & $^{+0.352}_{-0.262}$& 1&437 & $^{+0.375}_{-0.276}$ & $^{+0.035}_{-0.032}$\\ 
700 & 0&747 & $^{+0.188}_{-0.139}$& 0&754 & $^{+0.203}_{-0.148}$ & $^{+0.020}_{-0.018}$\\ 
800 & 0&407 & $^{+0.105}_{-0.077}$& 0&414 & $^{+0.115}_{-0.083}$ & $^{+0.012}_{-0.011}$\\
\hline 
\end{tabular} 

\end{center}
\caption{Leading order cross sections for $pp \rightarrow tW'^{-}$ at the 8 TeV LHC. \label{tab:lo-wp-minus-8}}
\end{table}

\begin{table}
\begin{center}
\begin{tabular}{|c||r@{.}lll|r@{.}lll|}
\hline
$M(Z')$\,(GeV) & \multicolumn{4}{|c|}{$\sigma^{\mathrm{CT10}}_{\mathrm{nlo}} (tZ'+\bar{t}Z')$\,(pb)}& \multicolumn{4}{|c|}{$\sigma^{\mathrm{MSTW\ 2008\ NLO}}_{\mathrm{nlo}} (tZ'+\bar{t}Z')$\,(pb)}\\ 
\hline 
200 & 134&7 & $^{+  8.8}_{-  9.1}$ & $^{+  3.2}_{-  3.6}$& 137&7 & $^{+  9.2}_{-  9.6}$ & $^{+  1.3}_{-  0.9}$\\ 
300 & 48&43 & $^{+ 3.08}_{- 3.40}$ & $^{+ 1.31}_{- 1.39}$& 49&48 & $^{+ 3.25}_{- 3.55}$ & $^{+ 0.48}_{- 0.32}$\\ 
400 & 20&63 & $^{+ 1.32}_{- 1.50}$ & $^{+ 0.69}_{- 0.67}$& 21&07 & $^{+ 1.39}_{- 1.57}$ & $^{+ 0.23}_{- 0.16}$\\ 
500 & 9&80 & $^{+0.64}_{-0.75}$ & $^{+0.40}_{-0.37}$& 10&00 & $^{+0.67}_{-0.78}$ & $^{+0.13}_{-0.10}$\\ 
600 & 5&03 & $^{+0.34}_{-0.40}$ & $^{+0.24}_{-0.22}$& 5&12 & $^{+0.35}_{-0.41}$ & $^{+0.08}_{-0.06}$\\ 
700 & 2&73 & $^{+0.19}_{-0.22}$ & $^{+0.16}_{-0.14}$& 2&78 & $^{+0.20}_{-0.23}$ & $^{+0.05}_{-0.04}$\\ 
800 & 1&549 & $^{+0.111}_{-0.131}$ & $^{+0.104}_{-0.091}$& 1&573 & $^{+0.116}_{-0.135}$ & $^{+0.033}_{-0.028}$\\
\hline 
\end{tabular} 

\end{center}
\caption{The sum of the next-to-leading order cross sections for $pp \rightarrow tZ'$ and $pp \rightarrow \bar{t}Z'$ at the 8 TeV LHC. \label{tab:nlo-zp-8}}
\end{table}

\begin{table}
\begin{center}
\begin{tabular}{|c||r@{.}lll|r@{.}lll|}
\hline
$M(Z')$\,(GeV) & \multicolumn{4}{|c|}{$\sigma^{\mathrm{CT10}}_{\mathrm{nlo}} (\bar{t}Z')$\,(pb)}& \multicolumn{4}{|c|}{$\sigma^{\mathrm{MSTW\ 2008\ NLO}}_{\mathrm{nlo}} (\bar{t}Z')$\,(pb)}\\ 
\hline 
200 & 13&2 & $^{+ 0.8}_{- 0.9}$ & $^{+ 1.1}_{- 1.2}$& 14&8 & $^{+ 1.0}_{- 1.1}$ & $^{+ 0.5}_{- 0.4}$\\ 
300 & 4&11 & $^{+0.26}_{-0.29}$ & $^{+0.39}_{-0.42}$& 4&63 & $^{+0.30}_{-0.34}$ & $^{+0.17}_{-0.16}$\\ 
400 & 1&54 & $^{+0.10}_{-0.12}$ & $^{+0.17}_{-0.18}$& 1&74 & $^{+0.12}_{-0.14}$ & $^{+0.08}_{-0.07}$\\ 
500 & 0&65 & $^{+0.04}_{-0.05}$ & $^{+0.09}_{-0.08}$& 0&74 & $^{+0.05}_{-0.06}$ & $^{+0.04}_{-0.04}$\\ 
600 & 0&299 & $^{+0.021}_{-0.025}$ & $^{+0.046}_{-0.043}$& 0&339 & $^{+0.025}_{-0.029}$ & $^{+0.020}_{-0.019}$\\ 
700 & 0&147 & $^{+0.011}_{-0.013}$ & $^{+0.026}_{-0.023}$& 0&167 & $^{+0.013}_{-0.015}$ & $^{+0.011}_{-0.010}$\\ 
800 & 0&076 & $^{+0.006}_{-0.007}$ & $^{+0.015}_{-0.013}$& 0&086 & $^{+0.007}_{-0.008}$ & $^{+0.007}_{-0.006}$\\
\hline 
\end{tabular} 

\end{center}
\caption{Next-to-leading order cross sections for $pp \rightarrow \bar{t}Z'$ at the 8 TeV LHC. \label{tab:nlo-zp-plus-8}}
\end{table}

\begin{table}
\begin{center}
\begin{tabular}{|c||r@{.}lll|r@{.}lll|}
\hline
$M(Z')$\,(GeV) & \multicolumn{4}{|c|}{$\sigma^{\mathrm{CT10}}_{\mathrm{nlo}} (tZ')$\,(pb)}& \multicolumn{4}{|c|}{$\sigma^{\mathrm{MSTW\ 2008\ NLO}}_{\mathrm{nlo}} (tZ')$\,(pb)}\\ 
\hline 
200 & 121&5 & $^{+  7.9}_{-  8.2}$ & $^{+  3.0}_{-  3.3}$& 122&9 & $^{+  8.3}_{-  8.5}$ & $^{+  1.4}_{-  1.0}$\\ 
300 & 44&33 & $^{+ 2.81}_{- 3.11}$ & $^{+ 1.21}_{- 1.29}$& 44&86 & $^{+ 2.94}_{- 3.21}$ & $^{+ 0.50}_{- 0.34}$\\ 
400 & 19&10 & $^{+ 1.21}_{- 1.39}$ & $^{+ 0.62}_{- 0.62}$& 19&33 & $^{+ 1.27}_{- 1.44}$ & $^{+ 0.22}_{- 0.16}$\\ 
500 & 9&15 & $^{+0.60}_{-0.69}$ & $^{+0.36}_{-0.34}$& 9&26 & $^{+0.62}_{-0.71}$ & $^{+0.12}_{-0.09}$\\ 
600 & 4&73 & $^{+0.32}_{-0.37}$ & $^{+0.22}_{-0.20}$& 4&78 & $^{+0.33}_{-0.38}$ & $^{+0.07}_{-0.06}$\\ 
700 & 2&58 & $^{+0.18}_{-0.21}$ & $^{+0.14}_{-0.13}$& 2&61 & $^{+0.18}_{-0.22}$ & $^{+0.04}_{-0.04}$\\ 
800 & 1&473 & $^{+0.106}_{-0.124}$ & $^{+0.094}_{-0.085}$& 1&486 & $^{+0.110}_{-0.128}$ & $^{+0.029}_{-0.025}$\\
\hline 
\end{tabular} 

\end{center}
\caption{Next-to-leading order cross sections for $pp \rightarrow tZ'$ at the 8 TeV LHC. \label{tab:nlo-zp-minus-8}}
\end{table}

\begin{table}
\begin{center}
\begin{tabular}{|c||r@{.}lll|r@{.}lll|}
\hline
$M(W')$\,(GeV) & \multicolumn{4}{|c|}{$\sigma^{\mathrm{CT10}}_{\mathrm{nlo}} (tW'^-+\bar{t}W^+)$\,(pb)}& \multicolumn{4}{|c|}{$\sigma^{\mathrm{MSTW\ 2008\ NLO}}_{\mathrm{nlo}} (tW'^-+\bar{t}W'^+)$\,(pb)}\\ 
\hline 
200 & 76&0 & $^{+ 4.9}_{- 5.3}$ & $^{+ 2.8}_{- 3.2}$& 78&1 & $^{+ 5.3}_{- 5.6}$ & $^{+ 0.9}_{- 0.8}$\\ 
300 & 26&08 & $^{+ 1.67}_{- 1.89}$ & $^{+ 1.13}_{- 1.22}$& 26&66 & $^{+ 1.77}_{- 1.99}$ & $^{+ 0.34}_{- 0.33}$\\ 
400 & 10&66 & $^{+ 0.69}_{- 0.80}$ & $^{+ 0.55}_{- 0.56}$& 10&83 & $^{+ 0.73}_{- 0.84}$ & $^{+ 0.16}_{- 0.16}$\\ 
500 & 4&88 & $^{+0.33}_{-0.38}$ & $^{+0.30}_{-0.29}$& 4&92 & $^{+0.34}_{-0.40}$ & $^{+0.09}_{-0.09}$\\ 
600 & 2&42 & $^{+0.17}_{-0.20}$ & $^{+0.17}_{-0.16}$& 2&42 & $^{+0.17}_{-0.20}$ & $^{+0.05}_{-0.05}$\\ 
700 & 1&270 & $^{+0.091}_{-0.108}$ & $^{+0.104}_{-0.094}$& 1&262 & $^{+0.094}_{-0.110}$ & $^{+0.031}_{-0.031}$\\ 
800 & 0&699 & $^{+0.052}_{-0.061}$ & $^{+0.066}_{-0.058}$& 0&689 & $^{+0.053}_{-0.062}$ & $^{+0.019}_{-0.020}$\\
\hline 
\end{tabular} 

\end{center}
\caption{The sum of the next-to-leading order cross sections for $pp \rightarrow tW'^-$ and  $pp \rightarrow \bar{t}W'^+$ at the 8 TeV LHC. \label{tab:nlo-wp-8}}
\end{table}

\begin{table}
\begin{center}
\begin{tabular}{|c||r@{.}lll|r@{.}lll|}
\hline
$M(W'^{+})$\,(GeV) & \multicolumn{4}{|c|}{$\sigma^{\mathrm{CT10}}_{\mathrm{nlo}} (\bar{t}W'^{+})$\,(pb)}& \multicolumn{4}{|c|}{$\sigma^{\mathrm{MSTW\ 2008\ NLO}}_{\mathrm{nlo}} (\bar{t}W'^{+})$\,(pb)}\\ 
\hline 
200 & 15&8 & $^{+ 1.0}_{- 1.1}$ & $^{+ 1.3}_{- 1.7}$& 17&2 & $^{+ 1.2}_{- 1.3}$ & $^{+ 0.5}_{- 0.5}$\\ 
300 & 4&94 & $^{+0.32}_{-0.36}$ & $^{+0.49}_{-0.61}$& 5&38 & $^{+0.36}_{-0.41}$ & $^{+0.20}_{-0.20}$\\ 
400 & 1&86 & $^{+0.12}_{-0.14}$ & $^{+0.22}_{-0.26}$& 2&01 & $^{+0.14}_{-0.16}$ & $^{+0.09}_{-0.09}$\\ 
500 & 0&78 & $^{+0.05}_{-0.06}$ & $^{+0.11}_{-0.12}$& 0&84 & $^{+0.06}_{-0.07}$ & $^{+0.04}_{-0.04}$\\ 
600 & 0&359 & $^{+0.026}_{-0.030}$ & $^{+0.057}_{-0.063}$& 0&383 & $^{+0.029}_{-0.034}$ & $^{+0.022}_{-0.022}$\\ 
700 & 0&175 & $^{+0.013}_{-0.015}$ & $^{+0.032}_{-0.034}$& 0&185 & $^{+0.015}_{-0.017}$ & $^{+0.012}_{-0.012}$\\ 
800 & 0&090 & $^{+0.007}_{-0.008}$ & $^{+0.019}_{-0.019}$& 0&093 & $^{+0.008}_{-0.009}$ & $^{+0.007}_{-0.007}$\\
\hline 
\end{tabular} 

\end{center}
\caption{Next-to-leading order cross sections for $pp \rightarrow \bar{t}W'^{+}$ at the 8 TeV LHC. \label{tab:nlo-wp-plus-8}}
\end{table}

\begin{table}
\begin{center}
\begin{tabular}{|c||r@{.}lll|r@{.}lll|}
\hline
$M(W'^{-})$\,(GeV) & \multicolumn{4}{|c|}{$\sigma^{\mathrm{CT10}}_{\mathrm{nlo}} (tW'^{-})$\,(pb)}& \multicolumn{4}{|c|}{$\sigma^{\mathrm{MSTW\ 2008\ NLO}}_{\mathrm{nlo}} (tW'^{-})$\,(pb)}\\ 
\hline 
200 & 60&1 & $^{+ 4.0}_{- 4.2}$ & $^{+ 2.6}_{- 2.6}$& 60&8 & $^{+ 4.2}_{- 4.4}$ & $^{+ 0.8}_{- 0.8}$\\ 
300 & 21&11 & $^{+ 1.36}_{- 1.53}$ & $^{+ 1.03}_{- 1.01}$& 21&25 & $^{+ 1.42}_{- 1.57}$ & $^{+ 0.31}_{- 0.30}$\\ 
400 & 8&80 & $^{+0.58}_{-0.66}$ & $^{+0.49}_{-0.46}$& 8&81 & $^{+0.60}_{-0.68}$ & $^{+0.14}_{-0.14}$\\ 
500 & 4&09 & $^{+0.27}_{-0.32}$ & $^{+0.26}_{-0.24}$& 4&07 & $^{+0.28}_{-0.33}$ & $^{+0.07}_{-0.07}$\\ 
600 & 2&05 & $^{+0.14}_{-0.17}$ & $^{+0.15}_{-0.14}$& 2&03 & $^{+0.15}_{-0.17}$ & $^{+0.04}_{-0.04}$\\ 
700 & 1&094 & $^{+0.078}_{-0.092}$ & $^{+0.091}_{-0.080}$& 1&076 & $^{+0.080}_{-0.093}$ & $^{+0.025}_{-0.025}$\\ 
800 & 0&609 & $^{+0.045}_{-0.053}$ & $^{+0.057}_{-0.050}$& 0&595 & $^{+0.046}_{-0.053}$ & $^{+0.016}_{-0.016}$\\
\hline 
\end{tabular} 

\end{center}
\caption{Next-to-leading order cross sections for $pp \rightarrow tW'^{-}$ at the 8 TeV LHC. \label{tab:nlo-wp-minus-8}}
\end{table}

\begin{table}
\begin{center}
\begin{tabular}{|c||r@{.}ll|r@{.}lll|}
\hline
$M(Z')$\,(GeV) & \multicolumn{3}{|c|}{$\sigma^{\mathrm{CTEQ6L1}}_{\mathrm{born}} (tZ'+\bar{t}Z')$\,(pb)}& \multicolumn{4}{|c|}{$\sigma^{\mathrm{MSTW\ 2008\ LO}}_{\mathrm{born}} (tZ'+\bar{t}Z')$\,(pb)}\\ 
\hline 
200 & 8&1 & $^{+1.3}_{-1.1}$& 10&1 & $^{+ 1.7}_{- 1.4}$ & $^{+ 0.4}_{- 0.4}$\\ 
300 & 2&32 & $^{+0.41}_{-0.33}$& 3&07 & $^{+0.58}_{-0.47}$ & $^{+0.14}_{-0.14}$\\ 
400 & 0&81 & $^{+0.15}_{-0.12}$& 1&12 & $^{+0.23}_{-0.18}$ & $^{+0.06}_{-0.06}$\\ 
500 & 0&317 & $^{+0.063}_{-0.050}$& 0&461 & $^{+0.101}_{-0.078}$ & $^{+0.028}_{-0.028}$\\ 
600 & 0&136 & $^{+0.028}_{-0.022}$& 0&207 & $^{+0.048}_{-0.036}$ & $^{+0.014}_{-0.014}$\\ 
700 & 0&063 & $^{+0.014}_{-0.010}$& 0&100 & $^{+0.024}_{-0.018}$ & $^{+0.008}_{-0.008}$\\ 
800 & 0&030 & $^{+0.007}_{-0.005}$& 0&050 & $^{+0.013}_{-0.009}$ & $^{+0.004}_{-0.004}$\\
\hline 
\end{tabular} 

\end{center}
\caption{The sum of the leading order cross sections for $pp \rightarrow tZ'$ and  $pp \rightarrow \bar{t}Z'$  at the 8 TeV LHC. Where $Z'$ couples to $tc$. \label{tab:lo-zp-t2-8}}
\end{table}

\begin{table}
\begin{center}
\begin{tabular}{|c||r@{.}lll|r@{.}lll|}
\hline
$M(Z')$\,(GeV) & \multicolumn{4}{|c|}{$\sigma^{\mathrm{CT10}}_{\mathrm{nlo}} (tZ'+\bar{t}Z')$\,(pb)}& \multicolumn{4}{|c|}{$\sigma^{\mathrm{MSTW\ 2008\ NLO}}_{\mathrm{nlo}} (tZ'+\bar{t}Z')$\,(pb)}\\ 
\hline 
200 & 11&8 & $^{+ 0.6}_{- 0.6}$ & $^{+ 1.3}_{- 1.1}$& 12&5 & $^{+ 0.7}_{- 0.7}$ & $^{+ 0.5}_{- 0.7}$\\ 
300 & 3&50 & $^{+0.18}_{-0.20}$ & $^{+0.48}_{-0.42}$& 3&70 & $^{+0.20}_{-0.22}$ & $^{+0.19}_{-0.24}$\\ 
400 & 1&26 & $^{+0.07}_{-0.08}$ & $^{+0.22}_{-0.19}$& 1&33 & $^{+0.07}_{-0.08}$ & $^{+0.08}_{-0.10}$\\ 
500 & 0&518 & $^{+0.028}_{-0.033}$ & $^{+0.107}_{-0.090}$& 0&541 & $^{+0.031}_{-0.036}$ & $^{+0.039}_{-0.047}$\\ 
600 & 0&232 & $^{+0.013}_{-0.016}$ & $^{+0.057}_{-0.047}$& 0&240 & $^{+0.015}_{-0.017}$ & $^{+0.020}_{-0.024}$\\ 
700 & 0&112 & $^{+0.007}_{-0.008}$ & $^{+0.032}_{-0.025}$& 0&114 & $^{+0.007}_{-0.008}$ & $^{+0.011}_{-0.012}$\\ 
800 & 0&057 & $^{+0.004}_{-0.004}$ & $^{+0.019}_{-0.014}$& 0&057 & $^{+0.004}_{-0.004}$ & $^{+0.006}_{-0.007}$\\
\hline 
\end{tabular} 

\end{center}
\caption{The sum of the next-to-leading order cross sections for $pp \rightarrow tZ'$ and  $pp \rightarrow \bar{t}Z'$ at the 8 TeV LHC. Where $Z'$ couples to $tc$. \label{tab:nlo-zp-t2-8}}
\end{table}

\begin{table}
\begin{center}
\begin{tabular}{|c||r@{.}ll|r@{.}lll|}
\hline
$M(W')$\,(GeV) & \multicolumn{3}{|c|}{$\sigma^{\mathrm{CTEQ6L1}}_{\mathrm{born}} (tW'^-+\bar{t}W'^+)$\,(pb)}& \multicolumn{4}{|c|}{$\sigma^{\mathrm{MSTW\ 2008\ LO}}_{\mathrm{born}} (tW'^-+\bar{t}W'^+)$\,(pb)}\\ 
\hline 
200 & 13&5 & $^{+ 2.6}_{- 2.0}$& 15&0 & $^{+ 3.0}_{- 2.3}$ & $^{+ 0.8}_{- 0.8}$\\ 
300 & 4&05 & $^{+0.84}_{-0.65}$& 4&58 & $^{+0.99}_{-0.76}$ & $^{+0.29}_{-0.30}$\\ 
400 & 1&46 & $^{+0.32}_{-0.25}$& 1&68 & $^{+0.39}_{-0.29}$ & $^{+0.12}_{-0.12}$\\ 
500 & 0&59 & $^{+0.14}_{-0.10}$& 0&69 & $^{+0.13}_{-0.13}$ & $^{+0.06}_{-0.06}$\\ 
600 & 0&263 & $^{+0.063}_{-0.047}$& 0&313 & $^{+0.079}_{-0.059}$ & $^{+0.028}_{-0.028}$\\ 
700 & 0&124 & $^{+0.031}_{-0.023}$& 0&150 & $^{+0.040}_{-0.029}$ & $^{+0.015}_{-0.015}$\\ 
800 & 0&062 & $^{+0.016}_{-0.012}$& 0&076 & $^{+0.021}_{-0.015}$ & $^{+0.008}_{-0.008}$\\
\hline 
\end{tabular} 

\end{center}
\caption{The sum of the leading order cross sections for $pp \rightarrow tW'^-$ and   $pp \rightarrow \bar{t}W'^+$ and at the 8 TeV LHC. Where $W'$ couples to $ts$. \label{tab:lo-wp-t2-8}}
\end{table}

\begin{table}
\begin{center}
\begin{tabular}{|c||r@{.}lll|r@{.}lll|}
\hline
$M(W')$\,(GeV) & \multicolumn{4}{|c|}{$\sigma^{\mathrm{CT10}}_{\mathrm{nlo}} (tW'^-+\bar{t}W^+)$\,(pb)}& \multicolumn{4}{|c|}{$\sigma^{\mathrm{MSTW\ 2008\ NLO}}_{\mathrm{nlo}} (tW'^-+\bar{t}W^+)$\,(pb)}\\ 
\hline 
200 & 20&3 & $^{+ 1.3}_{- 1.3}$ & $^{+ 4.4}_{- 2.5}$& 21&0 & $^{+ 1.3}_{- 1.4}$ & $^{+ 1.2}_{- 1.0}$\\ 
300 & 6&1 & $^{+0.4}_{-0.4}$ & $^{+1.4}_{-0.8}$& 6&4 & $^{+0.4}_{-0.4}$ & $^{+0.4}_{-0.4}$\\ 
400 & 2&23 & $^{+0.14}_{-0.16}$ & $^{+0.56}_{-0.35}$& 2&36 & $^{+0.15}_{-0.17}$ & $^{+0.17}_{-0.17}$\\ 
500 & 0&92 & $^{+0.06}_{-0.07}$ & $^{+0.25}_{-0.16}$& 0&98 & $^{+0.07}_{-0.08}$ & $^{+0.08}_{-0.08}$\\ 
600 & 0&412 & $^{+0.028}_{-0.033}$ & $^{+0.125}_{-0.081}$& 0&443 & $^{+0.031}_{-0.036}$ & $^{+0.040}_{-0.040}$\\ 
700 & 0&198 & $^{+0.014}_{-0.016}$ & $^{+0.066}_{-0.043}$& 0&213 & $^{+0.016}_{-0.018}$ & $^{+0.021}_{-0.021}$\\ 
800 & 0&100 & $^{+0.007}_{-0.009}$ & $^{+0.037}_{-0.024}$& 0&108 & $^{+0.008}_{-0.010}$ & $^{+0.012}_{-0.012}$\\
\hline 
\end{tabular} 

\end{center}
\caption{The sum of the next-to-leading order cross sections for $pp \rightarrow tW'^-$ and  $pp \rightarrow \bar{t}W'^+$ at the 8 TeV LHC. Where $W'$ couples to $ts$. \label{tab:nlo-wp-t2-8}}
\end{table}

\begin{table}
\begin{center}
\begin{tabular}{|c||r@{.}ll|r@{.}lll|}
\hline
$M(W')$\,(GeV) & \multicolumn{3}{|c|}{$\sigma^{\mathrm{CTEQ6L1}}_{\mathrm{born}} (tW'^-+\bar{t}W'^+)$\,(pb)}& \multicolumn{4}{|c|}{$\sigma^{\mathrm{MSTW\ 2008\ LO}}_{\mathrm{born}} (tW'^-+\bar{t}W'^+)$\,(pb)}\\ 
\hline 
200 & 4&94 & $^{+0.57}_{-0.54}$& 6&10 & $^{+0.76}_{-0.68}$ & $^{+0.23}_{-0.23}$\\ 
300 & 1&42 & $^{+0.17}_{-0.16}$& 1&83 & $^{+0.24}_{-0.22}$ & $^{+0.08}_{-0.08}$\\ 
400 & 0&491 & $^{+0.066}_{-0.059}$& 0&658 & $^{+0.101}_{-0.087}$ & $^{+0.033}_{-0.034}$\\ 
500 & 0&193 & $^{+0.028}_{-0.025}$& 0&268 & $^{+0.045}_{-0.038}$ & $^{+0.016}_{-0.016}$\\ 
600 & 0&083 & $^{+0.013}_{-0.011}$& 0&119 & $^{+0.022}_{-0.018}$ & $^{+0.008}_{-0.008}$\\ 
700 & 0&0378 & $^{+0.006}_{-0.005}$& 0&056 & $^{+0.011}_{-0.009}$ & $^{+0.004}_{-0.004}$\\ 
800 & 0&0182 & $^{+0.0032}_{-0.0026}$& 0&0282 & $^{+0.0058}_{-0.0046}$ & $^{+0.0023}_{-0.0022}$\\
\hline 
\end{tabular} 

\end{center}
\caption{The sum of the leading order cross sections for $pp \rightarrow tW'^-$ and $pp \rightarrow \bar{t}W'^+$ at the 8 TeV LHC. Where $W'$ couples to $tb$. \label{tab:llo-wp-tb-8}}
\end{table}

\begin{table}
\begin{center}
\begin{tabular}{|c||r@{.}lll|r@{.}lll|}
\hline
$M(W')$\,(GeV) & \multicolumn{4}{|c|}{$\sigma^{\mathrm{CT10}}_{\mathrm{nlo}} (tW'^- + \bar{t}W'^+)$\,(pb)}& \multicolumn{4}{|c|}{$\sigma^{\mathrm{MSTW\ 2008\ NLO}}_{\mathrm{nlo}} (tW'^-+ \bar{t}W'^+)$\,(pb)}\\ 
\hline 
200 & 6&93 & $^{+0.45}_{-0.40}$ & $^{+0.74}_{-0.65}$& 7&50 & $^{+0.48}_{-0.42}$ & $^{+0.30}_{-0.39}$\\ 
300 & 2&04 & $^{+0.12}_{-0.10}$ & $^{+0.28}_{-0.24}$& 2&20 & $^{+0.13}_{-0.10}$ & $^{+0.11}_{-0.14}$\\ 
400 & 0&736 & $^{+0.040}_{-0.031}$ & $^{+0.125}_{-0.106}$& 0&787 & $^{+0.043}_{-0.034}$ & $^{+0.047}_{-0.057}$\\ 
500 & 0&301 & $^{+0.016}_{-0.013}$ & $^{+0.061}_{-0.051}$& 0&319 & $^{+0.017}_{-0.015}$ & $^{+0.022}_{-0.026}$\\ 
600 & 0&134 & $^{+0.007}_{-0.006}$ & $^{+0.032}_{-0.026}$& 0&141 & $^{+0.007}_{-0.007}$ & $^{+0.011}_{-0.013}$\\ 
700 & 0&0643 & $^{+0.0030}_{-0.0032}$ & $^{+0.0179}_{-0.0139}$& 0&0667 & $^{+0.0031}_{-0.0036}$ & $^{+0.0059}_{-0.0068}$\\ 
800 & 0&0325 & $^{+0.0015}_{-0.0017}$ & $^{+0.0104}_{-0.0078}$& 0&0333 & $^{+0.0016}_{-0.0019}$ & $^{+0.0033}_{-0.0037}$\\
\hline 
\end{tabular} 

\end{center}
\caption{The sum of the next-to-Leading order cross sections for $pp \rightarrow tW'^-$ and  $pp \rightarrow \bar{t}W'^+$ at the 8 TeV LHC. Where $W'$ couples to $tb$. \label{tab:nlo-wp-tb-8}}
\end{table}

\clearpage

\bibliography{refs.bib}

\end{document}